\documentclass{emulateapj}
\usepackage{apjfonts}
\newcommand{\scaleup}{\epsscale{1.1}}

\newcommand{\plotter}{\plotone}

\newcommand{\breaker}{}

\newcommand{\Mdot}{\dot{M}}

\newcommand{\etal}{et al.}

\newcommand{\mstar}{M_{\ast}}
\newcommand{\msun}{M_{\sun}}
\newcommand{\lstar}{L_{\ast}}

\newcommand{\qeos}{q_{\rm eos}}
\newcommand{\fgas}{f_{\rm gas}}
\newcommand{\mdyn}{M_{\rm dyn}}
\newcommand{\re}{R_{\rm e}}
\newcommand{\fsb}{f_{\rm sb}}
\newcommand{\fextra}{f_{\rm extra}}
\newcommand{\tilt}{\alpha}
\newcommand{\mtrue}{M_{\rm true}}
\newcommand{\paperone}{Paper \textrm{I}}
\newcommand{\papertwo}{Paper \textrm{II}}
\newcommand{\paperthree}{Paper \textrm{III}}
\newcommand{\mdynnorm}{3.8}

\shorttitle{Dissipation and the FP}
\shortauthors{Hopkins \etal}
\slugcomment{Submitted to ApJ, June 5, 2008}
\begin{document}

\title{Dissipation and the Fundamental Plane: Observational Tests}
\author{Philip F. Hopkins\altaffilmark{1}, 
Thomas J. Cox\altaffilmark{1,2}, 
Lars Hernquist\altaffilmark{1}
}
\altaffiltext{1}{Harvard-Smithsonian Center for Astrophysics, 
60 Garden Street, Cambridge, MA 02138}
\altaffiltext{2}{W.~M.\ Keck Postdoctoral Fellow at the 
Harvard-Smithsonian Center for Astrophysics}

\begin{abstract}

We develop and implement an observational test of the theoretical
notion that dissipation in major mergers of gas-rich galaxies produces
the fundamental plane (FP) and related correlations obeyed by
ellipticals.  Observations have shown that the ``tilt'' of the FP
involves more than a simple non-homology or stellar population effect:
lower-mass ellipticals have a higher ratio of stellar to dark matter
within their stellar effective radii.  Theoretical models have
attempted to explain this via dissipation: if ellipticals are formed
in major mergers of disks, then mergers between disks having a larger
gas content (typically observed to be lower-mass disks) will yield
remnants with a larger mass fraction formed in a central, compact
starburst, giving a smaller stellar $R_{e}$ and lower $M_{\rm tot}/
\mstar$ within that $R_{e}$.  Such starbursts leave a characteristic
imprint in the surface brightness profiles of ellipticals, in the form
of a central excess above the outer profile established by the
dissipationless, violent relaxation of disk stars.  In previous work,
we implemented a purely empirical method to use such features in the
observed profiles of ellipticals to robustly estimate the amount of
dissipation involved in the original spheroid-forming merger.
Applying this to a large sample of ellipticals with detailed kinematic
and photometric observations, we demonstrate that the location of
ellipticals on the FP and its tilt are in fact driven by dissipation.
We show that at fixed mass, ellipticals formed in more dissipational
events, as indicated by their observed profiles, are smaller and have
a lower ratio $M_{\rm tot}/\mstar$. {\em At the same (fixed) degree of
  dissipation, there is no tilt in the FP} -- i.e.\ ellipticals formed
with a similar level of dissipation have the same ratio of enclosed
stellar to total mass within $R_{e}$.

We further demonstrate that observations and these models obey the
``homology assumption,'' i.e.\ that the true enclosed mass
$\mtrue(R_{e})\propto \sigma^{2}\,R_{e}$.  Measured at the radii of
disks of the same mass, we show that ellipticals have the same total
enclosed masses as those disks -- i.e.\ that the FP tilt can be
effectively removed. Therefore, the fundamental plane tilt cannot
primarily owe to non-homology or to changes in the dark matter
distribution: it {\em must} arise as a result of a contraction of the
baryonic component relative to the dark matter in the process that
transforms disks to ellipticals, as predicted by dissipational
mergers. If we allow for the observed cosmological dependence of disk
gas fraction on mass, the observed FP, size-mass, and velocity
dispersion-mass correlations are reproduced by our models, as are the
observed homology constraints and profile shapes. {\em Dissipation is
both necessary and sufficient to explain the observed FP
correlations of ellipticals.}  These observations all favor theories
in which ellipticals are formed in major mergers of disks with gas
fractions, sizes, and dark matter content similar to that observed as
a function of mass in low-redshift disks: unusually compact disks are 
{\em not} required to make $\sim0.01-10\,L_{\ast}$ ellipticals.  We
present a number of associated predictions that can be used to further
test these assertions.

\end{abstract}


\section{Introduction}
\label{sec:intro}

Understanding the scaling relations between the photometric and
kinematic properties of galaxy spheroids -- their masses, sizes,
velocity dispersions, and luminosities -- is fundamental to explaining
the origin of early-type galaxies.  \citet{fj76} demonstrated that
ellipticals obey a relatively tight correlation between optical
luminosity and central velocity dispersion, and
\citet{kormendy77:correlations} found an analogous relationship
between their effective surface brightness and radii. With improved
observations and the advent of stellar population modeling, these
observed trends can be translated into robust correlations between
physical parameters: a velocity dispersion-stellar mass
($\sigma-\mstar$) and a size-stellar mass ($R_{e}-\mstar$) relation.

\citet{dd87:fp} and \citet{dressler87:fp} demonstrated that the
scatter in either the \citet{fj76} or \citet{kormendy77:correlations}
relation could be reduced by adopting a three parameter correlation of
the form $\log{(R_{e})}= a\,\log{(\sigma)} - 0.4\,b\,\mu_{e} + c$
(equivalently $R_{e}\propto\sigma^{a}\,I_{e}^{b}$), with best fit
scalings $a\sim1.3-1.4$, $b\sim-0.8$ to $-0.9$.  This defines the
``fundamental plane'' (FP) of elliptical galaxies: a correlation
relating stellar mass or luminosity (implicit in the surface
brightness), effective radius, and velocity dispersion (effectively
the dynamical mass of the system). With a small observed scatter
$\sim0.1\,$dex, the FP has presented as a long-standing, and still
unresolved challenge to observations and theoretical models of
spheroid formation.

In developing a physical understanding of the FP and associated
elliptical scaling laws, \citet{djorgovski:fp.tilt,
jorgensen:fp.scatter} and others demonstrated
that the FP could be represented as a ``tilted'' virial plane. If
ellipticals were perfectly homologous systems with constant stellar
mass-to-light ratios $\mstar/L$, then a virial correlation $L \propto
\mstar = k\,\sigma^{2}\,R_{e}/G \equiv \mdyn$, with constant integral
factor $k$, would be expected.  Since $I_{e} \propto L/R_{e}^{2}$,
this translates to an expected ``virial FP''
$R_{e}\propto\sigma^{2}\,I_{e}^{-1}$.  The observed FP is similar to 
this, but not exactly so; it is equivalent to and can be 
represented as a ``tilted'' version of this correlation, namely
\begin{equation}
\mdyn \propto \mstar^{1+\tilt}
\label{eqn:tilt}
\end{equation}
with some small, but non-zero $\alpha$. 
Equivalently, the difference between the best-fit observed FP (in any
projection) and the virial FP can be expressed as a mass dependent
mass-to-light (or, for our purposes, total mass-to-stellar mass) ratio
\begin{equation}
\frac{\mdyn}{\mstar}\propto \mstar^{\tilt}
\end{equation}
where the quantity $\tilt$ quantifies the tilt, or deviation of the FP
from the virial relation.  Various independent measurements find
similar values of $\tilt\approx0.2$ \citep[e.g.][]{pahre:nir.fp,
  gerhard:giant.ell.dynamics, borriello03,
  padmanabhan:mdyn.mstar.tilt,gallazzi06:ages}.  Although this is not
strictly identical to the best-fit relation $R_{e}\propto
\sigma^{a}\,I_{e}^{b}$ if both $a$ and $b$ are fit as free parameters,
multiple observations have shown that it is statistically an
equivalent representation (i.e.\ has the same scatter in {\em
  physical} quantities), and that there is no additional information
in the best-fit FP beyond this tilt (i.e.\ once this tilt is
accounted for, there is no additional systematic scaling in $R_{e}$ or
$\sigma$ that can reduce the scatter in predicting the other
quantities).

It is now well-established that part of the observed tilt in optical
bands is a consequence of stellar population effects: lower-mass
ellipticals tend to be younger, yielding lower stellar mass-to-light
ratios \citep[see e.g.][]{trager:ages}.  However, various constraints
imply that only a small fraction 
of the optical tilt owes to these effects \citep[see e.g.][]
{pahre:nir.fp,gerhard:giant.ell.dynamics,bertin:weak.homology,
borriello03,padmanabhan:mdyn.mstar.tilt,trujillo:non-homology,
gallazzi06:ages,vonderlinden:bcg.scaling.relations}.  For example,
in the $K$-band, the tilt is still substantial: observations indicate
$\mdyn\propto L_{K}^{1.25\pm0.05}$ \citep{pahre:nir.fp}, whereas the
systematic dependence of $\mstar/L_{K}$ is quite weak \citep[most
  estimates suggest $\mstar/L_{K}\propto L_{K}^{0.03}$;][]{bell:mfs}.
It is now possible to combine high-resolution spectra and stellar
population synthesis models, allowing reliable stellar mass estimates,
and almost all such studies yield a similar relation:
\begin{equation}
\mdyn \propto \mstar^{1.2},
\end{equation}
i.e.\ $\alpha\approx0.2$ as described above.
It has been demonstrated that this result is robust to the 
details of the stellar population model, spectral 
coverage, or even simplifying assumptions such as the use of 
a single color to derive a mean $\mstar/L$. 

There are only two reasonable explanations for this finding (some
combination of the two is also possible).  First, the true mass
enclosed within the stellar effective radius ($\mtrue$) could in fact
be proportional to the stellar mass $\mstar$, but owing to e.g.\
changes in the profile shape or kinematics of galaxies with mass
(traditional non-homology), the relation between actual mass and the
dynamical mass estimator $\mdyn\propto\sigma^{2}\,R_{e}/G$ is a
changing function of mass.  In other words, $\mdyn \propto
\mtrue^{1.2}$.  However, observations appear to rule out this
possibility, at least as the origin of most of the tilt.  Integral
modeling of the mass distribution from two-dimensional kinematic maps
\citep[which should recover any systematic difference between $\mdyn$
and $\mtrue$ without reference to any homology assumptions,
e.g.][]{cappellari:fp}, as well as mass distributions estimated from
gravitational lensing \citep{bolton:fp,bolton:fp.update,nipoti:homology.from.mp}, 
independently give $\mdyn
\propto \mtrue^{1.00\pm0.03}$.  That is, the allowed contribution of
non-homology to the FP tilt is small.

The only remaining explanation is that the FP tilt reflects a
meaningful physical change, namely that the ratio of total enclosed
mass within $R_{e}$ ($\mtrue$, represented reasonably well to within a
normalization constant by $\mdyn$) to the stellar mass is an
increasing function of mass ($\mtrue\propto\mstar^{1.2}$).  In other
words, low-mass ellipticals are more baryon-dominated within their
stellar $R_{e}$, and high-mass ellipticals have higher dark matter
fractions.  We emphasize that these constraints apply {\em within the
  stellar effective radii}. The change in dark matter fraction is not
required to be global: if one were to contract the stellar mass
distribution but keep the dark matter halo relatively fixed, for
example, it would significantly decrease the dark matter fraction (and
correspondingly $\mdyn/\mstar$) within $R_{e}$.

This trend is {\em contrary} to that followed by disks.  For disk
galaxies, an opposite (negative) tilt $\mdyn\propto\mstar^{0.7-0.8}$
($\tilt \approx-0.2 $ to $-0.3$) is observed \citep[see
  e.g.][]{persic96,belldejong:tf,shen:size.mass,courteau:disk.scalings} -- low mass disks (and dwarf
spheroidals) are the most dark-matter dominated systems
\citep[][]{persic88,persic90,persic96:data,persic96,borriello01}. Such a
scaling is expected if the properties of disks track those of their
dark matter halos: lower-mass halos are more compact \citep[][and
  references therein]{neto:concentrations}, and it is also
well-established that lower-mass disks experience less efficient star
formation \citep{belldejong:disk.sfh,gallazzi:ssps}.  Consequently,
disks and ellipticals have similar ratios $\mdyn/\mstar$ at high
masses ($\sim$a few $L_{\ast}$), but disks are more dark-matter
dominated (have much higher $\mdyn(R_{e})$) at low (stellar) masses.
This difference in scaling laws also relates to their stellar
size-mass relations: disks obey a shallow relation
$R_{e}\propto\mstar^{0.25-0.35}$, roughly consistent with the scaling
of halo effective radii as a function of mass, whereas spheroids obey
a much steeper relation $R_{e}\propto\mstar^{0.6}$
\citep{shen:size.mass}.  Again, at $\sim$a few $L_{\ast}$, disks and
ellipticals have similar sizes and densities, but at low masses ($\ll
L_{\ast}$), ellipticals are smaller (in their stellar/baryonic mass
distributions) and more dense.

This difference has, for $\sim30$ years, represented a major challenge
for theory -- especially models which posit that ellipticals are
formed through the merger of disk galaxies \citep[the ``merger
  hypothesis'';][]{toomre72,toomre77}.  In particular, it has been
noted that purely dissipationless mergers of stellar disks cannot
raise the mass and phase-space densities of ellipticals, and so cannot
change the scaling laws of ellipticals to something different from
disks \citep{ostriker80,carlberg:phase.space,gunn87}.

However, these arguments do not pertain if ellipticals are formed from
mergers of {\em gas-rich} disks.  Particularly at low masses,
where ellipticals are more compact than spirals, disks have a large
fraction of their mass in gas, so {\em mergers must account for
  dissipation}.  In a merger of two disks containing both gas and
stars, the stars are dissipationless and they cannot increase their
phase space density and so violently relax to a distribution with an
$R_{e}$ similar to the progenitor disks.  Gas, on the other hand, can
radiate, and tidal torques excited during a merger can remove its
angular momentum \citep{hernquist.89,barnes.hernquist.91,barneshernquist96}. 
The resulting inflows
produce a dissipational, merger-induced starburst which is compact
(typical size scales $\sim0.5-1$\,kpc).  If a significant fraction of
the final stellar mass is formed in this manner, the scale length of
the stellar component will be much smaller than that of its
progenitor.

\citet{onorbe:diss.fp.model,robertson:fp} and \citet{dekelcox:fp}
argued that, because low mass disks are more gas rich (on average)
than high mass disks, dissipation will be more important in low-mass
systems.  That is, lower-mass ellipticals (the merger products of
low-mass spirals) should have smaller effective radii relative to
their progenitor disks, and, since the halo mass distribution is not
strongly affected by this process, a correspondingly smaller dark
matter fraction (lower $\mdyn/\mstar$) within the stellar $R_{e}$. On
the other hand, high-mass disks are observed to be gas-poor: at $\gg
L_{\ast}$ gas fractions become negligible ($\ll 10\%$) -- this is
precisely where ellipticals are {\em not} more compact than
disks.  Together, \citet{robertson:fp} and \citet{dekelcox:fp} argued
that the dependence of dissipational fraction on mass is sufficient,
in principle, to explain the tilt of the FP, the size-mass and
velocity-dispersion mass correlations of ellipticals.

The importance of gas dynamics and triggered star formation in mergers
is reinforced by observations of ultraluminous infrared galaxies
(ULIRGs) \citep[e.g.][]{soifer84a,soifer84b}, which are invariably
associated with mergers in the local Universe
\citep{joseph85,sanders96:ulirgs.mergers}.  The infrared emission from
ULIRGs is thought to be powered by intense starbursts in their nuclei,
originating in central concentrations of gas
\citep[e.g.][]{scoville86, sargent87,sargent89}, which will leave
dense stellar remnants
\citep{kormendysanders92,hibbard.yun:excess.light,rj:profiles}, as
predicted theoretically \citep{mihos:cusps}.  Moreover, observations
of merging systems and gas-rich merger remnants
\citep[e.g.,][]{LakeDressler86,Doyon94,ShierFischer98,James99}, as
well as post-starburst (E+A/K+A) galaxies
\citep{goto:e+a.merger.connection}, have shown that their kinematic
and photometric properties are consistent with them eventually
evolving into typical $\sim L_{\ast}$ elliptical galaxies. The
correlations obeyed by these mergers and remnants
\citep[e.g.,][and references above]{Genzel01,rothberg.joseph:kinematics,
rothberg.joseph:rotation}
are similar to e.g.\ the observed fundamental plane and
\citet{kormendy77:correlations} relations for relaxed ellipticals, and
consistent with evolution onto these relations as their stellar
populations age, as well as the clustering and mass density of
ellipticals \citep{hopkins:clustering}.

Unfortunately, the consequences of the models are less clear; as such,
there has been some ambiguity regarding whether or not recent
observations of the FP support or disagree with theory. In particular,
while merger remnants may fall on the FP, it is not obvious that a
differential role of dissipation is in fact responsible for the FP
tilt in the manner predicted by \citet{robertson:fp}, 
or that this applies to all ellipticals.  However, there is
hope: \citet{mihos:cusps} predicted that these dissipational
starbursts should leave an observable signature in the surface
brightness profiles of remnants, in the form of a steep departure from
the outer \citet{devaucouleurs} $r^{1/4}$-law distribution in the
inner regions: i.e. a central ``extra light'' component above the
inwards extrapolation of the outer profile. Observations have 
now uncovered distinctive evidence for this two-component structure in local 
ellipticals \citep{kormendy99,jk:profiles,ferrarese:profiles}, 
classical bulges \citep{balcells:bulge.xl}, and recent merger remnants 
\citep{hibbard.yun:excess.light,rj:profiles}. 
With the combination of 
HST and ground-based photometry, it now appears that such 
components are ubiquitous \citep{jk:profiles}, 
with mass ranges and spatial extents comparable to those
expected from observations of ongoing merger-induced 
starbursts and numerical simulations.

In a series of papers, \citet{hopkins:cusps.mergers,hopkins:cusps.ell,
hopkins:cores} (hereafter \paperone, \papertwo\ and \paperthree,
respectively), we used these simulations and data to 
develop and test a method to empirically
determine the degree of dissipation involved in the formation of a
particular elliptical galaxy -- i.e.\ the mass fraction in the stellar
remnant of a central, compact nuclear (dissipational) starburst. In
\paperone, we demonstrated that observed merger remnants can be
robustly decomposed into two components: an outer, dissipationless
(violently relaxed) component with a Sersic law-like profile,
comprising the pre-merger stars, and an inner, compact starburst
remnant, produced in a starburst.  Combining large ensembles of
observations with a library of simulations that enabled us to
calibrate various empirical methods, we developed a purely empirical
technique to separate the inner ``excess'' owing to the true physical
starburst component in the observed surface brightness profile from
the outer profile.  

In \papertwo, we showed that this method -- given photometry of
sufficient quality and covering a large dynamic range (from
$\lesssim100\,$pc to $\gtrsim20-50$\,kpc) -- could be extended to
observed ``cusp'' ellipticals (i.e.\ ellipticals with steep nuclear
profiles), commonly believed to be the direct remnants of gas-rich
mergers \citep{faber:ell.centers}.  Separating the observed surface
brightness profile in this manner, we demonstrated that simulations
and independent observations (e.g.\ distinctions in stellar
populations evident in kinematics, colors, stellar ages,
metallicities, or abundances) confirm that the component of the
elliptical formed via dissipation (in a nuclear starburst) could be
reliably (statistically) determined.

In \paperthree, we showed that the same methods can robustly recover
the dissipational starburst remnants in ``core'' ellipticals
(ellipticals with shallow nuclear profiles). In general, even if other
processes such as e.g.\ scattering of stars by a binary black hole
create the core, their impact on the overall starburst component is
negligible (by both mass and radius, the scales of the core, typically
$\lesssim30-50\,$pc, are much smaller than the mass and size of the
starburst). We also showed that even if core or other ellipticals have
subsequently been modified by spheroid-spheroid ``dry'' re-mergers,
profile shape is preserved to a sufficient degree that the original
nuclear excess (i.e.\ the indicator of the degree of dissipation in 
the original, spheroid-forming merger) remains.

By applying our methodology to observations of ellipticals over a wide
range in mass and size, we can, for the first time, empirically
compare the degree of dissipation (starburst or dissipational mass
fraction) in ellipticals to their global properties and locations on
the FP.  If a differential effect of dissipation as a function of mass
is the explanation for the FP and elliptical scaling relations, as the
models suggest, then we should be able to see and quantify the
signatures of this directly in their {\it observed}
profiles.  Therefore, in this
paper we present a critical examination of the relationship between
spheroid properties and FP correlations and ``extra light''
components, in both simulations and observed galaxies.  The
combination of a large number of observations, together with an
ensemble of hydrodynamic gas-rich merger simulations sampling the
entire observed range in e.g.\ mass, gas content, and other
properties, enables us to develop and apply new, detailed, empirical
tests of these models for the origin of the FP correlations.

In \S~\ref{sec:sims} we summarize our library of merger simulations,
and in \S~\ref{sec:data} we describe the compilation of observations
used to test the models.  In \S~\ref{sec:data:proxies} we review
different approaches to fit the surface density profile and recover
the physically distinct (dissipational versus dissipationless)
components in merger remnants.  We use a set of simulations to infer
how galaxy properties are predicted to scale with dissipation in
\S~\ref{sec:diss.fx}. We then compare with observed systems: examining
how observed sizes and masses scale with gas content in
\S~\ref{sec:obs}, and the scaling relations obeyed at fixed
dissipational content in \S~\ref{sec:obs.tests}.  We combine the
observed dependences on dissipation and gas content in
\S~\ref{sec:obs.tests.2} to determine whether this is sufficient to
explain the tilt and scatter of the FP and its projected correlations.
In \S~\ref{sec:obs.tests.mtot} we compare dynamical and total masses
and consider possible non-homology effects.  In \S~\ref{sec:remergers}
we examine the impact of subsequent re-mergers on the FP
correlations. Finally, in \S~\ref{sec:discuss} we discuss our results
and outline future explorations of these correlations.

Throughout, we adopt a $\Omega_{\rm M}=0.3$, $\Omega_{\Lambda}=0.7$,
$H_{0}=70\,{\rm km\,s^{-1}\,Mpc^{-1}}$ cosmology, and normalize all
observations and models accordingly.  We note that this has little
affect on our conclusions, however.  We also adopt a
\citet{chabrier:imf} stellar initial mass function (IMF), and convert
all stellar masses and mass-to-light ratios to this choice. The exact
form of the IMF systematically shifts the normalization of stellar
masses herein, but does not substantially influence our
comparisons. All magnitudes are in the Vega system, unless otherwise
specified.

\breaker
\section{The Simulations}
\label{sec:sims}

The merger simulations we 
analyze in this paper were performed with the parallel TreeSPH code
{\small GADGET-2} \citep{springel:gadget}, based on a fully conservative
formulation \citep{springel:entropy} of smoothed particle hydrodynamics (SPH),
which conserves energy and entropy simultaneously even when smoothing
lengths evolve adaptively \citep[see e.g.,][]{hernquist:sph.cautions,oshea:sph.tests}. 
The simulations account for radiative cooling and optional heating by
a UV background \citep[as in][although it 
is not important for the masses of interest here]{katz:treesph,dave:lyalpha}, and
incorporate a sub-resolution model of a multiphase interstellar medium
(ISM) to describe star formation and supernova feedback \citep{springel:multiphase}.
Feedback from supernovae is captured in this sub-resolution model
through an effective equation of state for star-forming gas, enabling
us to stably evolve disks with arbitrary gas fractions \citep[see, e.g.][]{springel:models,
springel:spiral.in.merger,robertson:disk.formation,robertson:msigma.evolution}. 
This is described by the parameter $\qeos$,
which ranges from $\qeos=0$ for an isothermal gas with effective
temperature of $10^4$ K, to $\qeos=1$ for our full multiphase model
with an effective temperature $\sim10^5$ K. We also compare with a subset of 
simulations which adopt the star formation and feedback prescriptions 
from \citet{mihos:cusps,mihos:starbursts.94,mihos:starbursts.96}, in which the ISM is treated as a
single-phase isothermal medium and feedback energy is deposited in a 
purely kinetic radial impulse (for details, 
see, e.g.\ \cite{mihos:method}).

Although we find that they make little difference to 
the extra light component, most of the simulations include 
supermassive black holes at the centers of both progenitor galaxies.  
The black holes are represented by ``sink'' particles
that accrete gas at a rate $\Mdot$ estimated from the local gas
density and sound speed using an Eddington-limited prescription based
on Bondi-Hoyle-Lyttleton accretion theory.  The bolometric luminosity
of the black hole is taken to be $L_{\rm bol}=\epsilon_{r}\dot{M}\,c^{2}$,
where $\epsilon_r=0.1$ is the radiative efficiency.  We assume that a
small fraction (typically $\approx 5\%$) of $L_{\rm bol}$ couples dynamically
to the surrounding gas, and that this feedback is injected into the
gas as thermal energy, weighted by the SPH smoothing kernel.  This
fraction is a free parameter, which we determine as in \citet{dimatteo:msigma}
by matching the observed $M_{\rm BH}-\sigma$ relation.  For now, we do
not resolve the small-scale dynamics of the gas in the immediate
vicinity of the black hole, but assume that the time-averaged
accretion rate can be estimated from the gas properties on the scale
of our spatial resolution (roughly $\approx 20$\,pc, in the best
cases). In any case, repeating our analysis for simulations with no black 
holes yields identical conclusions. 

The progenitor galaxy models are described in
\citet{springel:models}, and we review their properties here.  For each
simulation, we generate two stable, isolated disk galaxies, each with
an extended dark matter halo with a \citet{hernquist:profile} profile,
motivated by cosmological simulations \citep{nfw:profile,busha:halomass}, 
an exponential disk of gas and stars, and (optionally) a
bulge.  The galaxies have total masses $M_{\rm vir}=V_{\rm
vir}^{3}/(10GH[z])$ for an initial redshift $z$, with the baryonic disk having a mass
fraction $m_{\rm d}=0.041$, the bulge (when present) having $m_{\rm
b}=0.0136$, and the rest of the mass in dark matter.  The dark matter
halos are assigned a
concentration parameter scaled as in \citet{robertson:msigma.evolution} appropriately for the 
galaxy mass and redshift following \citet{bullock:concentrations}. We have also 
varied the concentration in a subset of simulations, and find it has little 
effect on our conclusions because the central regions of the 
galaxy are baryon-dominated. 
The disk scale-length is computed
based on an assumed spin parameter $\lambda=0.033$, chosen to be near
the mode in the $\lambda$ distribution measured in simulations \citep{vitvitska:spin},
and the scale-length of the bulge is set to $0.2$ times this. Modulo explicit 
variation in these parameters, these choices ensure that the initial disks 
are consistent with e.g.\ the observed baryonic 
Tully-Fisher relation and estimated halo-galaxy mass 
scaling laws \citep[][and references therein]{belldejong:tf,kormendyfreeman:scaling,
mandelbaum:mhalo}.

Typically, each galaxy initially consists of 168000 dark matter halo
particles, 8000 bulge particles (when present), 40000 gas and 40000
stellar disk particles, and one black hole
(BH) particle.  We vary the numerical
resolution, with many simulations using twice, and a subset up to 128
times, as many particles. We choose the initial seed
mass of the black hole either in accord with the observed $M_{\rm
BH}$-$\sigma$ relation or to be sufficiently small that its presence
will not have an immediate dynamical effect, but we have varied the seed
mass to identify any systematic dependencies.  Given the particle
numbers employed, the dark matter, gas, and star particles are all of
roughly equal mass, and central cusps in the dark matter and bulge
are reasonably well resolved. 
The typical gravitational 
softening in the simulations is $\sim20-50\,$pc in the 
$\lesssim L_{\ast}$ systems of particular interest here, 
with a somewhat higher $\sim50-100\,$pc in the most massive 
systems (yielding an effectively constant resolution $\sim 0.01\,R_{e}$
in terms of the effective radius). In \paperone\ and \papertwo\  
we demonstrate that this is sufficient to properly resolve not only the mass 
fractions but also the spatial extent of the extra light components of 
interest here (resolution may become an issue when attempting to 
model the very smallest galaxies, with $R_{e}\lesssim100$\,pc and 
$L<0.01\,L_{\ast}$, but this is well below the range of the observations of 
interest here). The hydrodynamic 
gas smoothing length in the peak starburst phases of interest is 
always smaller than this gravitational softening. 

We consider a series of several hundred simulations of colliding
galaxies, described in \citet{robertson:fp,robertson:msigma.evolution} and
\citet{cox:xray.gas,cox:kinematics}.  We vary the numerical resolution, the orbit of the
encounter (disk inclinations, pericenter separation), the masses and
structural properties of the merging galaxies, initial gas fractions,
halo concentrations, the parameters describing star formation and
feedback from supernovae and black hole growth, and initial black hole
masses. 

The progenitor galaxies have virial velocities $V_{\rm vir}=55, 80, 113, 160,
226, 320,$ and $500\,{\rm km\,s^{-1}}$, and redshifts $z=0, 2, 3, {\rm
and}\ 6$, and the simulations span a range in final spheroid stellar mass
$M_{\ast}\sim10^{8}-10^{13}\,M_{\sun}$, covering essentially the
entire range of the observations we consider at all redshifts, and
allowing us to identify any systematic dependencies in the models.  We
consider initial disk gas fractions by mass of $\fgas = 0.05,\ 0.1,\ 0.2,\ 0.4,\ 0.6,\ 
0.8,\ {\rm and}\ 1.0$ (defined as the fraction of disk baryonic mass which is gas) 
for several choices of virial velocities,
redshifts, and ISM equations of state. The results described in this
paper are based primarily on simulations of equal-mass mergers;
however, by examining a small set of simulations of unequal mass
mergers, we find that the behavior does not change dramatically for
mass ratios to about 3:1 or 4:1. The mass ratios we study are appropriate for the
observations of ellipticals used in this paper, which are only formed 
in our simulations in major merger events. At higher mass ratios, 
the result is a small bulge in a still disk-dominated galaxy 
\citep[see e.g.][]{younger:minor.mergers,hopkins:disk.survival,
hopkins:disk.heating}, which we do not study here 
(although in general our conclusions should still apply, so long as the 
bulges of interest are ``classical'' bulges formed in mergers). 

We also briefly consider in \S~\ref{sec:remergers} a subset of
spheroid-spheroid ``re-mergers,'' representative of gas-poor or
``dry'' spheroid-spheroid mergers of elliptical galaxies. In these
mergers, we collide two remnants of previous disk-disk mergers, in
order to explore how their properties are modified through
re-merging. We typically merge two identical remnants (i.e.\ two
identical copies of the remnant of a given disk-disk merger), but have
also explored re-mergers of various mass ratios (from 1:1 to
$\approx$ 4:1), and mixed morphology re-mergers (i.e.\ merging an 
elliptical remnant with an un-merged gas-rich disk).  In the former
case, we generally find a similar division in mass ratio at which a
major merger is significant. In the latter, we find that the
properties are more akin to those of other gas-rich (disk-disk)
mergers, and the remnant should for most purposes should still be
considered the direct product of a gas-rich merger.  In the re-merger
series, we vary the orbital parameters, both of the initial gas-rich
merger and re-merger, and consider systems with 
a similar range of initial gas fractions in the progenitor disks (of the 
original gas-rich merger), $\fgas=0.05,\ 0.2,\ 0.4,\ 0.8$. The re-mergers 
span a similar range in virial velocities and final stellar masses to 
the gas-rich mergers. 

Each simulation is evolved until the merger is complete and the remnants are 
fully relaxed, typically $\sim1-2$\,Gyr after the final merger 
and coalescence of the BHs. We then analyze the 
remnants following \citet{cox:kinematics}, in a manner designed to mirror 
the methods typically used by observers. For each remnant, we project the 
stars onto a plane as if observed from a particular direction, and consider 
100 viewing angles to each remnant, which uniformly sample the unit sphere. 
Given the projected stellar mass distribution, we calculate the iso-density contours 
and fit ellipses 
to each (fitting major and minor 
axis radii and hence ellipticity at each iso-density contour), 
moving concentrically from $r=0$ until the entire stellar mass 
has been enclosed. This is designed to mimic observational isophotal fitting 
algorithms \citep[e.g.][]{bender:87.a4,bender:88.shapes}. The radial deviations 
of the iso-density contours from the fitted ellipses are 
expanded in a Fourier series in the standard fashion to determine 
the boxyness or diskyness of each contour (the $a_{4}$ parameter). 
Throughout, we show profiles and quote our results in 
terms of the major axis radius. For further details, we refer to \citet{cox:kinematics}.

We directly extract the effective radius $\re$ as the projected half-mass stellar 
effective radius, and the velocity dispersion $\sigma$ as the average 
one-dimensional velocity dispersion within a circular 
aperture of radius $\re$. This differs from what is sometimes adopted 
in the literature, where $\re$ is determined from the best-fitting
Sersic profile, but because 
we are fitting Sersic profiles to the observed systems we usually quote both the 
true effective radius of the galaxy and effective radii of the fitted Sersic components. 
Throughout, the stellar mass $M_{\ast}$ refers to the total stellar mass of the galaxy, and 
the dynamical mass $\mdyn$ refers to the 
traditional dynamical mass estimator 
\begin{equation}
\mdyn\equiv k\,\frac{\sigma^{2}\,\re}{G},
\end{equation}
where we adopt $k=3.8$ (roughly what is 
expected for a \citet{hernquist:profile} profile, and the choice that most accurately 
matches the true enclosed stellar plus dark matter mass within $\re$ in the 
simulations; 
although this choice is irrelevant as long as we apply it 
uniformly to both observations and simulations). 
When we plot quantities such as $\re$, $\sigma$, and $\mdyn$, we 
typically show just 
the median value for each simulation across all $\sim100$ sightlines. The sightline-to-sightline 
variation in these quantities is typically smaller than the 
simulation-to-simulation scatter, but we explicitly note where it is large.

\breaker
\section{The Data}
\label{sec:data}

We compare the simulations to and test our predictions on an ensemble
of observed surface brightness profiles of ellipticals, described in
\papertwo\ and \paperthree. Specifically, we consider three samples of
ellipticals and a compilation of remnants of recent gas-rich mergers. The first
is the $V$-band Virgo elliptical survey of \citet{jk:profiles}, based
on the complete sample of Virgo galaxies down to extremely faint
systems in \citet{binggeli:vcc} 
\citep[the same sample studied in][]{cote:virgo,ferrarese:profiles}. 
\citet{jk:profiles} combine observations from a
large number of sources
\citep[including][]{bender:data,bender:06,caon90,caon:profiles,davis:85,kormendy:05,
lauer:85,lauer:95,lauer:centers,liu:05,peletier:profiles} 
and new photometry from McDonald Observatory, the HST archive, and 
the SDSS 
for each of their objects which (after careful conversion to a single
photometric standard) enables accurate surface brightness measurements
over a wide dynamic range (with an estimated 
zero-point accuracy of $\pm0.04\,V\,{\rm mag\, arcsec^{-2}}$). 
Typically, the galaxies in this sample have
profiles spanning $\sim12-15$ magnitudes in surface brightness,
corresponding to a range of nearly four orders of magnitude in
physical radii from $\sim10\,$pc to $\sim100\,$kpc, permitting the
best simultaneous constraints on the shapes of both the outer and
inner profiles of any of the objects we study.  The profiles include
e.g.\ ellipticity, $a_{4}/a$, and $g-z$ colors as a function of
radius.  
Unfortunately, since this is restricted to Virgo ellipticals,
the number of galaxies is limited, especially at the intermediate and high end of the
mass function.

We therefore add surface brightness profiles from \citet{lauer:bimodal.profiles}, 
further supplemented by \citet{bender:data}. 
\citet{lauer:bimodal.profiles} compile $V$-band measurements of a
large number of nearby systems for which HST imaging of the galactic
nuclei is available.  These include the 
\citet{lauer:centers} WFPC2 data-set, the \citet{laine:03} WFPC2 BCG
sample (in which the objects are specifically selected as brightest
cluster galaxies from \citet{postmanlauer:95}), and the \citet{lauer:95}
and \citet{faber:ell.centers} WFPC1 compilations.
Details
of the treatment of the profiles and conversion to a single standard
are given in \citet{lauer:bimodal.profiles}. 
The sample includes ellipticals over
a wide range of luminosities, down to $M_{B}\sim-15$, but is dominated
by intermediate and giant ellipticals, with typical magnitudes $M_{B}
\lesssim -18$. This therefore greatly extends our sampling of the
intermediate and 
high-mass end of the mass function, but at the cost of some dynamic
range in the data. The HST images alone,
while providing information on the central regions, typically extend
to only $\sim1$\,kpc outer radii, which is insufficient to fit the
outer profile. \citet{lauer:bimodal.profiles} 
therefore combine these data with ground-based
measurements from a number of sources (see the references for the
\citet{jk:profiles} sample) to construct profiles that typically span
physical radii from $\sim10\,$pc to $\sim10-20$\,kpc. Although the 
composite profiles 
were used in \citet{lauer:bimodal.profiles} to estimate effective radii, they were not 
actually shown in that paper. 
It should also be noted that there is
no single criterion that characterizes galaxies included in this
sample, but they generally
comprise luminous nearby ellipticals and S0 galaxies for which
detailed imaging is available.  We emphasize that issues of completeness
and e.g.\ environment are not important for any of our conclusions.

We occasionally 
supplement the profiles from \citet{lauer:bimodal.profiles} with additional 
profiles used in \citet{bender:data,bender:ell.kinematics,bender:ell.kinematics.a4,
bender:velocity.structure}, and in some cases subsequently updated. 
These are more limited: typically the profiles cover $\sim7$ magnitudes in
surface brightness, extending from $\sim30-50\,$pc out to $\sim$ a few
kpc (typically $\sim3$\,kpc in low-luminosity systems, and $\sim
15$\,kpc in the brightest systems, sufficient for acceptable, but not
strong constraints on the outer profile shapes).
However, the measurements are usually
in each of the $V$, $R$, and $I$ bands, and hence allow us to
construct multicolor surface brightness, ellipticity, and $a_{4}/a$
profiles. We use this to estimate e.g.\ the 
sensitivity of the fitted parameters 
and galaxy profiles on the observed waveband and on the 
quality and dynamic range of the photometry.

We also consider the sample of local remnants of recent gas-rich
mergers from \citet{rj:profiles} with which we compare our simulations
in \paperone. For these objects, \citet{rj:profiles} compile
$K$-band imaging, surface brightness, ellipticity, and $a_{4}/a$
profiles, where the profiles typically range from $\sim100\,$pc
to $\sim10-20$\,kpc. These span a moderate range in luminosity
(including objects from $M_{K}\sim-20$ to $M_{K}\sim-27$, but with
most from $M_{K}\sim-24$ to $M_{K}\sim-26$) and a wide range in merger
stage, from ULIRGs and (a few) unrelaxed systems to shell
ellipticals. As demonstrated in \citet{rj:profiles} and 
argued in \paperone, these systems will almost all
become (or already are, depending on the classification scheme used)
typical $\sim \lstar$ ellipticals, with appropriate phase space
densities, surface brightness profiles, fundamental plane relations,
kinematics, and other properties. For a detailed discussion of the modeling 
of these systems and the profiles themselves, we refer to \paperone\ 
(all of the results shown for these systems are derived therein). We 
show the results from \paperone\ here in order to test whether observed 
merger remnants (not just our simulations) obey the same correlations 
we study in ellipticals. In particular, this allows us to provide an additional 
empirical check of the simulations and the argued 
continuity of merger and elliptical populations.

Because the generally accepted belief is that core ellipticals were
not directly formed in gas-rich major mergers but were subsequently
modified by dry re-mergers, we have repeated our 
analysis considering just those ellipticals which
are confirmed via HST observations as being either cusp or core
ellipticals, and designate the two populations as two separate
sub-samples throughout. However, as we discuss below, treating them as a single
population (or including all ellipticals, even those without HST
observations, in our sample) makes no difference to our conclusions.
We include all the confirmed gas-rich merger remnants, but note there
are a small number of extreme unrelaxed cases for which sharp features
in the surface brightness profiles prevented derivation of meaningful
quantities (note, however, as shown in \paperone, that almost all of
the objects in this sample are sufficiently well-relaxed at the radii
of interest for our fitting). We exclude spheroidals, as they are not
believed to form in major mergers as are ellipticals \citep[e.g.][]{kormendy:spheroidal1,
kormendy:spheroidal2,jk:profiles}, 
and in any case they dominate at
extremely low masses where our simulations do not sample the
population (they also predominate as satellite galaxies, whose effects
we do not model).

We also exclude S0 galaxies. This is not because of a physical
distinction: observations suggest that these likely form a continuous
family with the low-luminosity cusp ellipticals, and in fact a number
of our simulated gas-rich merger remnants would, from certain viewing
angles, be classified as S0s (although we exclude disk-dominated 
simulation remnants from our comparisons here). However, in order to derive e.g.\ the
parameters of the outer, violently relaxed profile and central extra
light, it would be necessary to remove the contribution of the
large-scale disk from the surface brightness profiles of these
objects.  Our two-component (outer dissipationless and inner
dissipational) Sersic models (described in \S~\ref{sec:data:proxies}) then
become three-component fits, and the degeneracies involved with three
independent components, even with our best data and simulations, are
so large as to render the results meaningless. We have, however,
re-visited all of the S0s in these samples in light of our results,
and find that they are, in all cases, consistent with our predicted
and observed trends.  However, it is too difficult to infer these
trends directly from the S0s themselves without ideal disk
subtraction.

This yields a final sample of $\approx 180$ unique elliptical
galaxies, and $\approx 50$ confirmed remnants of gas-rich mergers. Most of the
sample spans a range of three orders of magnitude in stellar mass,
from $\lesssim0.1\,\mstar$ to $\sim10\,\mstar$, and a wide range in
extra light properties.  There is, of course, some overlap in the 
samples that define our compilation; we have $\sim600$
surface brightness profiles for our collection of unique ellipticals, 
including (for many objects) repeated measurements in 
multiple bands and with various instruments. 
This turns out to be quite useful, as
it provides a means to quantify error estimates in fits to these
profiles. The variations between fit parameters 
derived from observations 
in different bands or made using different
instruments are usually much larger than the formal statistical errors in the
fits to a single profile. There are no obvious systematic effects
(i.e.\ systematic changes in profile fits from $V$ to $I$ band), but as
demonstrated in \paperone\ the effects of
using different bands or changing
dynamic range (from different instruments) can be complex, depending
on the structure and degree of relaxation of the outer regions of a
system. On the other hand, there are well-relaxed objects for which
almost no significant change in the fits occurs from band to band.  
It is therefore useful to have multiple observations of the same system, 
as it allows us to get some idea of how sensitive our fits are to 
differences in e.g.\ the choice of observed wavelength or dynamic 
range from instrument to instrument. 

In \papertwo\ and \paperthree, we 
present the results of our fits to each elliptical in our sample; we 
use these values throughout this paper. For sources with 
multiple independent observations, we define error bars for each 
fit parameter representing 
the $\sim1\,\sigma$ range in 
parameters derived from various observations, typically from three
different surface brightness profiles but in some cases from as many
as $\approx 5-6$ sources (where there are just $2$ sources, the ``error'' 
is simply the range between the two fits). 
In many cases the different observations are comparable; 
in some there are clearly measurements 
with larger dynamic range and better resolution: the errors derived 
in this manner should in such cases be thought of as the 
typical uncertainties introduced by lower dynamic range or less 
accurate photometry. 

In terms of direct comparison with our
simulations, the data often cover a dynamic range and have resolution
comparable to our simulations, provided we do not heavily weight the
very central ($\lesssim30\,$pc) regions of HST nuclear profiles. 
Experimenting with different smoothings and imposed dynamic range
limits, we find it is unlikely that resolution or seeing differences
will substantially bias our comparisons. They can introduce
larger scatter, however: the robustness of our results increases considerably 
as the dynamic range of the observed profiles is increased.

We have converted all the observations to physical 
units given our adopted cosmology, 
and compile global parameters (where not available in the original papers) 
including e.g.\ kinematic properties, luminosities, and black hole masses
from the literature. 
We determine stellar masses ourselves in a uniform manner for 
all the objects, based on their total $K$-band luminosities and 
$(B-V)$ color-dependent mass-to-light ratios from \citet{bell:mfs}, 
corrected for our adopted IMF. We have 
also repeated our analysis using stellar masses derived from a 
mean $M/L$ as a function of luminosity or from fitting 
the integrated $UBVRIJHK$ photometry of each object to a 
single stellar population with the models of \citet{BC03}, and 
find this makes no difference to our conclusions. 

Throughout, we will usually refer interchangeably to the observed surface
brightness profiles in the given bands and the surface stellar mass
density profile. Of course, stellar light is not exactly the same as
stellar mass, but in \paperone\ and \papertwo, 
we consider the differences between
the stellar light and the stellar mass density profiles as a function
of time, wavelength, and properties of the merger remnant, and show
that the optical ($V$) and $K$-band results introduce little bias 
(i.e.\ are good tracers of the stellar mass). 
It is important to note that while we are not concerned about the
absolute normalization of the profile (i.e.\ mean $M/L$), since we
derive total stellar masses separately from the integrated photometry,
we must account for systematics that might be induced by change in
$M/L$ as a function of radius.  The profiles in optical bands such as
$V$ require more care when the system is very young ($\lesssim
1-2$\,Gyr after the major merger-induced peak of star formation),
and there can be considerable bias or uncertainty owing to stellar
population gradients and dust. However, once the system is relaxed,
the optical bands become good proxies for the stellar mass
distribution, with $\lesssim20\%$ variation in our $M/L$ over the entire 
fitted range of radii in our simulations, in good agreement with the 
simple expectation based on the observed weak color gradients in 
most of the observed systems.

In fact, in \paperone\ and \papertwo\ 
we demonstrate that once the system reaches
intermediate age, the bias in e.g.\ $B$ or $V$ band is often less than
that in $K$ band, because systems tend to be both younger and more
metal rich in their centers.  In $K$-band,
these both increase $L/M$, leading to a (small) systematic bias. 
Our simulation results indicate that our merger remnant 
sample (observed in $K$-band) are, on average, robust in 
this sense, but they 
should be treated with care, especially in the most 
extreme cases (namely the few LIRGs and ULIRGs in the sample). In
optical bands, however, age and metallicity gradients have opposite effects (younger age
increases $L/M$, but higher metallicity decreases $L/M$), and they
tend to mostly cancel.  Since essentially all of our ellipticals
observed in the optical bands are older than this stellar population
age (even in their centers), and they have been carefully vetted and
either corrected or (where correction was too difficult) excluded for
the effects of e.g.\ dust lanes in the sources
\citep{jk:profiles,lauer:bimodal.profiles}, we are not concerned that
significant bias might persist. Furthermore, comparison of systems
observed in different bands demonstrates that our conclusions are
unchanged (modulo small systematic offsets) regardless of the observed
bands in which we analyze these systems.

\breaker
\section{Inferring the Degree of Dissipation from Observed Profiles}
\label{sec:data:proxies}

In order to test the theoretical models from \citet{robertson:fp} and others 
using observed systems, we
require some empirical means to estimate the amount of dissipation
involved in forming elliptical galaxies.  The simulations imply
that the light profile should be considered to be the sum of two
physically distinct components.  The outer, violently relaxed,
dissipationless component is made from stars that formed in the disks
prior to the final stages of a merger, and therefore evolve in a
collisionless manner.  The inner, ``dissipational'' component is
comprised of stars produced in a central starburst, from gas which
loses its angular momentum in the merger.  This gas is channeled into
the center of the remnant and is converted into stars on a short
timescale (effectively in an almost fixed background potential set by
the already nearly relaxed outer, dissipationless component). The
``degree of dissipation'' is, therefore, effectively measured by the
mass fraction of the central, dissipational starburst component, which
causes $R_{e}$ to contract and gives rise to the effects described
above.  In \paperone, \papertwo, and \paperthree, we develop and test
a methodology for decomposing observed surface brightness profiles
into these two components.  We refer to those papers for details and
verification that this approach is reliable when applied to either
simulations or observations, but we briefly review the methodology
here.  We consider two different observable proxies for the amount of
dissipation.

\begin{figure}
    \centering
    \scaleup
    \plotone{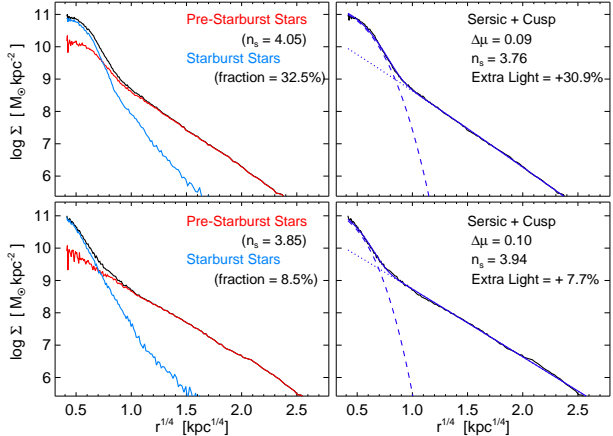}
    \caption{{\em Left:} Surface mass density of a typical merger 
    remnant from our simulation library (black), 
    decomposed into stars formed prior to the final merger (which 
    are then violently relaxed; red) and stars formed in the 
    dissipational starburst (blue). The Sersic index fitted to the pre-starburst component 
    alone is shown, with the mass fraction of the 
    starburst component. We show two examples, one very gas-rich ({\em top}) 
    and one less gas-rich ({\em bottom}). 
    {\em Right:} Two-component (Sersic plus 
    cusp or extra light) fit (inner exponential 
    and outer Sersic) to the total light profile, with the Sersic index of the outer component 
    and mass fraction of the inner component, and rms scatter ($\Delta\mu$) about the 
    fit. Our two-component, 
    cusp plus Sersic function fit ({\em right}) accurately 
    recovers the profile of the violently relaxed component and mass fraction 
    of the starburst component. 
    \label{fig:demo.fit.methods}}
\end{figure}

First, we directly fit the surface brightness profile to determine the
mass fraction in a central ``extra light'' or starburst component,
which we denote as $\fextra$. We demonstrate in \paperone\ that this
quantity can be measured by fitting the total observed surface
brightness profile of merger remnants to the following two component 
model:
\begin{equation} 
I_{\rm tot} =
I_{\rm extra}\,\exp{{\Bigl\{}-b_{1}\,{\Bigl(}\frac{r}{r_{\rm extra}}{\Bigr)}{\Bigr\}}}+
I_{\rm o}\,\exp{{\Bigl\{}-b_{n}\,{\Bigl(}\frac{r}{r_{\rm o}}{\Bigr)}^{1/n}{\Bigr\}}},
\label{eqn:two.component.law}
\end{equation} 
with the inner part ($I_{\rm extra}$, $r_{\rm extra}$) serving as a
proxy for the central dissipational component and the outer component
($I_{\rm o}$, $r_{\rm o}$) representing the dissipationless outer
profile. The ``extra light,'' in other words, is the mass fraction in
a high surface brightness central component that rises above the
extrapolation of the outer, intermediate $n_{s}$ Sersic law-like profile (which in any
reasonable model will be dominated by the violently relaxed stars).
In \paperone\ and \papertwo, we show that this form provides, on average, a reliable
physical decomposition of the mass distribution in remnants of
simulated mergers between gas-rich galaxies, including mass fractions,
effective radii, and profile shapes of the two components. Analogous decompositions 
have been applied to e.g.\ observed bulges in disk galaxies \citep{balcells:bulge.xl} and 
to Virgo ellipticals, motivated by purely empirical considerations including 
surface brightness profile shapes, isophotal shape and kinematic profiles, and 
central stellar populations \citep[see][and references therein]{kormendy99,jk:profiles}, 
as well as recent merger remnants \citep{rj:profiles}; 
we refer to \papertwo\ for a detailed discussion and comparison with 
similar methodologies from the literature.

Figure~\ref{fig:demo.fit.methods} illustrates this procedure for two
typical merger remnants from in our simulation library. We show the
total surface brightness profile, as well as the known physical
decomposition of that profile into a pre-starburst, violently relaxed
component and a dissipational starburst component (here, we use a cut
in time about the peak starburst to identify the ``starburst'' stars,
but as shown in \paperone, our results are not sensitive to the exact
choice). We compare this with the results obtained by fitting the
total profile (temporarily ignoring the known physical decomposition)
to a two-component model of the form given by
Equation~(\ref{eqn:two.component.law}). In both cases, the inner
component of our fit is a good proxy for the physical starburst light
distribution, accurately capturing its mass fraction, effective
radius, and shape where it is important; i.e. where it contributes
significantly to the light profile.  Likewise, the shape (Sersic
index), radius, and mass fraction of the dissipationless component are
accurately recovered by this purely empirical method.

We apply this to observed merger remnants in \paperone\ and cusp
ellipticals in \papertwo, and show that it yields good fits with
reliable results. In particular, we find that where there is
independent evidence for a distinct stellar population (i.e.\ a
starburst superimposed on a more extended, older stellar population),
this method recovers the empirically estimated starburst
population.  We demonstrate that other indicators, including e.g.\
stellar population 
gradients, kinematics, and isophotal shapes all support the
physical nature of these decompositions.

In \paperthree, we apply this methodology to remnants of gas-poor,
spheroid-spheroid re-mergers, and show that, while the profiles are
somewhat smoothed out, the general profile shape is preserved, and
applying this procedure will still reliably recover the {\em original} 
breakdown between 
dissipational and dissipationless components (i.e.\ the starburst
component formed in the original, spheroid-forming merger). We extend
our analysis to a large sample of ``core'' ellipticals, and similarly
demonstrate the accuracy of this decomposition and agreement with
stellar population models (where available); even if ``scouring'' by a
binary black hole has flattened the profiles on $\lesssim30-50$\,pc
scales, the bulk of the ``extra light'' has not been strongly affected
and can still be recovered.

This emphasizes an important caveat: 
the merger history and series of induced dissipational events in a 
given galaxy may be 
more complex than a single or couple of idealized major mergers 
\citep[see e.g.][]{kobayashi:pseudo.monolithic,naab:etg.formation}. 
Moreover, 
merger-induced starbursts may not be the only source of 
dissipation; for example, stellar mass loss may replenish the gas supply and 
lead to new dissipational bursts \citep[see e.g.][]{ciottiostriker:recycling}. 
For our purposes, however, all dissipational star formation will appear similar when observed 
and have the same effects -- it is convenient to simulate idealized cases, but our results 
should most appropriately be considered a measurement of the integrated amount of 
dissipation (regardless of other details of the merger and dissipational history). 
Experiments with e.g.\ more complex merger histories, simulations of multiple 
simultaneous mergers, and series of dissipational events after ``rejuvenation'' 
suggest that in this sense, our results are robust and independent of the 
detailed history. Furthermore, the agreement between these estimators 
and independent observational constraints from stellar populations, kinematics, 
and isophotal shapes suggest that the recovery is robust.

Second, we consider the physical starburst mass fraction $\fsb$
implied by directly fitting simulated surface brightness profiles to
the observations.  In detail, we demonstrate in \paperone, \papertwo,
and \paperthree\ that in our large library of simulations (with
varying masses, gas fractions, orbital parameters, stellar and black
hole feedback prescriptions, and other properties) there are remnant
surface brightness profiles which (modulo small offsets in the exact
normalization) agree well with the observed profiles over the entire
dynamic range of the data.  Considering the best fit simulation in
each case, we find that there are almost invariably simulations with
similar profiles -- variance less than $\Delta\mu=0.1\,{\rm
mag\,arcsec^{-2}}$ of the simulated profile with respect to the
observed profile (comparable to the inherent point-to-point scatter
obtained with arbitrary spline fits to the simulated or observed
profiles).  

These matches are insensitive to e.g.\ orbital parameters, disk
initial conditions, or prescriptions for feedback and star formation 
(as expected given the relative independence of the inferred dissipational 
fractions on details of a given merger history),
but good fits to a given observed profile are obtained for only a
narrow range in the physical starburst mass fraction $\fsb$ (what
dependence there is on other parameters tends to indirectly reflect
this -- varying orbital parameters, for example, can alter the time
required for a merger and therefore the mass of gas still available at
the time of the final merger to participate in the central
starburst). We demonstrate that we can define a robust and tightly
constrained best-fit starburst mass fraction, in a $\chi^{2}$ sense,
from fitting each observed profile to the entire set of simulations.
These estimates agree well with the ``extra light'' mass fractions
$\fextra$ directly determined by profile fitting, and with the other
indicators we consider, lending further confidence to the view that
$\fextra$ does indeed represent a physically meaningful indicator of
dissipation.

\breaker
\section{The Impact of Dissipation: A Case Study}
\label{sec:diss.fx}

In previous papers
\citep{robertson:fp,dekelcox:fp} 
argued and \citet{hopkins:cusps.mergers,hopkins:cusps.ell} 
demonstrated that increasing the quantity of gas available during the
final stages of a merger (i.e.\ increasing the dissipative component
of the merger which can collapse to small scales and form stars in a
central starburst) leads to more compact remnants.  Because of its
importance for our analysis here, we use a subset of our simulations
to highlight the physical significance of this result.

\begin{figure}
    \centering
    \scaleup
    \plotter{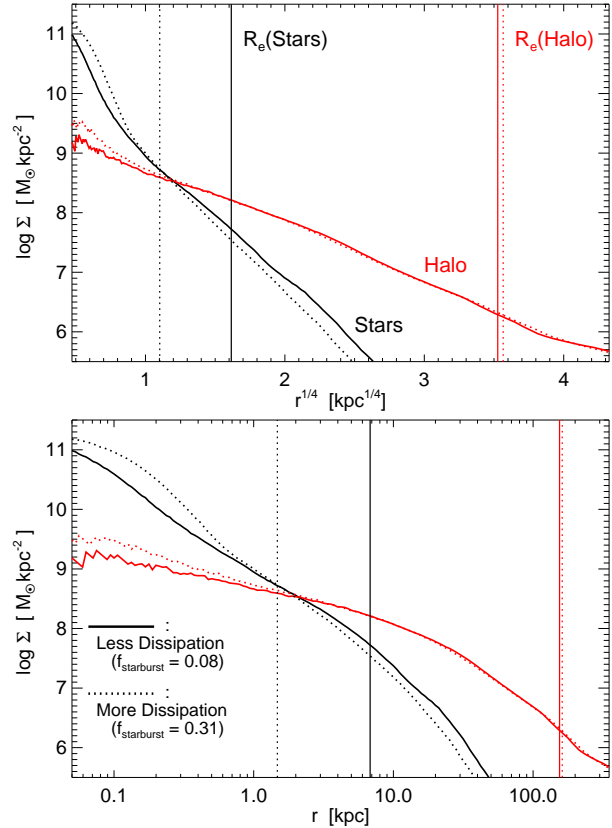}
    \caption{Surface density profiles of two simulated gas-rich merger remnants
    (the same from Figure~\ref{fig:demo.fit.methods}), as a function of 
    radius $r$ ({\em bottom}) or $r^{1/4}$ ({\em top}). We show both the stellar profile 
    and the dark matter (halo) profile, and label the effective radii of each (vertical lines). 
    Solid line shows a case with a moderate degree of dissipation -- a mass fraction in the 
    final, centrally concentrated and dissipational merger-induced starburst $\fsb=0.08$. 
    Dotted line shows a very gas-rich merger remnant, with $\fsb=0.31$. The profile 
    shapes are similar (the non-homology effects are 
    weak), and the halos are nearly identical, but the effective radii of the stellar distributions 
    are quite different, owing to the dense central concentration of mass from the starburst 
    in the latter case. 
    \label{fig:sb.vs.gas}}
\end{figure}

Figure~\ref{fig:sb.vs.gas} shows two illustrative surface density
profiles of major merger remnants -- the same cases shown in
Figure~\ref{fig:demo.fit.methods}. The initial galaxies were otherwise
identical $\sim \lstar$ (Milky Way-like) spirals placed on a random
orbit, except that in the first example, the progenitor disks had
initial gas fractions ($\sim1.5$\,Gyr before the final merger) of
$20\%$, and in the other case, of $80\%$. In the two panels, we show
the stellar mass density profile of the final (relaxed) merger
remnant, and separately decompose this into two {\em physical}
components following \citet{mihos:cusps}: namely an outer component
comprised of those stars that formed prior to the final coalescence of
the two disks, and were therefore violently relaxed in the merger, and
those stars which formed in the (by then roughly static) center of the
galaxy in the final, merger-induced starburst. They correspond to
$\sim8\%$ and $\sim30\%$ of the final total stellar mass in the two
simulations, respectively.

In both cases, the violently relaxed stars produce an extended profile
that is well-described by a Sersic law with index
$n_{s}\sim4$. Unsurprisingly, in the example with an initial gas
fraction $\sim80\%$, the central light excess reflecting the final
starburst component is much more massive -- however, the profile shape
is similar, and this has relatively little effect on the outer light
profile (it is slightly more compact, because the dense central mass
concentration means stars scattered to an orbit of a given energy will
remain at slightly smaller radii in this potential, but this effect is
small enough that it is not important).  It is clear from
Figure~\ref{fig:demo.fit.methods} that the differences in the two
profiles primarily reflect the mass fraction of the central,
dissipational component, rather than any change in the dissipationless
component.  We also show the surface mass density of the final dark
matter halo.  As expected, baryons dominate the density within the
central $\sim1-2\,$kpc of the system, but then trail off.  Clearly,
the halo mass profile is not very different in the two cases.

Despite the fact that the central mass concentration does not strongly
influence the shape and size of the dissipationless remnants of the
merger (both the pre-merger, violently relaxed stars which constitute
the dissipationless stellar component of the merger, and the dark
matter halo), it nevertheless represents a non-negligible component of
the stellar mass, and therefore the effective radius is smaller in the
case with $80\%$ gas ($\sim30\%$ of its mass in the final,
merger-induced compact starburst component).  The resulting difference
in $R_{e}$ (for such a large difference in gas fraction) is dramatic:
the effective radius in the case with $\fsb=0.08$ is
$\approx6.8\,$kpc, and for the case with $\fsb=0.31$ it is
$\approx1.4$\,kpc.  In the latter example, nearly half the mass is
actually in the dissipational component, so the effective radius
becomes small.

\begin{figure}
    \centering
    \plotter{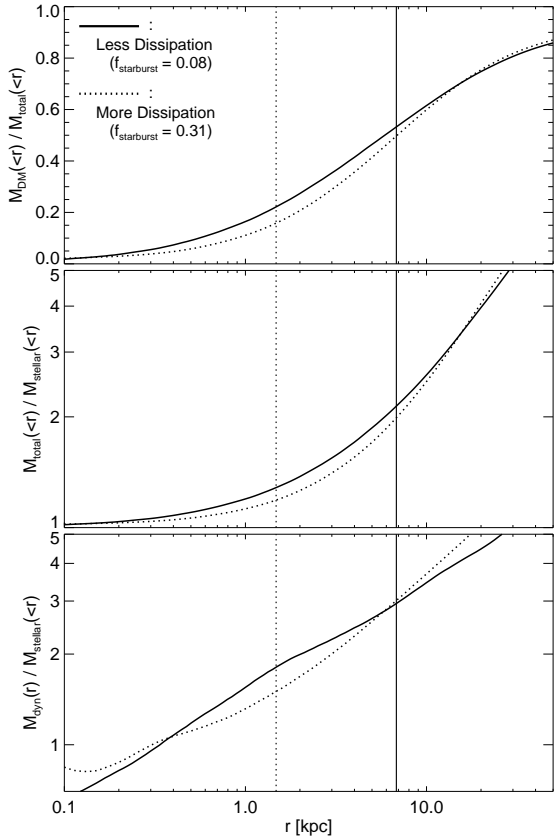}
    \caption{{\em Top:} Ratio of dark matter mass to total mass enclosed within the 
    radius of the elliptical isophotal aperture with major-axis radius $r$, for the 
    two simulations shown in Figure~\ref{fig:sb.vs.gas}. Vertical lines 
    denote the effective radii of each simulation. 
    {\em Middle:} The corresponding ratio of dark matter mass to stellar mass enclosed. 
    {\em Bottom:} The ratio of the radius-dependent dynamical mass estimator 
    at $r$, $3.8\,r\,\sigma(r)^{2}/G$ (where $\sigma(r)$ is measured in a narrow 
    annulus at $r$) to the stellar mass enclosed. 
    There is almost no difference in the scaling of the two simulations -- i.e.\ 
    the systems appear (in this sense) homologous, but the change in $R_{e}$ is such that 
    evaluating these quantities at $R_{e}$ will give rather different answers for the 
    two systems. 
    \label{fig:dm.frac}}
\end{figure}

How does this effect relate to the ratio of dark matter to stellar
mass? Figure~\ref{fig:dm.frac} plots the cumulative dark matter mass
fraction as a function of radius,
\begin{equation}
f_{\rm DM} = \frac{M_{\rm DM}(<r)}{M_{\rm DM}(<r) + M_{\ast}(<r)} 
\end{equation}
(for convenience we ignore the negligible mass of gas which survives
the merger). We also plot the (trivially related) ratio of total
enclosed mass to total enclosed stellar mass,
\begin{equation}
\frac{M_{\rm tot}}{M_{\ast}} = \frac{M_{\rm DM}(<r) + M_{\ast}(<r)}{M_{\ast}(<r)}. 
\end{equation}
It is clear that the difference as a function of radius, at least from
$\sim100\,$pc to $\gtrsim50\,$kpc, is not large between the two
simulations (which should not be surprising: the difference in their
stellar mass density profiles is primarily evident at small $r$, where
a relatively small fraction of the total mass is enclosed).  However,
if we take the value of these ratios within the effective radii
$R_{e}$ of the stellar light, we obtain two quite different
results. In the gas-rich case, the large central light concentration
yields a small effective radius for the stellar light, within which,
for {\em both} remnants, the baryons dominate the mass.  In the case
with less dissipation, and therefore a weaker central component and
larger effective radius, the ratio is measured further out (at the
larger $R_{e}$), where in both simulations there are about equal
masses enclosed of both dark matter and baryons.

To make a better analogy with observations, we can construct a 
dynamical mass estimator
\begin{equation}
M_{\rm dyn}(<r) \equiv k\,\sigma(r)^{2}\,r / G \, ,
\label{eqn:mdyn}
\end{equation}
where we adopt the factor of $k=\mdynnorm$ for convenience,
as noted in \S~\ref{sec:sims}
(making the estimator not so far from the true mass), and $\sigma(r)$
is the mean (light-weighted) projected velocity dispersion within a
narrow annulus at $r$. We could also use the mean (light-weighted)
dispersion within $r$, $\langle\sigma(<r)\rangle$, which gives a
similar result (but traces the enclosed mass less faithfully).
Figure~\ref{fig:dm.frac} compares the ratio $M_{\rm dyn}/M_{\ast}$ as
a function of radius, which yields a similar (albeit noisier) result
as considering $M_{\rm tot}/M_{\ast}$.  Again, it is primarily the
difference in the effective radii at which this quantity is evaluated,
rather than the absolute value at fixed $r$, that drives the
difference between the gas rich and gas poor mergers.

\begin{figure}
    \centering
    \scaleup
    \plotone{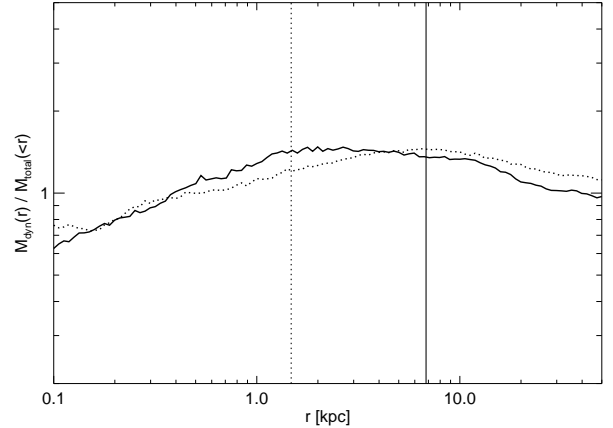}
    \caption{Ratio of the radius-dependent dynamical mass estimator 
    at $r$, $3.8\,r\,\sigma(r)^{2}/G$ (where $\sigma(r)$ is measured in a narrow 
    annulus at $r$) to the true total (stellar plus dark matter) mass enclosed 
    ($\mtrue(r)$), 
    for the simulations in Figure~\ref{fig:dm.frac} (shown in the same style; 
    solid and dotted denote low and high degrees of dissipation). 
    Vertical lines show the effective radii of each simulation. 
    The amount of dissipation makes almost no difference in $\mdyn/\mtrue$ 
    at any radius, nor is there a significant difference when evaluated 
    at the effective radii of either system. The ``homology assumption'' 
    that $\mdyn\propto\mtrue$ is valid in these simulations. 
    \label{fig:mdyns}}
\end{figure}

We obtain the same result using either the true total enclosed mass or
a dynamical mass estimator.  This implies that there is not substantial
kinematic or ``traditional'' structural non-homology between our two
merger remnants. We can, however, check this directly.  

We compare the ratio $M_{\rm dyn}(r)/M_{\rm tot}(r)$ as a function of
radius in Figure~\ref{fig:mdyns}. The quantity $M_{\rm dyn}$ provides
a good tracer of $M_{\rm tot}$, their ratio changing by less than a
factor of $\sim2$ over three orders of magnitude in radius, and the
ratio varies only slightly between the two simulations. The overall
structure of the system has not been significantly modified -- and in
particular over the range of the effective radii of the two mergers,
there is almost no dependence of $M_{\rm dyn}(r)/M_{\rm tot}(r)$ on
either the simulation gas content or radius, consistent with 
``isothermality'' implied by observational 
constraints from gravitational lensing \citep[e.g.][]{rusin03:lensing.structure,
rusin05:lensing.structure,jiang:lensing.baryon.frac}.  In other words, the
increase in the central density of the system owing to the different
degrees of dissipation in the two mergers, while apparent in the
nuclear profile in Figure~\ref{fig:sb.vs.gas}, does not represent a
sufficiently dramatic change in the total mass profile shape to
significantly alter the virial constant (i.e.\ to significantly change
the ratio $\mtrue(r)/\mdyn(r)$).  That the estimator $\mdyn$ is
similar across these extreme cases demonstrates that, despite the
subtle difference in the strength of the central (dissipational or
starburst) component in the stellar light profile, the homology
assumption is valid in a purely empirical sense.

The role of dissipation in driving tilt in the FP is not, then, to
introduce substantial structural non-homology (in the sense of
changing the profile sufficiently to alter the coefficient of the
virial scalings) or to produce kinematic non-homology, but primarily
to decrease the effective radius of the stellar light or mass
distribution. At a smaller radius, the more stellar-dominated
component of the system is sampled, and so when the traditional
dynamical mass estimator is constructed, ($M_{\rm dyn} =
k\,\sigma^{2}\,R_{e} / G$) and compared to the stellar mass, the ratio
$\mdyn/\mstar$ is lower for systems with more dissipation, 
as argued in \citet{robertson:fp} and \citet{dekelcox:fp}.  Note that
here $\sigma$ is either the central velocity dispersion or the average
dispersion within $R_{e}$ (it makes little difference which we
consider as long as we do so uniformly, since the dispersion changes
weakly with radius).

\begin{figure}
    \centering
    \scaleup
    \plotone{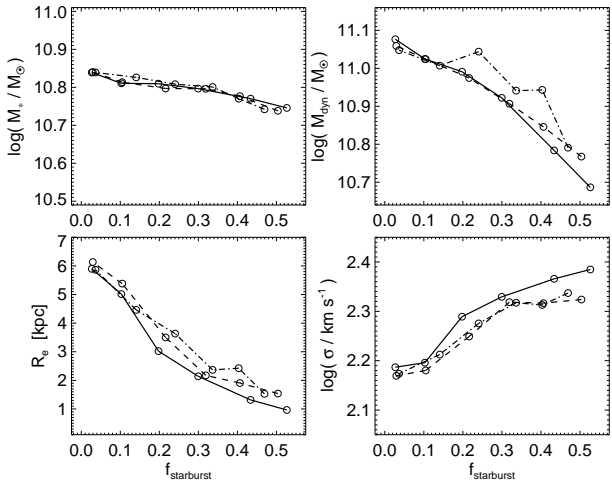}
    \caption{Final stellar mass, dynamical mass 
    $\mdyn=k\,R_{e}\,\sigma^{2}/G$, effective radius, 
    and central velocity dispersion as a function of the 
    dissipational or starburst mass fraction, 
    for three sets of otherwise 
    identical $\sim\lstar$ disk-disk major mergers with 
    varying initial gas content.  Different line styles denote 
    varying orbital parameters. 
    The stellar mass is only weakly affected, but there is a 
    strong scaling of size and dynamical mass 
    (towards smaller sizes and correspondingly 
    lower dynamical masses) with increasing dissipation 
    (in the sense seen in Figure~\ref{fig:dm.frac}). 
    \label{fig:re.fgas}}
\end{figure}

Of course, $\fsb=0.31$ is a rather extreme case, but
Figure~\ref{fig:re.fgas} shows how the effective radius scales with
gas content in another otherwise identical set of simulations of Milky
Way-like spiral mergers that span a range in dissipational
fraction. The effective radii decrease systematically with
dissipational/starburst fraction in a continuous manner.  The velocity
dispersions do become slightly larger as larger central mass
concentrations are assembled, but the effect is much weaker
($\Delta{\log{\sigma}}\sim0.15$\,dex, compared to
$\Delta{\log{R_{e}}}\sim0.8$\,dex, for $\fsb=0-0.5$).  The result is
that the dynamical mass estimated at $R_{e}$ decreases substantially
with gas fraction -- again owing to the effective radius of the
stellar light being drawn further in -- by $\Delta{\log{M_{\rm
dyn}}}\sim0.4$\,dex from low to high gas fraction. Meanwhile, the
stellar mass is almost completely unchanged, since in any case most of
the gas is eventually turned into stars, whether in the pre-merger
disks or in the final starburst. So, driven by the increasing
dissipation yielding smaller effective radii, the ratio $M_{\rm
dyn}/M_{\ast}$ decreases by a factor $\gtrsim2-3$ as the gas fraction
increases.

\breaker
\section{Assessing The Role of Dissipation in Observed Systems}
\label{sec:obs}

Having illustrated how dissipation can alter the sizes of stellar
spheroids and the ratio of dynamical to stellar mass in simulations,
we now proceed to apply our analysis to observed ellipticals.

\subsection{How Do Observed Sizes Scale with Dissipation?}
\label{sec:obs:sizes}

\begin{figure*}
    \centering
    \scaleup
    \plotone{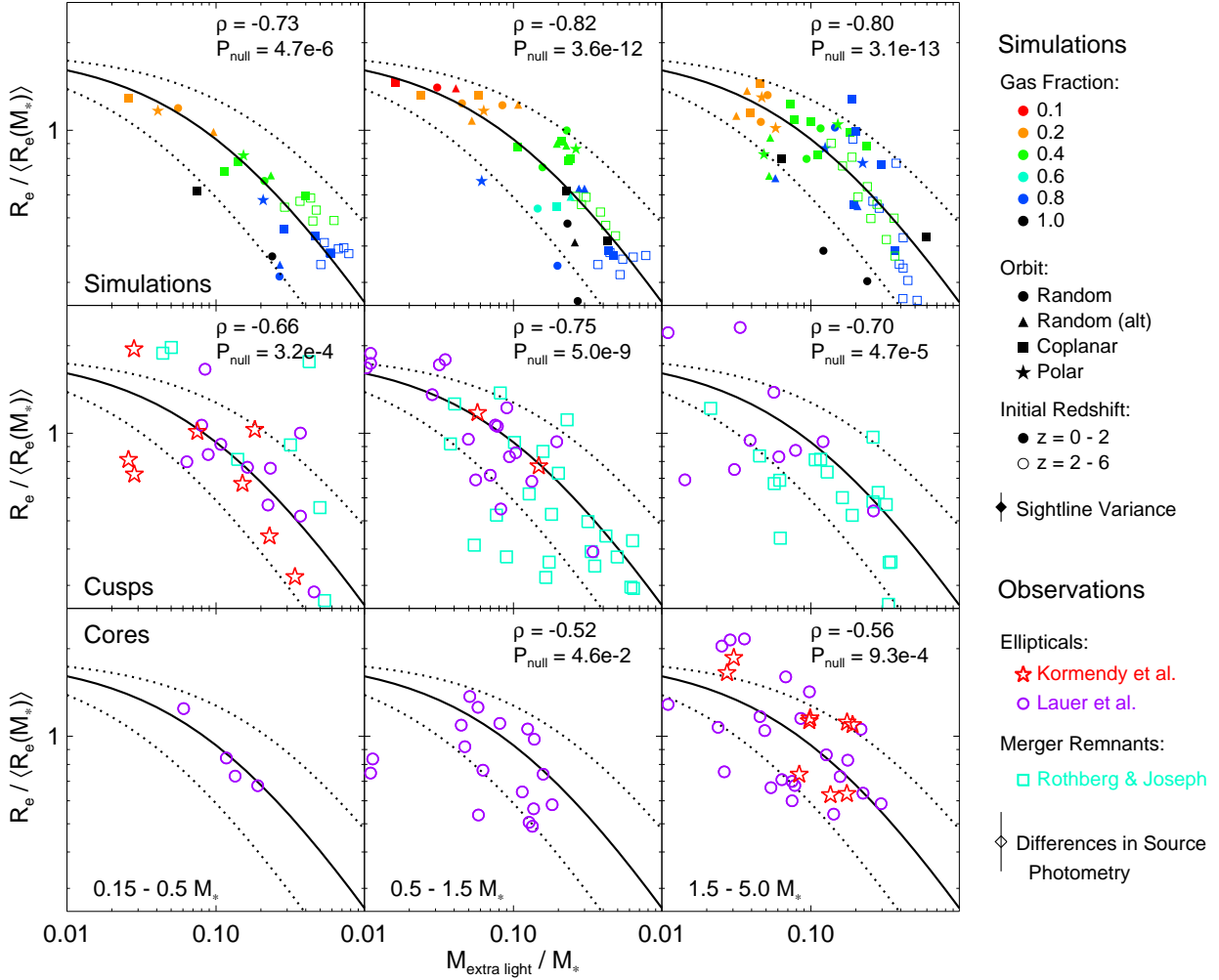}
    \caption{Effective radius $R_{e}$ 
    relative to the 
    median value for all ellipticals of the same stellar mass, 
    as a function of our 
    fitted extra light fractions (the empirical 
    tracer of the dissipational/starburst mass fraction). 
    We show simulated gas-rich 
    merger remnants ({\em top}),  
    observed cusp ellipticals and gas-rich merger remnants 
    ({\em middle}), and observed core ellipticals ({\em bottom}); we 
    use this 
    point notation ({\em key}) throughout. 
    We show this in three bins of stellar mass
    (relative to $\mstar\approx10^{11}\,\msun$, or $M_{V}^{\ast}=-21$).
    Solid (dashed) lines show the mean ($\pm1\,\sigma$) correlation, 
    following the analytic solution for dissipational mergers and 
    fits to our simulation in \citet{covington:diss.size.expectation}. 
    We show the Spearman rank correlation coefficient and probability of the 
    null hypothesis $P_{\rm null}$ (no correlation) in each panel. 
    Simulations and observations exhibit the same 
    behavior: systems with smaller $R_{e}$ at fixed mass have 
    systematically larger extra light fractions (in the sense predicted 
    by Figure~\ref{fig:re.fgas} for the starburst/dissipational mass fractions).
    This implies that, at fixed mass, 
    systems are driven along the fundamental plane by the relative amount of 
    dissipation involved in their formation. 
    This behavior is true regardless of cusp/core 
    status. 
    \label{fig:re.sigma.cusp}}
\end{figure*}

Figure~\ref{fig:re.sigma.cusp} shows how effective radius scales with
extra light mass $f_{\rm extra}$ at fixed mass. We consider three mass
bins, below, at, and above $\sim\mstar$.  For each, we plot the
residual $R_{e}$ relative to that expected for the given stellar mass,
as a function of the fitted extra light fraction. Specifically, we
determine $\langle{R_{e}(M_{\ast})}\rangle$ from the sample of
\citet{shen:size.mass}, and take the ratio of the half mass radius
of each system (determined directly from the light profile, or from 
the fits, it does not change the comparison) to that value.  Our mass 
bins are small enough, however, that this makes little difference 
compared to just e.g.\ considering $R_{e}$ in a given bin.
We show results separately for central cusp and core
ellipticals, as well as our simulations, although the three
distributions are similar.

There is a significant trend: at a given stellar mass, systems with
fractionally more extra light have systematically smaller $R_{e}$
($\sigma$ also rises slightly, but the
effect is weaker and there is more scatter at a given $f_{\rm extra}$,
as expected owing to the role of the extended stellar and dark matter
distribution in setting the central potential and $\sigma$).  In each
case, the simulations and observed systems occupy a similar locus. We
can also construct this plot with the starburst mass fraction $\fsb$
of the best-fitting simulation as the independent variable (instead of
the fitted extra light mass fractions $\fextra$), and find a
correlation of the same nature.
Given two progenitors of known size and mass, it is straightforward to 
predict the size of the remnant of a dissipationless merger, simply 
assuming energy conservation \citep[see e.g.][]{barnes:disk.halo.mergers}; 
in the case of a dissipative merger, 
\citet{covington:diss.size.expectation} use the impulse approximation to 
estimate the energy loss in the gaseous component, followed 
by collapse in a self-gravitating starburst. 
This yields a detailed approximation as a function 
of e.g.\ initial structural and orbital parameters, but if we assume typical initial 
disk structural scalings and parabolic orbits, it reduces to the 
remarkably simple approximation
\begin{equation}
R_{e} \approx \frac{R_{e}({\rm dissipationless})}{1+(f_{\rm sb}/f_{0})}, 
\end{equation}
where $f_{0}\approx0.25-0.30$ and $R_{e}({\rm dissipationless})$ 
is the radius expected for a gas-free ($f_{\rm sb}=0$) remnant. 
(A similar estimate is obtained by \citet{ciotti:dry.vs.wet.mergers} 
assuming, instead of the impact approximation, a constant fractional 
dissipational energy loss). 
We plot this in Figure~\ref{fig:re.sigma.cusp}, with the scatter seen 
in the simulations. At all masses, in both simulated and observed 
cusp and core ellipticals, more dissipational ellipticals are smaller 
in the manner predicted.

This result directly implies that a structural difference -- albeit
not traditional structural non-homology -- plays a key role in
establishing the fundamental plane tilt. At fixed mass, smaller
systems are so because a larger fraction of their mass is formed in a
central dissipational starburst (to the extent that our extra light
fractions recover this dissipational component). This dissipational
starburst is compact, so even though the pre-existing stars are
distributed to large radii, the effective radius is smaller.  In
\papertwo, we considered the light profiles of observed systems along
the $R_{e}-\fextra$ correlation, at fixed stellar mass: it is clear
that observed systems at fixed mass with the largest $R_{e}$ show
profiles close to a pure Sersic law, with little evidence for any
extra light component at $>0.01\,R_{e}$ (indeed, they have
$\fsb\lesssim0.02$). Observed systems in this regime can still be
classified as ``cusp'' ellipticals, but the cusps tend to be prominent
at small radii and (in several cases) somewhat shallow, and contribute
negligibly to the stellar mass. However, moving to smaller $R_{e}$ and
higher $f_{\rm extra}$ at fixed stellar mass, deviations from a Sersic
law at $r\ll\re$ become more noticeable. This is not to say that these
deviations are universal (that the extra light always takes the same
shape/form), but there are increasingly prominent central light
concentrations. If the systems were perfectly homologous (in the
strict sense of the definition), there should be no differences, and
there should be no trend whatsoever in Figure~\ref{fig:re.sigma.cusp}.

Figure~\ref{fig:re.sigma.cusp} also shows that there is little
difference, in this sense, between the size scalings with dissipation
of cusp and core ellipticals.  We demonstrate this in \paperthree, in
both simulations and observations.  In short, this is expected: even
if the systems with cores have expanded via re-mergers, they should
(so long as there is not a wide range in number of re-mergers or
strong systematic dependence of the number of re-mergers on starburst
fraction at fixed mass) grow uniformly, and preserve these trends.  We
might expect some normalization offset: if the mean size-mass relation
after a gas-rich merger is a power law $R_{e}\propto
M_{\ast}^{\alpha}$, and two such systems with mass ratio $f$ (where
$f$ is the mass ratio of the secondary to the primary) are involved in
a dry merger from a parabolic orbit (and preserve profile shape), then
the remnant will increase in radius by a factor
$(1+f)^{2}/(1+f^{2-\alpha})$ relative to the primary. However, it has
also grown in mass, so compared to ellipticals of the same (final)
mass, its relative increase in size is only
$(1+f)^{2-\alpha}/(1+f^{2-\alpha})$. For observationally suggested
values $\alpha=0.56$ \citep{shen:size.mass}, this predicts a $\sim
30\%$ ($\sim0.1$\,dex) relative size increase for major re-mergers
with mass ratios of 1:3 through 1:1. This is easily dwarfed by the
effects of dissipation seen in Figure~\ref{fig:re.sigma.cusp}, which
can alter the sizes of systems by nearly an order of magnitude at
fixed mass.  That there is not a substantial normalization offset
between the trends for cusp and core ellipticals in
Figure~\ref{fig:re.sigma.cusp} therefore implies that typical core
ellipticals have been involved in only a modest number of re-mergers 
\citep[see also][]{ciotti:dry.vs.wet.mergers},
but it is not a strong constraint.  The important point is that
observations demonstrate the dominant factor controlling the sizes of
a ellipticals is the amount of dissipation, even for systems which
have undergone re-mergers.

\subsection{How Does This Change Spheroid Dynamical Masses?}
\label{sec:obs.mdyn}

\begin{figure}
    \centering
    \plotter{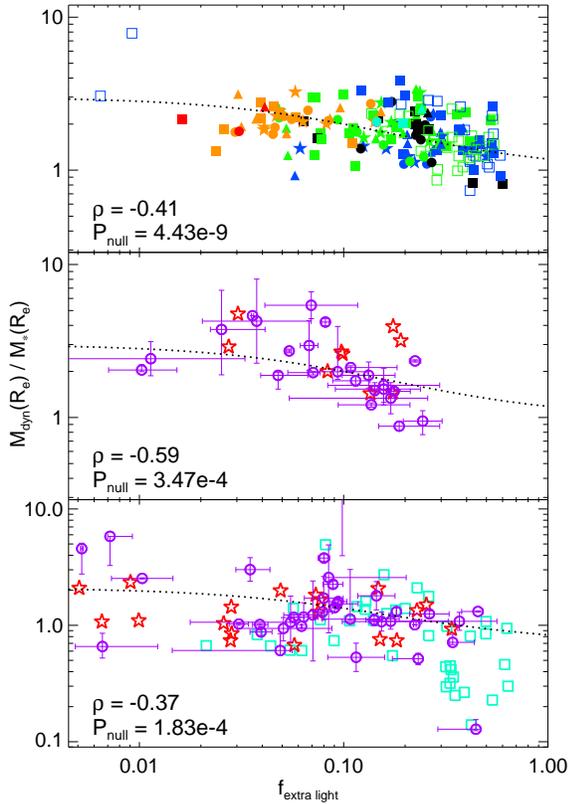}
    \caption{Ratio of dynamical to stellar mass within $R_{e}$, as a 
    function of extra light fraction derived from surface brightness profiles. 
    Results are shown for simulated gas-rich merger 
    remnants ({\em top}), cusp ellipticals and gas-rich merger 
    remnants ({\em middle}), and core ellipticals ({\em bottom}). 
    Dotted line shows an approximation to the 
    mean simulation trend.  
    We show the Spearman rank correlation coefficient ($\rho$) 
    for each sub-sample and the probability $P_{\rm null}$ of 
    the null hypothesis that there is no correlation between 
    $\mdyn/\mstar$ and $\fextra$. 
    Systems with larger degrees of dissipation (larger $\fextra$) 
    have larger ratios $\mdyn/\mstar$, with high 
    significance, in the sense expected 
    from Figure~\ref{fig:re.fgas}. 
    \label{fig:mdyn.fextra}}
\end{figure}

Figure~\ref{fig:mdyn.fextra} plots the quantity $M_{\rm dyn} /
M_{\ast}$ as a function of the amount of dissipation, quantified by
$f_{\rm extra}$. The expected trend based on our fiducial cases in
\S~\ref{sec:diss.fx} and the dissipational models 
\citep{onorbe:diss.fp.model,robertson:fp,dekelcox:fp}
is borne out in our library of simulations over a
wide dynamic range with fairly small scatter, despite a wide range of
total masses, orbital parameters, initial gas fractions, assumptions
regarding the gas pressurization, and other progenitor properties. The
same trend obtains for the observed systems, whether cusp or core
ellipticals.  The scatter is significantly larger among the observed
systems, but this is not surprising, given both the measurement errors
involved and the possibility for more complex growth histories that
our simulations do not completely model.  In a formal sense, the
inverse correlation (i.e.\ smaller $M_{\rm dyn}/M_{\ast}$ at larger
dissipational fraction $f_{\rm extra}$) is highly significant: we plot
the Spearman rank correlation coefficient $\rho$ of each sub-sample,
together with the significance of its deviation from zero. This
$P_{\rm null}$ represents the probability of the null hypothesis of no
correlation between $M_{\rm dyn}/M_{\ast}$ and $f_{\rm extra}$, which
is what would be expected if systems were perfectly homologous or if
extra light did not contract systems and yield smaller $R_{e}$.

\begin{figure*}
    \centering
    \scaleup
    \plotone{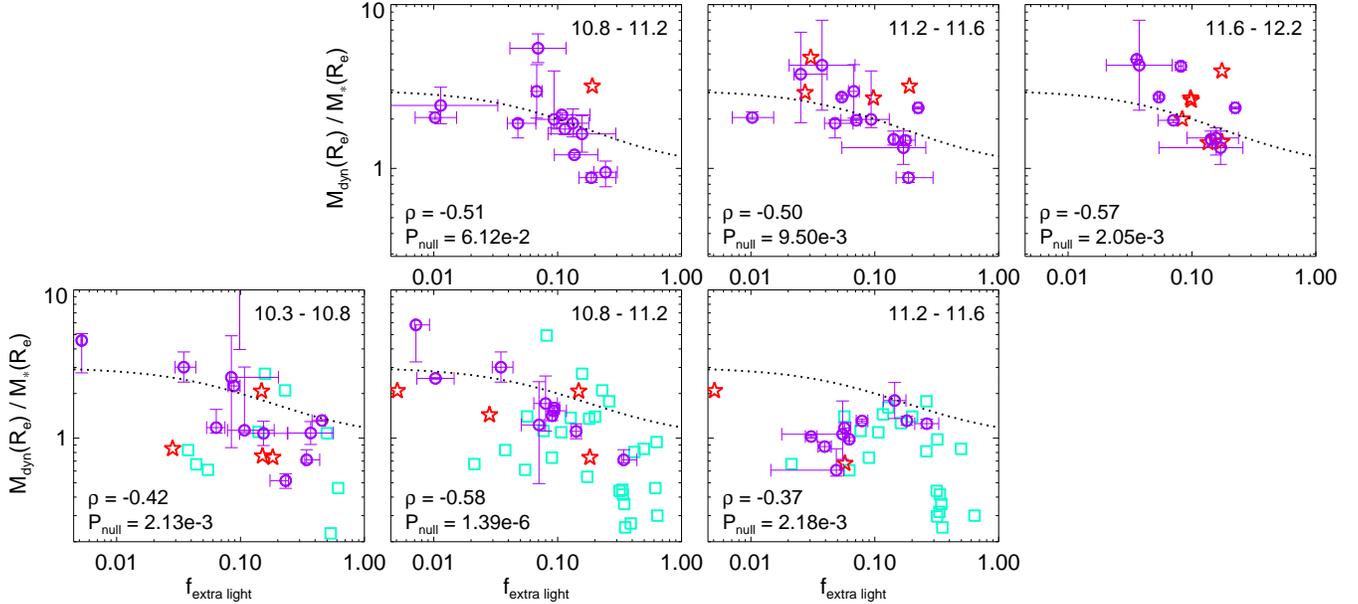}
    \caption{As Figure~\ref{fig:re.fgas}, but in narrow bins of 
    stellar mass (range in $\mstar$ for each bin shown in the upper 
    right of each panel). Cusp ({\em bottom}) and core ({\em top}) 
    ellipticals are shown. The same trend -- towards 
    lower $\mdyn/\mstar$ is seen at fixed $\mstar$. 
    The significance is lower in any one bin owing to the smaller 
    number of points, but the cumulative significance is 
    comparable to or higher than that in Figure~\ref{fig:mdyn.fextra}. 
    \label{fig:mdyn.fextra.mbins}}
\end{figure*}

It is possible, in principle, that the observed correlation could be
indirect.  For example, if lower mass systems happened to have higher
$f_{\rm extra}$, but it had no causal connection to their lower
$M_{\rm dyn}/M_{\ast}$.  In Figure~\ref{fig:mdyn.fextra.mbins} we
therefore consider this same comparison in narrow bins of stellar
mass. We note that, while there is still some width to our bins, they
are sufficiently narrow that we obtain the same answer if we assume
that all systems in the bin have the same stellar mass. We also check
a series of Monte Carlo experiments where we assume that
$\mdyn/\mstar$ and $f_{\rm extra}$ are both pure functions of $\mdyn$
or $\mstar$, but have no dependence on each other, and we find that
the mass bins plotted in Figure~\ref{fig:mdyn.fextra.mbins} are
sufficiently narrow that if this null hypothesis were true, then the
maximal indirect correlation would have a significance of only
$\rho\sim-0.1$ to $-0.3$ (with $P_{\rm null}\approx 0.2-0.8$); so a
significance $P_{\rm null}\ll 0.1$ is indeed meaningful.  We caution
that at a narrow range in stellar mass, there is naturally less
observed dynamic range in dissipation fraction (as expected, since
e.g.\ disks of a given mass have a reasonably narrow range of observed
gas fractions), and the number of objects in each bin is smaller, so
the significance of the inverse correlation in any one bin will be
correspondingly reduced. However, we still see an inverse correlation
in all bins, with significance stronger than expected from the null
hypothesis. Near $\sim \mstar$ where the sample size is large, the
inverse correlation is still highly significant.  Further, while the
significance in any one of the plotted bins may be lower owing to the
binning, the cumulative significance (i.e.\ likelihood of obtaining
such consistent inverse correlations, considering each mass bin as an
independent subsample) is actually quite high ($P_{\rm null} \ll
10^{-6}$).

\begin{figure}
    \centering
    \plotter{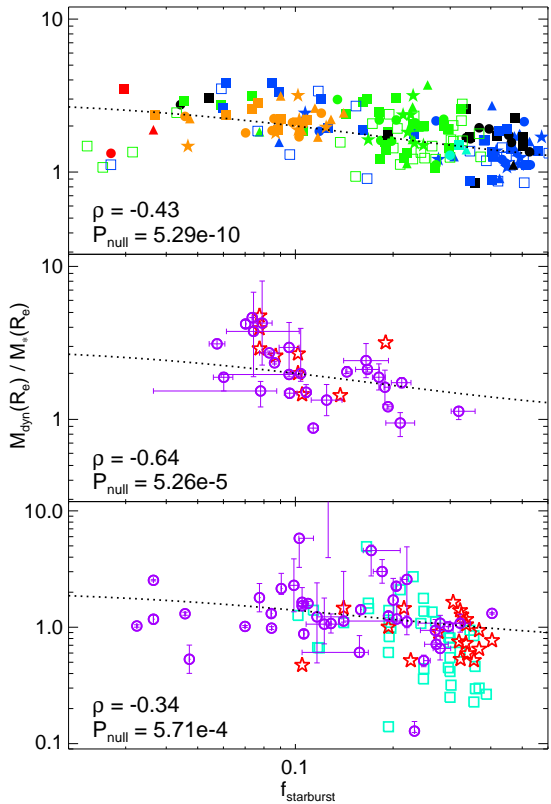}
    \caption{As Figure~\ref{fig:mdyn.fextra}, but using the starburst mass 
    fraction $\fsb$ (the physical mass fraction in our simulations, or that inferred 
    from directly fitting simulations to the observed light profiles for the observations) 
    instead of the purely empirical $\fextra$ as our estimator of the dissipational 
    mass fraction. The results are similar. 
    \label{fig:mdyn.fsb}}
\end{figure}
\begin{figure*}
    \centering
    \scaleup
    \plotone{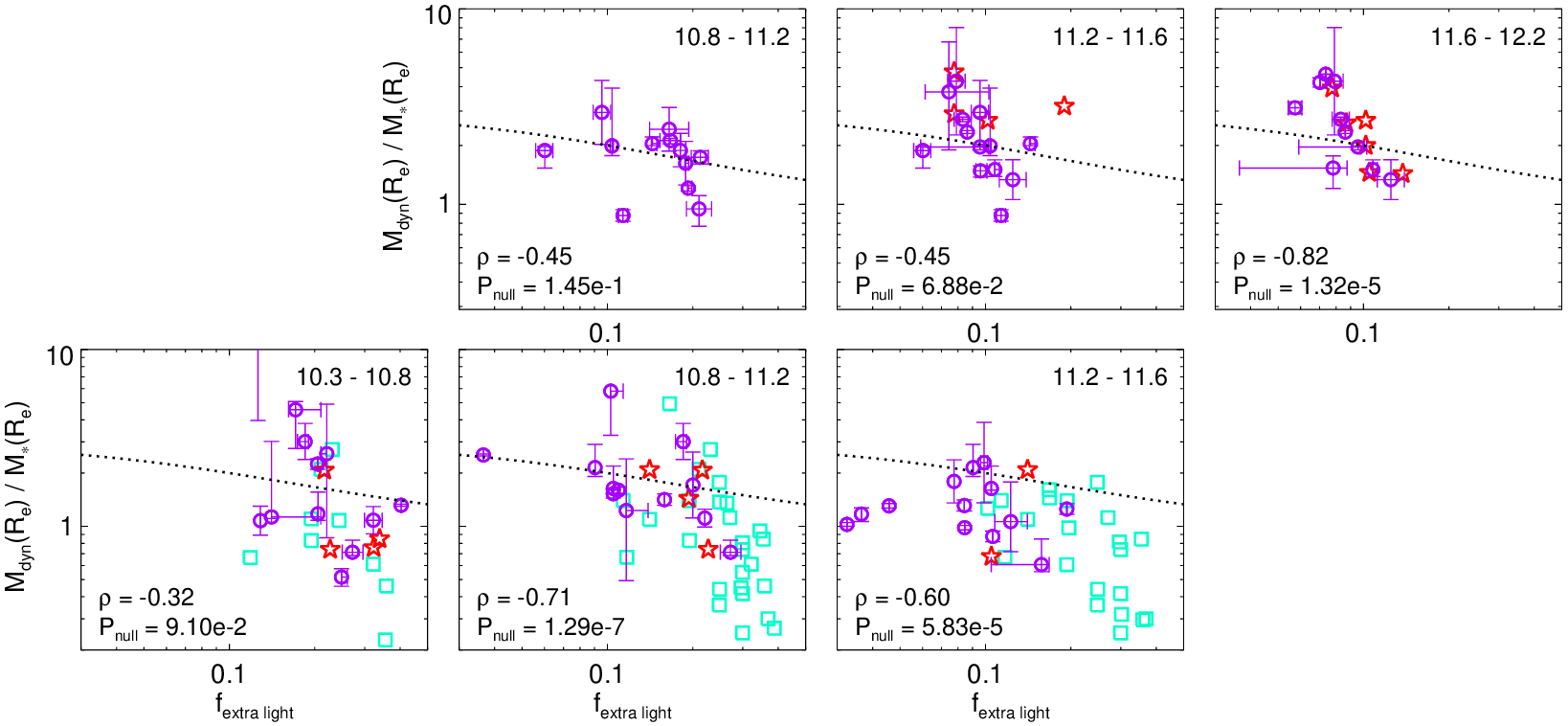}
    \caption{As Figure~\ref{fig:mdyn.fextra.mbins}, but again using the 
    physical starburst mass fraction $\fsb$ instead of the fitted extra 
    light mass fraction $\fextra$. The results are similar in either case. 
    \label{fig:mdyn.fsb.mbins}}
\end{figure*}

Figures~\ref{fig:mdyn.fsb} \&\ \ref{fig:mdyn.fsb.mbins} repeat this
test, using the starburst mass fraction $\fsb$ estimated from fitting
simulations to observed light profiles as the proxy for the
dissipational fraction.  We obtain similar answers to our previous
comparison using the fitted $\fextra$, suggesting that our comparison
is not peculiar to the exact estimator used, so long as it robustly
recovers the physical dissipational component of the galaxy. Again,
the cumulative significance of the inverse correlation -- namely that
$\mdyn/\mstar$ is smaller with increasing $\fsb$ at fixed mass -- is
very high, $P_{\rm null}\sim10^{-8}$.

\breaker
\section{Is Dissipation Necessary to Explain the FP?}
\label{sec:obs.tests}

\begin{figure}
    \centering
    \scaleup
    \plotone{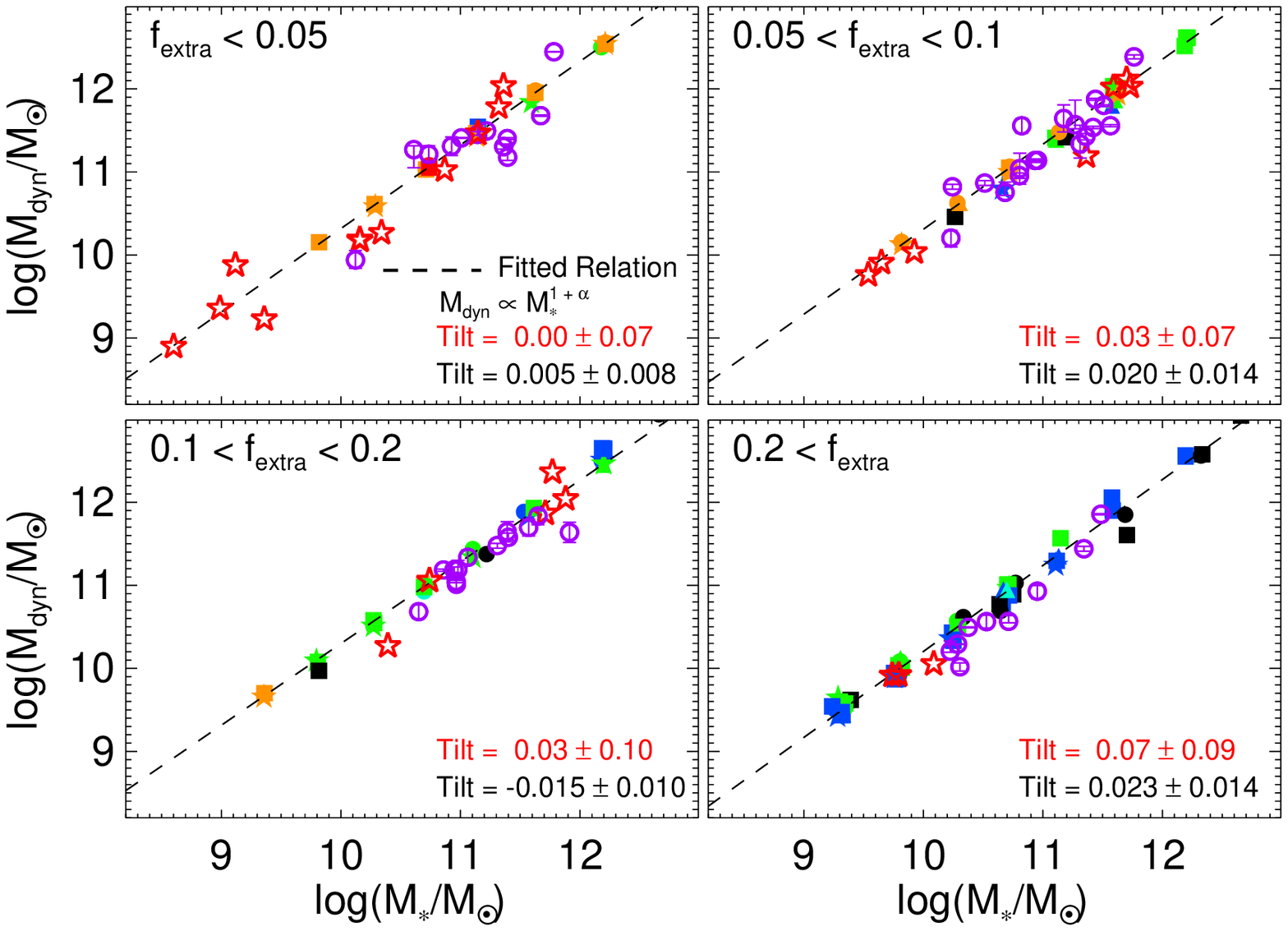}
    \caption{The stellar-mass FP (i.e.\ $\mdyn-\mstar$ correlation), 
    for simulated merger remnants (black solid) and observed ellipticals (open colored points), 
    restricted to systems with a narrow range of extra light/dissipational mass 
    fractions $\fextra$ (as labeled). In each bin of $\fextra$, we fit the simulations 
    and observations separately to a correlation of the form $\mdyn \propto \mstar^{1+\tilt}$, 
    where $\tilt$ quantifies the FP ``tilt.'' The resulting tilts are shown (red upper and black lower 
    values are for observations and simulations, respectively). The tilt is consistent with 
    zero in every bin, and is lower than the expected tilt of the entire population ($\tilt\sim0.2$) 
    by $\sim3\sigma$ -- i.e.\ there is no ``tilt'' in 
    either observed or simulated populations at the same, fixed value of $\fextra$. 
    \label{fig:fp.by.fextra}}
\end{figure}

These comparisons, while demonstrating that the dissipational mass
fraction does indeed correlate with the sizes and ratio $M_{\rm
dyn}/M_{\ast}$ in observed ellipticals (and therefore that it {\em
could} be the source of the tilt in the FP) do not necessarily
indicate {\em how much} of the FP tilt derives from dissipation.  To
test this, we construct the fundamental plane in
Figure~\ref{fig:fp.by.fextra} -- specifically the correlation between
$M_{\rm dyn}$ and $M_{\ast}$ -- in bins of similar fitted ``extra
light'' mass fractions $\fextra$ (i.e.\ we consider the FP for systems
only with similar amounts of dissipation).  Because we are binning by
extra light mass and still attempting to fit a correlation, we include
both cusp and core ellipticals in the observed sample. Separating the
two gives identical results, but the significance is reduced owing to
the limited dynamic range from further splitting the sample at fixed
$f_{\rm extra}$. We fit a power-law to the data (and separately, to
our simulations) in each bin, of the form $\mdyn\propto
\mstar^{1+\alpha}$ (Equation~\ref{eqn:tilt}), where $\tilt$ is the FP
``tilt.'' Because neither $\mdyn$ nor $\mstar$ should properly be
considered an independent variable, we quote the results from fitting
the least-squares bisector to the correlation. However, this makes
little difference (the least-squares best-fit $\mdyn(\mstar)$ and
$\mstar(\mdyn)$ relations are both consistent and agree).
We show this using the typical estimator $\mdyn$ here, but
have also considered the true total enclosed mass $\mtrue$ within
$\re$ and find identical results (as expected, since we show in
\S~\ref{sec:diss.fx} and \S~\ref{sec:obs.tests.mtot} that $\mdyn$ is a
good proxy for $\mtrue$).

The results are unambiguous in both our simulations and the observed systems. 
The tilt in any given bin is negligible -- 
within $1\,\sigma$ of $\tilt=0$ in every case. In other words, at fixed $f_{\rm extra}$, 
we recover the virial correlation with 
constant $\mdyn/\mstar$. 
Note that in several of these bins the dynamic range in mass is 
still large ($\sim2-3\,$orders of magnitude in $\mstar$), and 
the number of observed systems is sufficiently large that {\em if} a 
substantial tilt like that observed for the entire population 
($\tilt\sim0.2$) were present in the bin, we should see it. This 
is reflected in the quoted errors, which show that the tilt in most bins 
is inconsistent with that observed for the global population at 
$\sim3\,\sigma$ significance 
(in the sense that it is much smaller, consistent with a pure virial 
correlation). 
If we combine the data (i.e. normalize out 
the mean $\mdyn/\mstar$ at each bin in $f_{\rm extra}$ and fit, or take the cumulative 
significance of the bins shown) we obtain 
$\tilt = 0.028\pm 0.040$, consistent with zero and $\sim4-5\,\sigma$ below 
the observed tilt for the entire population.

\begin{figure}
    \centering
    \scaleup
    \plotone{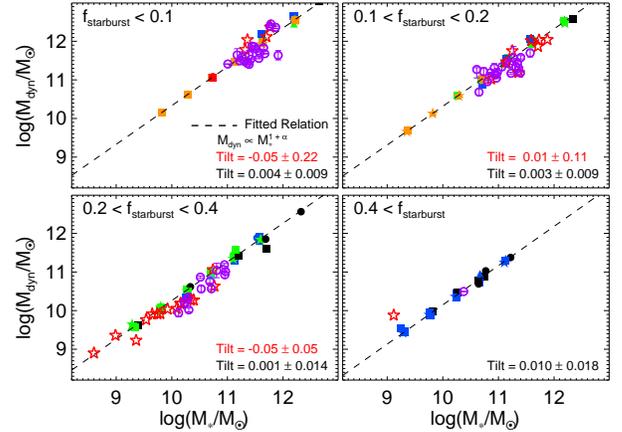}
    \caption{As Figure~\ref{fig:fp.by.fextra}, but again using the 
    physical starburst mass fraction $\fsb$ instead of the fitted extra 
    light mass fraction $\fextra$. The results are similar in either case -- 
    there is a highly significant lack of ``tilt'' at fixed dissipational fraction, despite 
    the observations and simulations still spanning a large baseline ($\sim2\,$dex) 
    in mass. 
    \label{fig:fp.by.fsb}}
\end{figure}


Figure~\ref{fig:fp.by.fsb} repeats this exercise, using the starburst
mass fraction $\fsb$ (estimated from the physical starburst mass
fractions in the best-fitting simulations) as the proxy for
dissipational mass fraction, instead of the fitted extra light
fraction $\fextra$.  In either case we reach identical conclusions for
both the simulations and observations. The inferred tilt at fixed
dissipational fraction is always consistent with zero, in some cases
scattering to $\tilt<0$ (i.e.\ the opposite sense of what is observed;
but again these are all consistent with $\tilt=0$).  In short, we have
demonstrated empirically that without invoking some systematic
dependence of dissipational fraction on mass, there is no FP
tilt. Dissipation is observationally {\em necessary} to explain the FP
tilt.

\begin{figure}
    \centering
    \plotone{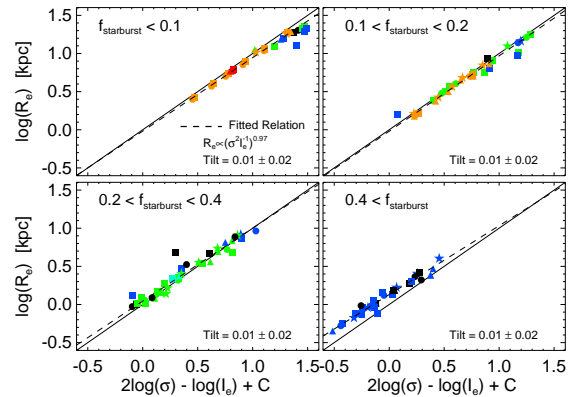}
    \caption{The FP in narrow bins of $\fsb$, for our simulations, as Figure~\ref{fig:fp.by.fsb}, 
    but showing an alternative projection of the FP ($R_{e} \propto (\sigma^{2}\,I_{e}^{-1})^{1+\tilt}$). 
    The result (no tilt at fixed $\fsb$) is the same; but in such a projection 
    (with a small sample) small errors in 
    $R_{e}$ will tend to amplify the observed tilt (and the physical meaning of the 
    tilt is less clear). 
    \label{fig:fp.by.fextra.reff}}
\end{figure}

Figure~\ref{fig:fp.by.fextra.reff} also shows this for the effective radius projection of 
the fundamental plane. We obtain identical answers in this case (for both 
simulations and observations) but note that the sense of the correlation is such 
that, if the baseline in $R_{e}$ is comparable to the scatter in observed points, 
severe biases can be introduced. We see this reflected in the fact that, in this case, 
fitting $R_{e}(R_{\rm pred})$ versus $R_{\rm pred}(R_{e})$ 
(where $R_{\rm pred}\propto \sigma^{2}\,I_{e}^{-1}$ is the 
virial expectation) yield rather different slopes. 
Therefore, 
while we show our simulations (which have small dispersion in $R_{e}$ 
across sightlines), we refrain from fitting the observations in such narrow bins 
in this particular projection of the FP. 

\breaker
\section{Predicting the FP: Is Dissipation Sufficient?}
\label{sec:obs.tests.2}

\subsection{Systematic Dependence of Dissipation on Mass: What Is 
Expected and Observed}
\label{sec:obs.tests.diss.frac}

We have shown that $\mdyn/\mstar$ depends on the degree of
dissipation, reflected in $\fextra$ or $\fsb$, and that at {\em fixed}
$\fextra$ or $\fsb$ there is essentially no FP tilt. This already
implies that dissipation {\em must} be responsible for the majority of
the tilt.  But if we knew how much dissipation was expected as a
function of galaxy mass (i.e.\ the mean expected mass fraction in a
dissipational starburst, for mergers of systems of a given mass), we
could entirely {\em predict} the FP and its tilt.

\begin{figure*}
    \centering
    \scaleup    
    \plotone{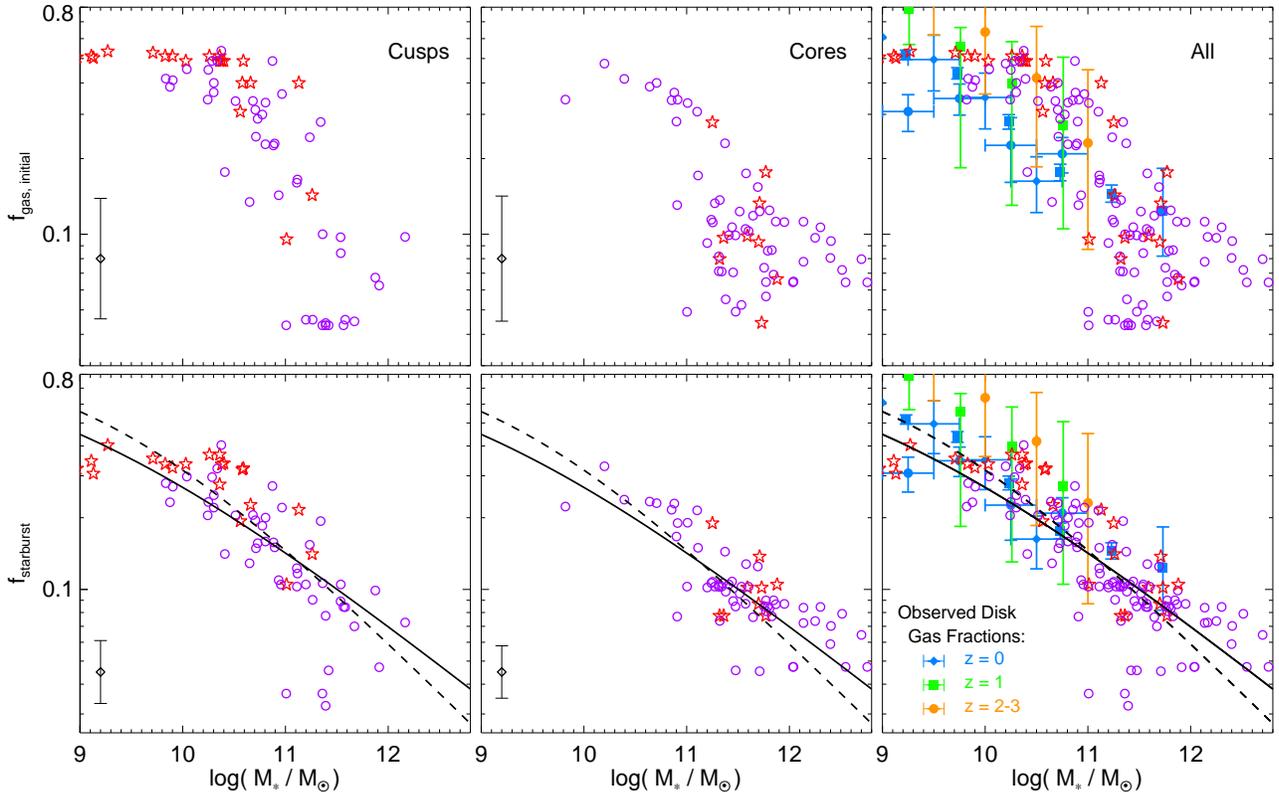}
    \caption{Inferred gas content (dissipational/starburst fraction) of 
    elliptical-producing mergers as a function of stellar mass. 
    Initial gas fraction ({\em top}) and physical final starburst mass 
    fraction ({\em bottom}) corresponding to the best-fit simulations to 
    each observed system in the samples of 
    \citet{lauer:bimodal.profiles} (circles) 
    and \citet{jk:profiles} (stars) are shown, 
    with the typical $25-75\%$ allowed range (error bar).
    We show results separately for cusp ellipticals ({\em left}), core ellipticals 
    ({\em center}), and both together ({\em right}). 
    Dashed (solid) line shows the fit to the data (Equation~\ref{eqn:fgas.m}) 
    in cusp (core) ellipticals. 
    Colored points with error bars indicate the mean (and $\pm1\,\sigma$ 
    range in) disk gas fractions at the same stellar mass, at 
    $z=0$ \citep[][blue diamonds, squares, and circles, respectively]
    {belldejong:tf,kannappan:gfs,mcgaugh:tf}, 
    $z=1$ \citep[][green squares]{shapley:z1.abundances}, and 
    $z=2$ \citep[][orange circles]{erb:lbg.gasmasses}. There is a clear trend of increasing 
    dissipation 
    required to explain elliptical profiles at lower masses 
    (significant at $>8\,\sigma$ in either core or cusp subsamples 
    separately), 
    in good agreement with the observed trend in progenitor disk 
    gas fractions over the redshift range where 
    cusp ellipticals are formed, and with what is invoked to explain  
    the observed densities and fundamental plane correlations of ellipticals 
    \citep[e.g.][]{kormendy:dissipation,hernquist:phasespace}.
    The best-fit trends in cusp and core populations are statistically 
    identical: i.e.\ the dissipational/extra light component is 
    preserved regardless of possible re-mergers. 
    \label{fig:fsb.mstar}}
\end{figure*}

Figure~\ref{fig:fsb.mstar} shows the mean dissipational mass fractions 
of both cusp and core ellipticals, as a function of mass, 
derived in \papertwo\ and \paperthree. There is a clear 
systematic trend (discussed in those papers):
lower-mass systems have systematically higher 
dissipational mass fractions. 
To the extent that $M_{\rm dyn}/M_{\ast}$ depends on 
dissipational fraction ($\fextra$ or $\fsb$), 
then, the existence of a systematic dependence of $\fextra$ and $\fsb$ 
on mass will yield a systematic dependence of 
$M_{\rm dyn}/M_{\ast}$ on mass. We can use 
the observed dependence of the dissipational mass fraction 
on mass to predict and empirically estimate the amount of 
tilt in the FP contributed by systematic trends in dissipation as a 
function of mass. 
For convenience, we noted in \papertwo\ that the dependence of 
dissipational fraction $\fsb$ on mass in cusp ellipticals 
(or observed gas-rich merger remnants) can be well-fitted by 
\begin{equation}
\langle f_{\rm starburst} \rangle \approx 
{\Bigl[}1+{\Bigl(}\frac{M_{\ast}}{M_{0}}{\Bigr)}^{\alpha}{\Bigr]}^{-1}, 
\label{eqn:fgas.m}
\end{equation}
with  $(M_{0},\ \alpha)=(10^{9.2\pm0.2}\,\msun,\ 0.43\pm0.04)$ 
(shown in Figure~\ref{fig:fsb.mstar}), 
and a roughly constant factor $\sim1.5-2$ scatter at each $\mstar$. 
In \paperthree\ we demonstrated that an essentially identical scaling 
(a statistically equivalent $(M_{0},\ \alpha)=(10^{8.8\pm0.3}\,\msun,\ 0.35\pm0.05)$) applies 
to core ellipticals, as expected if cusp ellipticals are indeed their 
progenitors (since identical re-mergers will conserve dissipational or ``extra light'' 
mass fractions). 

We also demonstrated in \papertwo\ and \paperthree\ that this
empirically measured systematic dependence of dissipational mass
fraction on stellar mass agrees well with the observed dependence of
disk galaxy gas fractions on mass. This is exactly what is expected if
ellipticals are (at least originally) formed in gas-rich, disk-disk
mergers (regardless of whether or not they subsequently experience
re-mergers).  We show this in Figure~\ref{fig:fsb.mstar} with the
observed disk gas fractions as a function of stellar mass from
\citet{belldejong:tf,kannappan:gfs,mcgaugh:tf} at $z=0$, as well as at
$z=1$ \citep{shapley:z1.abundances} and $z=2$
\citep{erb:lbg.gasmasses}.  At all these redshifts, the trend of gas
fraction is similar, and the observed gas fractions evolve relatively
weakly with redshift (by a factor $\sim1-2$), bracketing the range of
dissipational mass fractions observed in ellipticals at each mass.

In other words, regardless of the exact times of formation, a
systematic trend similar to that observed in the dissipational
fractions of ellipticals is inevitable if disks are the progenitors of
ellipticals.  Note that even if there is a systematic dependence of
the time of first gas-rich merger on stellar mass in this range, the
mixing of formation redshifts would be less important than the mean
dependence of disk gas fraction on mass (if we assume the stellar
population ages date the merger times -- which is an upper limit to
the magnitude of this effect, since the stellar ages at least in part
reflect the similar trend in stellar ages of disks as a function of
mass -- then the systematic effects added by this age scaling are
not important).  The dependence of disk gas fractions on mass is
therefore the primary driver of the observed dependence of
dissipational mass fraction on elliptical mass.

Among other things, this implies that we could {\em predict} the
dissipational mass fractions of ellipticals as a function of their
mass, based on the observed disk gas fractions at each mass.  We
therefore pursue the following exercise: we consider the consequences
of adopting the observed scaling of disk gas fractions with mass as
our expectation for the dissipational fractions in ellipticals as a
function of mass.  Since we have just shown they are equivalent, we
obtain identical results if we select the empirically fitted
dissipational fractions of ellipticals as a function of mass; in
either case, the important thing is that there is a well-defined,
physically motivated and observationally confirmed systematic scaling
of dissipational fraction with mass.

\subsection{The Consequence for the FP}
\label{sec:obs.tests.diss.tilt}

Given the known scaling of disk gas fractions as a function of mass,
we can then focus on the subsample of simulations that obey this
scaling.  That is, at a particular mass scale, we know the range of
disk gas fractions seen empirically. We therefore consider only
simulations within that range of gas fractions, as opposed to others
in our library that are more gas rich or gas poor than typical
observed disks of the same stellar mass.  We have considered the
effects of redshift evolution (using the mean age of ellipticals as a
function of their stellar mass as a proxy for their merger time, we
adopt the gas fraction of disks of the appropriate mass {\em at that
redshift}), and find it makes relatively little difference (although
some secondary correlations along the FP relating to stellar
populations and the age of systems will be discussed in
\S~\ref{sec:discuss:pred}). 
This restriction imposes a mean
dependence of extra light or starburst mass fraction on stellar mass
similar to that observed in ellipticals and anticipated from disk gas
fractions.  Given this constraint, we can consider the FP and other
correlations obeyed by just these objects: i.e.\ we can ask the
question: What are the scalings obeyed by systems that have the same
systematic dependence of dissipational fraction on mass as expected
and observed in the real Universe?

\begin{figure*}
    \centering
    \plotone{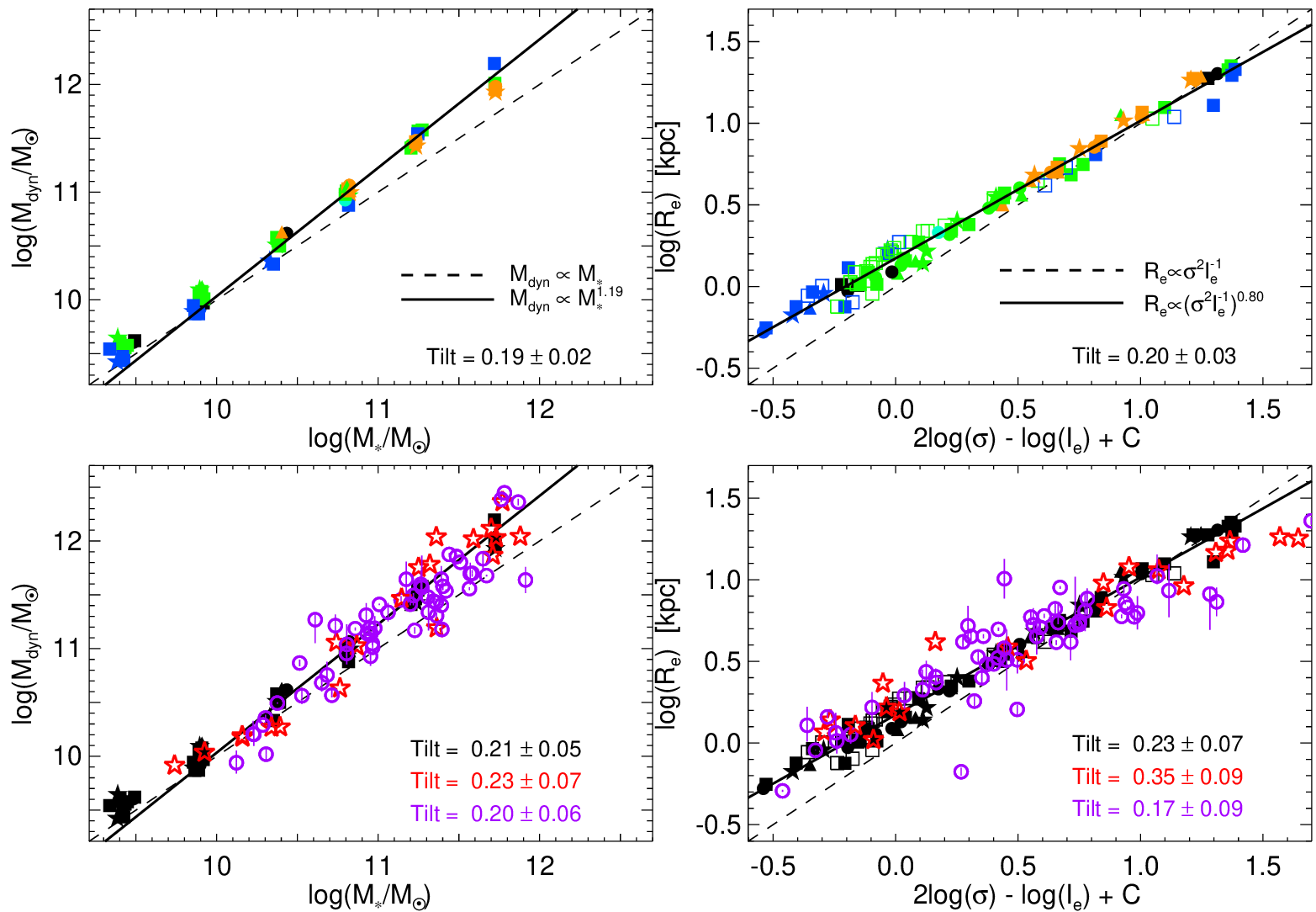}
    \caption{The FP for all observed ellipticals ({\em bottom}), 
    in the style of Figure~\ref{fig:fp.by.fextra} ({\em left}) and 
    Figure~\ref{fig:fp.by.fextra.reff} ({\em right}), with the measured tilt 
    (black for all observed points; red and violet show the tilt fitted for the 
    \citet{jk:profiles} and \citet{lauer:bimodal.profiles} subsamples, respectively). 
    The standard tilt $\tilt\sim0.2$ is now apparent. 
    We compare with the predicted correlation for our simulations ({\em top}), where we consider 
    only simulations that match (within the observed factor $\sim2$ scatter) 
    the observed systematic correlation between e.g.\ progenitor disk 
    gas fractions and stellar mass, or (equivalently) between dissipational/starburst 
    mass fraction and stellar mass (shown in Figure~\ref{fig:fsb.mstar}). 
    Accounting for the systematic dependence of dissipation on mass (owing 
    to the dependence of gas content on mass in the progenitor disks), we obtain a 
    FP tilt in good agreement with that observed, with comparable small scatter. 
    \label{fig:fp.pred}}
\end{figure*}

Figure~\ref{fig:fp.pred} shows the results of this exercise. Now that
we account for the cosmologically expected and observed dependence of
dissipational content on mass, the FP is ``tilted.'' Fitting a power
law, we obtain a value for this tilt of $\tilt=0.19\pm0.02$.  We
compare this to the observed systems, plotting the data on the same
footing, since we have constructed a cosmologically representative
comparison sample in our simulations.  We obtain an observed tilt of
$\tilt=0.21\pm0.05$, the canonical value in the literature. More
important (since any formal fitted tilt will vary if fit over
different dynamic range), we compare the predicted scaling directly to
actual ellipticals, and find that the simulations and observed systems
occupy a statistically identical locus in $\mdyn-\mstar$ space. As a
check, we fit the \citet{jk:profiles} and
\citet{lauer:bimodal.profiles} subsamples independently, and obtain
consistent values for the tilt.

We show the tilt again in terms of $\re$, and obtain a similar result
(for this projection, the formal tilt value is expected to be similar
to that for $\mdyn-\mstar$).  Fitting the observed samples, we find
that they trace an identical locus -- but the power-law fits in this
case are more sensitive to dynamic range (apparent by the fact that
the formal tilt from \citet{jk:profiles} and
\citet{lauer:bimodal.profiles} are slightly discrepant, despite the
points tracing each other where they overlap). As discussed above,
caution should be used fitting in this FP projection (here, the
dynamic range is much larger than the fit errors in $R_{e}$, so there
are not large biases introduced in such observational fits, but in a
more limited subsample, results fitted in this fashion are less robust
than those fitted in the more physical $\mdyn-\mstar$ space, where the
axes are independent).  Given the larger uncertainties in this case,
the agreement is reasonable.  Note that the absolute values of
$\mstar$ and $\mdyn$ and zero-points here are sensitive to our adopted
virial coefficients and IMF assumptions, but the tilt and our
meaningful comparisons are not.

\begin{figure}
    \centering
    \plotone{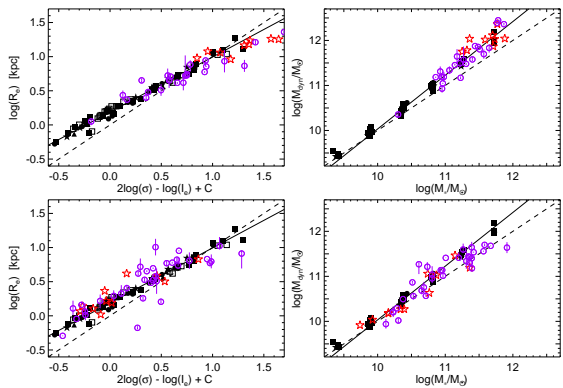}
    \caption{The observed FP, compared with our simulation predictions 
    as in Figure~\ref{fig:fp.pred}, but divided into core ({\em top}) and 
    cusp ({\em bottom}) ellipticals. The simulation predictions agree 
    well with the data in both
    cases, consistent with theoretical modeling that 
    even a modest number of re-mergers do not move systems significantly
    off the FP correlation established by the original, gas-rich mergers. 
    \label{fig:fp.pred.split}}
\end{figure}

Figure~\ref{fig:fp.pred.split} compares this result with the samples
divided into cusp and core ellipticals.  For both classes of objects,
the agreement with the simulations is good, and the two appear to
trace a continuous FP correlation. This is not surprising: it has been
shown that massive, boxy ellipticals with cores trace a continuous FP
with less massive, disky, cuspy ellipticals \citep[at least at the
massive end, where most of the core population resides;
see][]{gerhard:giant.ell.dynamics,
vonderlinden:bcg.scaling.relations}, and we have shown that the
scaling of $M_{\rm dyn}/M_{\ast}$ with dissipational fraction (and
systematic dependence of dissipational fraction on mass) are similar
in both cusp and core ellipticals (and, in simulations in \paperthree,
we explicitly demonstrate that this should be true for both the
original gas-rich merger remnants and dry re-merger
remnants). Furthermore, a number of studies
\citep{capelato:dry.mgr.fp,dantas:dry.mgr.fp,nipoti:dry.mergers,
boylankolchin:mergers.fp,robertson:fp} have shown that dissipationless
spheroid-spheroid re-mergers (popular as a mechanism for producing
cores in ellipticals) tend to preserve the FP, provided that the
number of re-mergers is modest; we discuss these issues further in
\S~\ref{sec:remergers}.

\begin{figure*}
    \centering
    \plotone{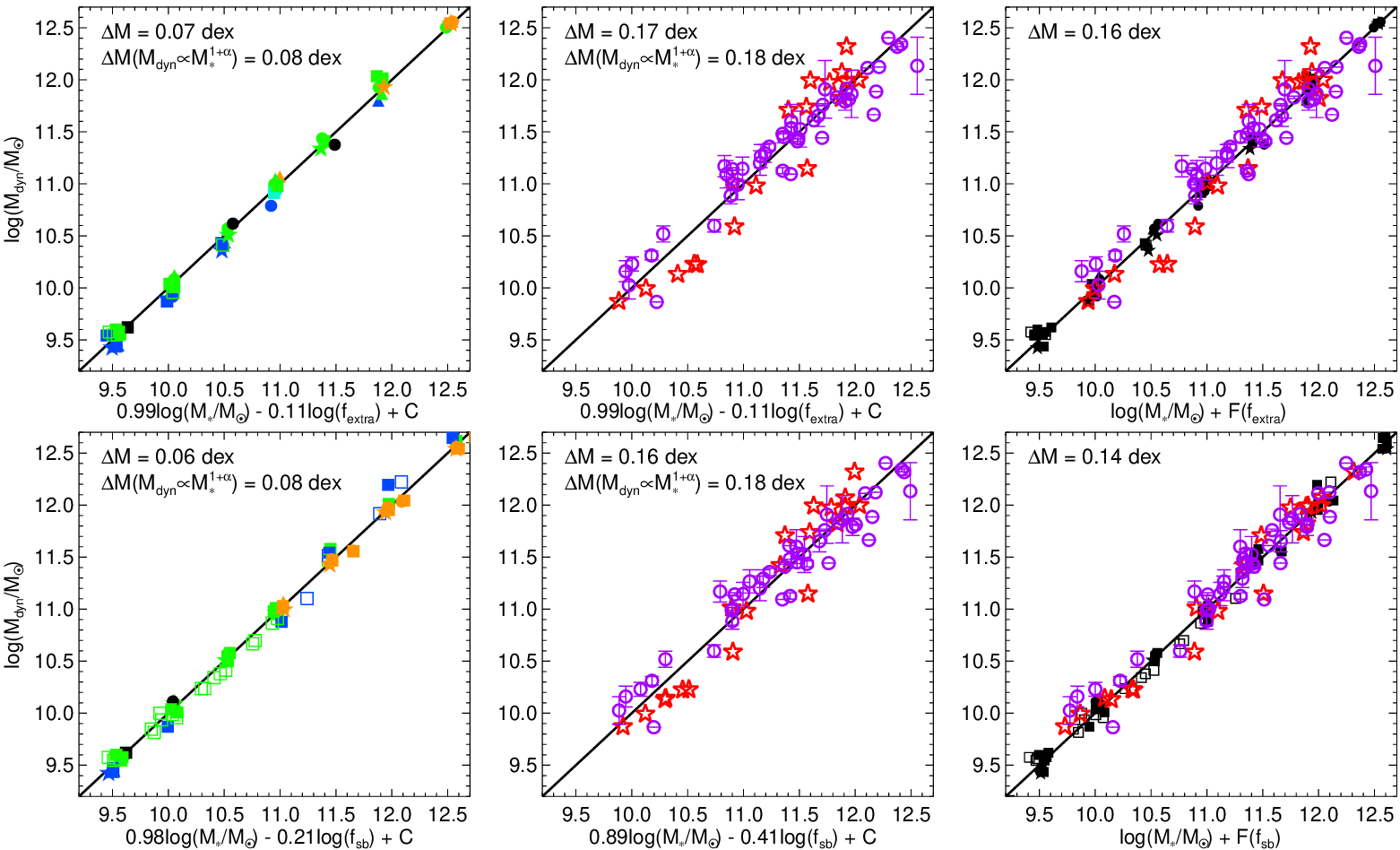}
    \caption{A ``more fundamental'' representation of the 
    FP in simulations ({\em left}; points from Figure~\ref{fig:fp.pred}) 
    and observations ({\em center, right}): $\mdyn$ 
    as a bivariate function of $\mstar$ and fitted extra light fraction ($\fextra$, {\em top}) or 
    inferred starburst fraction ($\fsb$, {\em bottom}). The best-fit correlation 
    recovers $\mdyn\propto\mstar$ with the residual dependence -- i.e.\ the ``tilt'' -- 
    owing entirely to the degree of dissipation.    
    Compared to the FP in terms of just $\mdyn$ and $\mstar$, 
    the scatter is slightly reduced (but not much, owing to the uncertainties introduced 
    in estimating $\fextra$ or $\fsb$). 
    Adopting $\mdyn\propto\mstar$ with the additional functional dependence of 
    $\mdyn/\mstar$ on $\fextra$ from Figure~\ref{fig:mdyn.fextra} yields a
    similar correlation ({\rm right}) with further slightly reduced scatter.     
    \label{fig:fp.w.f.m}}
\end{figure*}

Given that the FP tilt in Figure~\ref{fig:fp.pred} arises owing to 
dissipation -- where we have shown in Figure~\ref{fig:fp.by.fextra} 
that at the same degree of dissipation, 
the virial correlation (no tilt) is recovered -- it should be possible 
to explicitly factor out at least some of this tilt. 
Figure~\ref{fig:fp.w.f.m} attempts this exercise: we 
determine the best-fit bivariate function 
$\mdyn(\mstar,\,\fextra)$. For simplicity and to minimize the 
free parameters involved, we (for now) adopt a simple 
power-law parameterization $\mdyn\propto\mstar^{a}\,\fextra^{b}$, 
and determine the best-fit parameters $a$ and $b$ such that the 
perpendicular scatter (i.e.\ scatter about the least-squares bisector) 
is minimized; we then repeat this using the estimated starburst 
mass fraction $\fsb$ instead of the fitted extra light fraction $\fextra$. 
In both cases, and regardless of whether we fit to the 
observations or to the simulations, the best-fit correlation 
is essentially the virial correlation (i.e.\ $a\approx1$) coupled with a 
dependence on $\fextra$ or $\fsb$. In other words, the 
best-fit bivariate FP assigns the FP ``tilt'' entirely to the 
role of dissipation -- at fixed degree of dissipation, there is 
no tilt. Explicitly including the degree of dissipation in this manner 
demonstrates that it is indeed ``as fundamental'' as any bivariate correlation 
(it is preferred with respect to a purely mass-dependent $\mdyn/\mstar$). 
These fitted FP projections are nearly identical to the simplest predicted 
FP in dissipation-driven models (also shown in 
Figure~\ref{fig:fp.w.f.m}): namely, that $\mdyn\propto\mstar\,F(\fextra)$, 
where the function $F(\fextra)\propto\mdyn/\mstar$ has the form seen 
in Figure~\ref{fig:mdyn.fextra} (note that this is 
more complex than a simple power-law). 

In principle, this explicit inclusion of an $\fextra$ or $\fsb$ dependence 
should be able to account for some of the scatter in the FP, and 
therefore might provide a plane with smaller scatter. In practice, 
the scatter in e.g.\ $\mdyn(\mstar,\,\fextra)$ is only slightly reduced 
relative to that in just $\mdyn(\mstar)$. The reasons for this are twofold.
First, as discussed in \S~\ref{sec:obs.tests.scatter}, the majority of the 
scatter in the FP owes not to different degrees of dissipation but 
to the combination of 
sightline-to-sightline variations, dispersion in progenitor disk properties, 
and measurement errors. Second, as discussed in \S~\ref{sec:data:proxies} 
and \papertwo, an estimator such as $\fsb$ or $\fextra$ is of course 
not a perfect tracer of the 
true starburst mass fraction; with typical factor $\sim2$ uncertainties. 
In practice, then, most of the reduction in scatter from factoring out 
$\fextra$ is negated by scatter introduced in the variation between 
$\fextra$ and the ``true'' starburst mass fraction. 
The result is that the scatter in the FP in terms of $\mstar$ and 
$\fextra$ is comparable to or slightly less than that 
in terms of $\mstar$ alone. Owing to the uncertainties in estimating 
$\fextra$, then, this FP representation is not necessarily a 
substantially improved observational predictor of $\mdyn$ or e.g.\ 
$R_{e}$ at fixed $\sigma$ and $\mstar$. 
However, the FP in these terms is certainly 
comparable to the ``traditional'' FP in how tightly it relates 
$\mdyn$ and $\mstar$, and it is ``more fundamental'' in terms of 
a physical explanation for the FP tilt. 

Because our primary focus in this paper is the understanding of the 
observed $z=0$ FP correlations, we reserve a more detailed modeling 
of the evolution of the physical FP (in terms of e.g.\ stellar and dynamical 
mass) with redshift (as e.g.\ typical merger histories and progenitor 
gas content evolve) for future work \citep{hopkins:cusps.evol}. For now, 
we simply note 
that the evolution in physical parameters considered here is predicted to 
be weak, in agreement with direct 
observational constraints from high redshift weak lensing \citep{heymans:mhalo-mgal.evol} 
and optical studies \citep[note that there is expected 
evolution in optical bands owing to stellar population effects; these represent 
complimentary constraints on elliptical formation histories; see e.g.][]{alighieri:fp.evolution,
treu:fp.evolution,vanderwel:fp.evolution,vandokkum:fp.evol}. 
To the extent that these constraints agree with our expectations based on 
empirical estimates for the evolution of disk gas fractions, they represent 
independent support for the scenario outlined here.

\subsection{Projections of the FP}
\label{sec:obs.tests.projections}

\begin{figure}
    \centering
    \scaleup
    \plotone{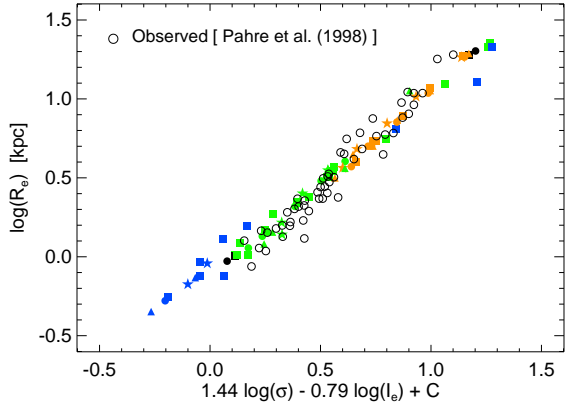}
    \caption{Observed FP, in the best-fit projection from \citet{pahre:nir.fp}; i.e.\ 
    freeing the exponents in $R_{e}\propto\sigma^{\beta}\, I_{e}^{\gamma}$. 
    Open points are the observed systems, filled points are our simulation 
    results from Figure~\ref{fig:fp.pred}. If the observed FP was more complex 
    than a tilt (e.g.\ if $\mstar$ depended substantially on $\re$ or $\sigma$ 
    separately at fixed $\mdyn$), our results would disagree here. The agreement is 
    good, as expected based on observations that the FP is, essentially, a tilted 
    virial plane. 
    \label{fig:pred.pahre}}
\end{figure}

\begin{figure}
    \centering
    \scaleup
    \plotone{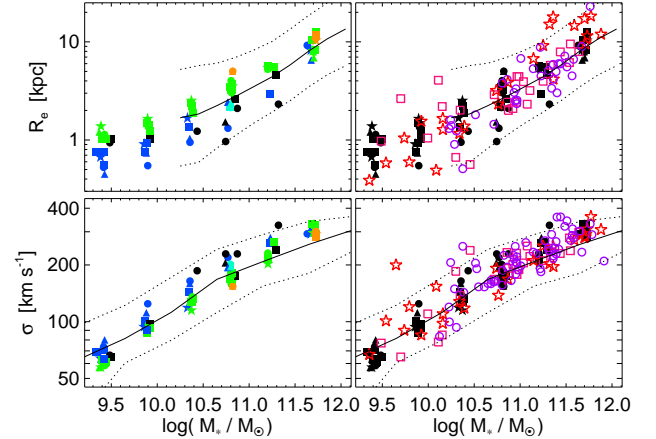}
    \caption{Projections of the FP. We show the 
    size-mass ({\em top}) and velocity dispersion-mass (Faber-Jackson; {\em bottom}) 
    correlations, for the same 
    subset of simulations used in 
    our FP predictions in Figure~\ref{fig:fp.pred} ({\em left}) and observations ({\em right}). 
    The observed mean relations for SDSS ellipticals (solid lines; with 
    $\sim95\%$ range as dashed lines) are shown over the range of 
    data used in the observed fits \citep[][for size-mass 
    and velocity dispersion-mass, respectively]{shen:size.mass,gallazzi06:ages}. 
    Note the curvature in both relations, owing to the dependence of 
    dissipation on mass and its role in setting both $R_{e}$ and $\sigma$. 
    The simulations and observations agree with each other and with the 
    mean correlations observed in much larger samples; i.e.\ once 
    systems with the appropriate range of dissipational fraction for their 
    mass are selected, they follow the observed size-mass and other 
    correlations. The relations would not agree if we used systems with 
    significantly different dissipational fractions as a function of mass.  
    \label{fig:fp.projections}}
\end{figure}

\begin{figure*}
    \centering
    \scaleup
    \plotone{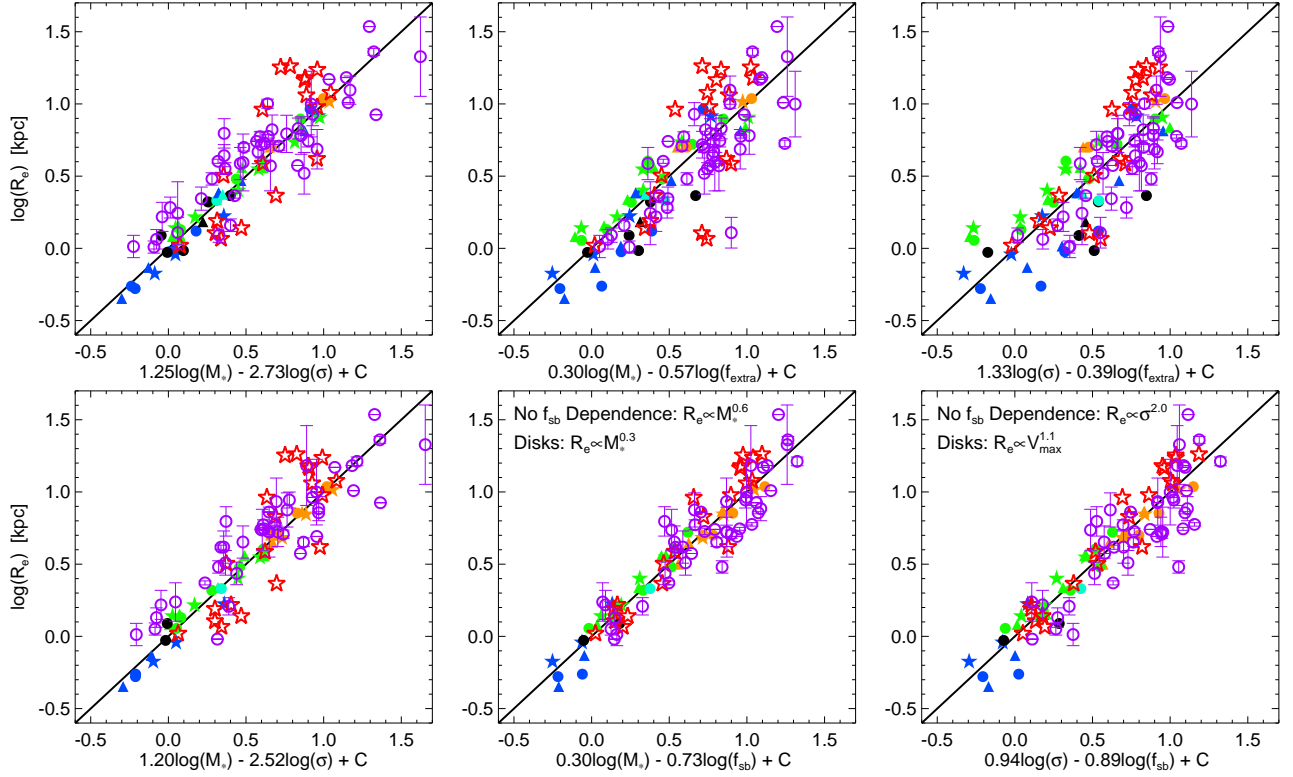}
    \caption{FP predictors of $R_{e}$. {\em Left:} $R_{e}$ as a function of 
    the ``traditional'' global parameters $\mstar$ (stellar mass) and $\sigma$ 
    (velocity dispersion). Simulations (the same as 
    used for the predicted FP in Figure~\ref{fig:fp.pred}) and observations are shown 
    (as in Figure~\ref{fig:mdyn.fextra}). 
    The scatter in $R_{e}$ in this projection is $\sim0.2\,$dex. 
    {\em Center:} $R_{e}$ as a bivariate function of $\mstar$ and 
    fitted extra light fraction ($\fextra$, {\em top}) or 
    inferred starburst fraction ($\fsb$, {\em bottom}). This 
    is an equivalent or better observational predictor of $R_{e}$ (scatter 
    $\sim0.15\,$dex). {\em Right:} $R_{e}$ as a function of 
    $\sigma$ and $\fextra$ ({\em top}) or $\fsb$ ({\em bottom}). Again, 
    this is an equivalent predictor (scatter $\sim0.2$\,dex). 
    Ignoring the effects of dissipation, we recover the 
    $\re-\mstar$ and $\sigma-\mstar$ relations in Figure~\ref{fig:fp.projections}, 
    with very different slopes from the relations for disks (comparison shown here). 
    However, 
    with the effects of dissipation explicitly included here, the relation between 
    $R_{e}$ and $\mstar$ or $R_{e}$ and $\sigma$ is 
    equivalent to that obeyed by spirals -- i.e.\ the difference between 
    the ``projected'' 
    elliptical and spiral scaling relations is entirely attributable to 
    the degree of dissipation ($\fextra$ or $\fsb$). 
    \label{fig:fp.w.f}}
\end{figure*}

As another check, we consider our results in the observed FP space of
\citet{pahre:nir.fp}, in Figure~\ref{fig:pred.pahre}. Here, the
coefficients of $\sigma$, $R_{e}$, and $I_{e}$ are independently free
in determining the best-fit projection of the observations.  If, in
principle, the FP were much more complex than a simple tilt (i.e.\ if
$M/L$ depended substantially on $\re$ or $\sigma$ at fixed $\mdyn$),
then the simulations might match the observations in
Figures~\ref{fig:fp.pred}-\ref{fig:fp.pred.trueM} but not in this
representation.  In fact, we find good agreement with the
observations.  This is not surprising, since it has been established
that the FP can be represented as a tilt under some homology
assumptions \citep[e.g.][]{jorgensen:fp,padmanabhan:mdyn.mstar.tilt}.

Figure~\ref{fig:fp.projections} plots projections of the FP;
specifically, the size-mass relation and the velocity dispersion-mass
(Faber-Jackson) relation. We show these correlations for the same
simulations used to construct the FP in
Figures~\ref{fig:fp.pred}-\ref{fig:pred.pahre}; since we showed in
\S~\ref{sec:obs:sizes} that e.g.\ effective radii scale systematically
with dissipational fraction, a robust prediction of the size-mass
relation requires that the simulations considered have appropriate
dissipational fractions for systems of their stellar mass.  We show
sizes and velocity dispersions as a function of stellar mass because
this is how these correlations are generally presented in the
literature, but differences are small and the comparison with
observations is similar if we consider them as a function of dynamical
mass.

We compare with the same correlations in the observed samples for
which we consider the FP, and with the mean correlations measured for
ellipticals in the SDSS. The simulations agree well with the
observations -- both those that we consider in our FP analysis, and
the trends in the general population (it is also reassuring that our
observed FP samples obey the same projected scalings, with similar
scatter). The scatter in these correlations is larger than
that between $\mdyn$ and $\mstar$, since at fixed $\mdyn$ there is a
tradeoff between $R_{e}$ and $\sigma$.  It is also interesting to note
that the trends show significant curvature -- their combination
($\mdyn-\mstar$) is reasonably approximated by a power-law scaling,
but either the $\re-\mstar$ or $\sigma-\mstar$ slope will vary with
the fitted mass interval (for this reason, we refrain from quantifying
a power-law fit to either the simulations or observations, but we note
that if we do so over the same mass interval for both, we obtain the
same result).

This has been discussed in the literature, for both the $\re-\mstar$
and $\sigma-\mstar$ relations \citep[see e.g.][]{lauer:massive.bhs,
vonderlinden:bcg.scaling.relations,desroches:scaling.law.curvature}
(note that even \citet{shen:size.mass} see tentative evidence for this
effect in their lowest-mass bins, but their dynamic range at the low
mass end is limited). We note here that such curvature can arise even
from pure gas-rich merger remnants, owing to the dependence of the
amount of dissipation on mass and corresponding scalings of size and
velocity dispersion. This is supported by the fact that both our cusp
and core sub-samples independently exhibit similar curvature. It has
also been observed that non-BCG and BCG galaxy samples show similar
curvature \citep[despite the latter being much more likely to have
experienced dry mergers; see e.g.][]{delucia:sam}. This is not to say
that re-mergers will not introduce curvature in these relations or
move systems considerably with respect to them \citep[see
e.g.][]{boylankolchin:dry.mergers}, but rather to emphasize that dry
re-mergers are not the {\em only} source of curvature in the
relations, and curvature should not necessarily be interpreted as
evidence for dry mergers (further note that systems cannot move too
far off their initial correlations, as they are constrained by the
observed scatter).

It is well-established that these correlations are different from those 
obeyed by spiral galaxies. If we do crudely approximate the correlations 
as power-laws (for illustrative purposes) 
over the dynamic range shown in 
Figure~\ref{fig:fp.projections}, we obtain  
correlations $R_{e}\propto\mstar^{0.56}$ \citep[as in][]{shen:size.mass} 
and $\sigma\propto\mstar^{0.25}$ \citep[see e.g.][]{bernardi:correlations}, 
also yielding $R_{e}\propto\sigma^{2.0-2.2}$. Compare 
these to the observed relations in disks: 
$R_{e}\propto\mstar^{0.30-0.35}$, 
$V_{\rm max}\propto\mstar^{0.27-0.30}$, and 
$R_{e}\propto V_{\rm max}^{1.0-1.2}$ \citep[see e.g.][]{persic96,courteau:disk.scalings, avilareese:baryonic.tf}. 
These differences relate to the tilt of the FP and the fundamental distinctions 
at issue here: at high masses, ellipticals and disks have similar sizes, 
but at low masses, ellipticals are much smaller. 
The difference, in our simulations, arises because of changes in the 
degree of dissipation, reflected in the extra light content. In principle then, 
it should be possible to consider the resulting 
correlations between $R_{e}$ and $\mstar$ or $\sigma$ 
in terms of the combination of their dependence on 
that variable and the additional dependence on $\fextra$ or $\fsb$. 

Figure~\ref{fig:fp.w.f} shows the results of this exercise: 
we consider $R_{e}$ as a bivariate function of 
combinations of $\mstar$, $\sigma$, and $\fextra$. 
The correlation $R_{e}(\mstar,\,\sigma)$ represents the 
``traditional'' FP correlation -- that the correlation is 
not simply $R_{e}\propto\mstar/\sigma^{2}$ reflects the ``tilt'' of the FP. 
If instead we fit $R_{e}$ to a 
combination of $\mstar$ and $\fextra$, we obtain a 
fit of similar quality (i.e.\ this combination is an equivalent predictor of 
$R_{e}$), but with the dependence $R_{e}\propto\mstar^{0.3}\,\fextra^{-0.57}$. 
This is an equivalent FP in terms of its usefulness as a 
predictor of $R_{e}$, with a scatter $\sim0.15-0.20\,$dex. 
More interesting, the explicit dependence of $\re$ on $\mstar$ fitted in this manner 
is identical to that observed in disks ($\re\propto\mstar^{0.3}$), with the 
remaining dependence owing to the dependence of $\re$ on 
$\fextra$. Given the cosmological scaling of 
$\fextra$ with $\mstar$, we obtain the steeper 
$\re\propto\mstar^{0.6}$ relation for ellipticals when $\fextra$ is ignored, but 
at {\em fixed} $\fextra$ or $\fsb$ -- i.e.\ fixed degree of dissipation -- 
a relation $\re\propto\mstar^{0.3}$, similar to that observed in disk 
galaxies, is observed \citep[in line with the predictions from][]{robertson:fp}. 
We obtain the same results for 
the size-velocity dispersion relation: a best-fit 
$\re\propto\sigma^{1.33}\,\fextra^{-0.39}$ (or $\re\propto\sigma^{0.94}\,\fextra^{-0.89}$); 
the dependence of $\re$ on $\sigma$ at fixed $\fextra$ or $\fsb$ 
is very similar to the dependence of 
$\re$ on rotational velocity $V_{\rm max}$ in disks (to lowest order, 
our simulations and other 
numerical experiments demonstrate that $\sigma$ traces the 
pre-merger $V_{\rm max}$, modulo a roughly constant normalization 
offset owing to e.g.\ the profile shape and kinematics). 
In other words, {\em the difference between the observed 
size-mass-velocity dispersion relations in disks and ellipticals can be entirely accounted for 
by dissipation}.

\subsection{The Small Scatter in the FP}
\label{sec:obs.tests.scatter}

A final requirement for any model of the FP is that it account for the
small observed scatter.  Analyzing the relations predicted by the
simulations, which broadly sample orbital parameters and stellar
masses, and fully cover the observed range of dissipational fractions
at each mass, we can see directly in
Figures~\ref{fig:fp.pred}-\ref{fig:fp.pred.trueM} that there is small
scatter. Formally, we find a nearly symmetric $1\sigma$ scatter of
$\sim0.07-0.08\,$dex in $\mdyn(\mstar$) or $\sim0.065-0.075$\,dex in
$\mstar(\mdyn)$ ($0.055\,$dex scatter perpendicular to the best-fit
correlation -- or equivalently in the $R_{e}$ projection of the FP --
depending weakly on how we define our expected $\fextra(\mstar)$ and
whether we use $\mtrue$ or $\mdyn$).
This is the scatter obtained in these correlations using the median
values of $\mstar$ and $\mdyn$ for each galaxy (across
$\sim100\,$sightlines): if we include the sightline-to-sightline
variance in the simulations, we obtain $\sim 0.14$\,dex total scatter
in $\mdyn(\mstar)$ ($0.09$\,dex scatter perpendicular to the
correlation).  This is still less than that observed in each case: for
the observed samples herein, we find $\sim0.18$\,dex scatter in
$\mdyn(\mstar)$ and $\sim0.15$\,dex scatter in $\mstar(\mdyn)$; i.e.\
a perpendicular scatter of $0.12$\,dex, with additional observational
errors in $R_{e}$ and $M_{\ast}$ (estimated from our experiments in
\papertwo) likely contributing most of the difference (for typical
$\sim0.1$\,dex stellar population model errors in $M_{\ast}$, this
yields roughly the observed scatter).

The origin of the small intrinsic scatter in the FP is straightforward
to understand in this scenario.  Some intrinsic variation will come
from the scatter in the total baryon-to-dark-matter content of the
progenitor galaxies. However, observations of the baryonic
Tully-Fisher relation suggest that the scatter in $\mstar$ at fixed
maximum circular velocity (a proxy for halo mass) is small,
$\approx0.1\,$dex \citep[at least over the mass range of interest
here;][]{belldejong:tf}.  Even considering the total baryon-to-halo
mass ratios, which extend well beyond $R_{e}$ and are therefore of
less interest, observations imply quite small scatter $\sim0.15\,$dex
\citep{wang:sdss.hod, conroy:monotonic.hod,weinmann:obs.hod}.  This
effectively subsumes a number of quantities, including e.g.\ scatter
in the initial radii of the disks at fixed mass, in halo
concentrations, and in other parameters (note that the scatter in any
one of these quantities is not important, given that the final
Tully-Fisher relation scatter is small). The contribution of this
initial scatter will actually be reduced after a gas-rich merger,
because the quantity of interest is the ratio $[\mstar/2 + M_{\rm
dm}(<R_{e})]/(M_{\ast}/2)$.  Since the merger channels gas into the
center and raises the central density, the final effective radius is
smaller (explaining the smaller sizes of ellipticals relative to
spirals at low mass).  Thus, while $\mstar$ is just the baryonic mass
of the progenitor, the contraction of the remnant makes the relative
value of $M_{\rm dm}(<R_{e})$ smaller, reducing the importance of
initial scatter in $M_{\rm dm}/M_{\rm baryon}$ to the final FP
scatter. Analysis of the simulations suggests that a realistic initial
$\sim0.1\,$dex scatter in the baryonic Tully-Fisher relation
contributes $\lesssim0.04-0.05$\,dex scatter directly to the final FP
prediction.

However, there will be additional intrinsic scatter contributed by
different degrees of dissipation in systems of the same
mass. Figures~\ref{fig:re.sigma.cusp} \&\ \ref{fig:mdyn.fextra.mbins}
shows that at fixed stellar mass, systems with different degrees of
dissipation have correspondingly different effective radii and ratios
$\mdyn/\mstar$.  However, the scatter in the expected (and
observationally inferred) dissipational fraction at each mass is
small, a factor $\lesssim2$, and the dependence of $\mdyn/\mstar$ (see
Figure~\ref{fig:mdyn.fextra}) on dissipational fraction over this
relatively narrow range is not strong.  Combining this scatter with
the dependence of $\mdyn/\mstar$, we expect it to contribute about
$\sim0.06-0.08$\,dex of intrinsic scatter.  Together, this yields the
relatively small $\sim0.08$\,dex intrinsic scatter in $\mdyn(\mstar)$
(before sightline-to-sightline and measurement errors).

\breaker
\section{Total versus Dynamical Mass: Invariance of the FP}
\label{sec:obs.tests.mtot}

\begin{figure}
    \centering
    \plotter{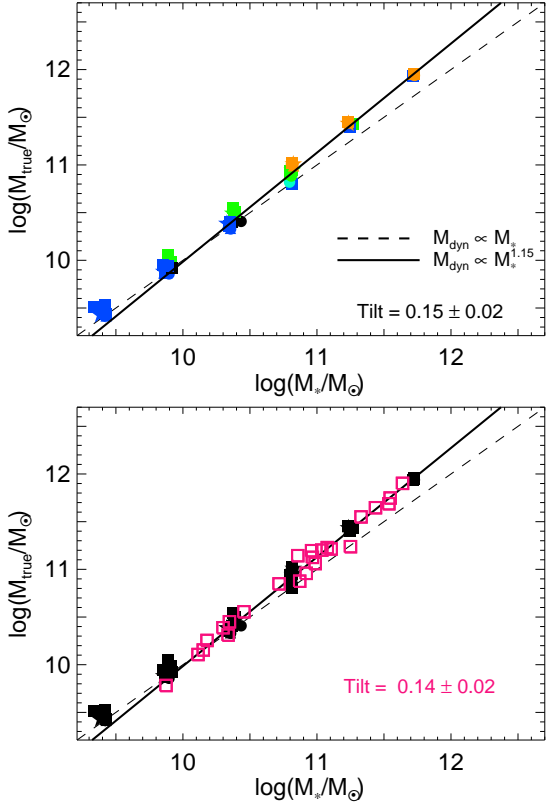}
    \caption{Predicted and observed FP (as in Figure~\ref{fig:fp.pred}), 
    but using the true total enclosed mass $\mtrue$ (stellar plus dark matter) within 
    $R_{e}$ instead of $\mdyn$. We compare with the observed systems from 
    \citet{cappellari:fp}, who use integral Jeans/Schwarzchild modeling 
    to determine a ``true'' mass 
    independent of a homology or constant $M/L$ assumption. 
    A nearly identical tilt is predicted and observed -- i.e.\ the predicted and observed 
    FP is really a reflection of the tilt in $\mtrue/\mstar$ within $R_{e}$, 
    not a tilt in $\mtrue/\mdyn$. 
    \label{fig:fp.pred.trueM}}
\end{figure}

In Figure~\ref{fig:fp.pred.trueM} we repeat our experiment from
Figure~\ref{fig:fp.pred}; i.e.\ we construct the FP from the 
simulations, given the observed and expected dependence of
dissipational fraction on mass. However, instead of plotting the
traditional dynamical mass estimator $\mdyn\propto\sigma^{2}\,R_{e}$,
we plot the {\em true} total projected mass (stellar plus dark matter
plus gas, although gas is generally negligible) within $R_{e}$, which
we extract directly from the simulations and refer to as $\mtrue$. The
predicted FP is nearly identical to that predicted using the $\mdyn$ estimator --
i.e.\ the FP predicted does, in fact, arise from a change of the
dark-to-luminous-matter ratio within $R_{e}$, owing to dissipation
contracting the stellar effective radius (i.e.\ a change in
$\mtrue/\mstar$ with mass), rather than to traditional structural or
kinematic non-homology (which would imply a constant ratio of
$\mtrue/\mstar$, with a changing apparent mass from the dynamical mass
estimator; i.e.\ a varying $\mtrue/\mdyn$ with mass).

This is despite the systematic variation of extra light fraction with
mass (reflecting the systematic dependence of dissipational or initial
disk gas fraction). In other words, while there is technically a
subtle non-homology implicit in the fact that the dissipational
fraction depends on mass (and, as a result, the fitted $\fextra$ and
$\fsb$ change with mass, whereas they would be constant for true
perfectly self-similar systems), it does not contribute any
significant or observationally meaningful structural or kinematic
non-homology.  The ``homology assumption,'' namely that the ratio of
the dynamical mass estimator to the true enclosed mass within the
effective radius ($\mtrue/\mdyn$) is constant, is predicted to hold,
while the FP is satisfied.

\begin{figure}
    \centering
    \plotter{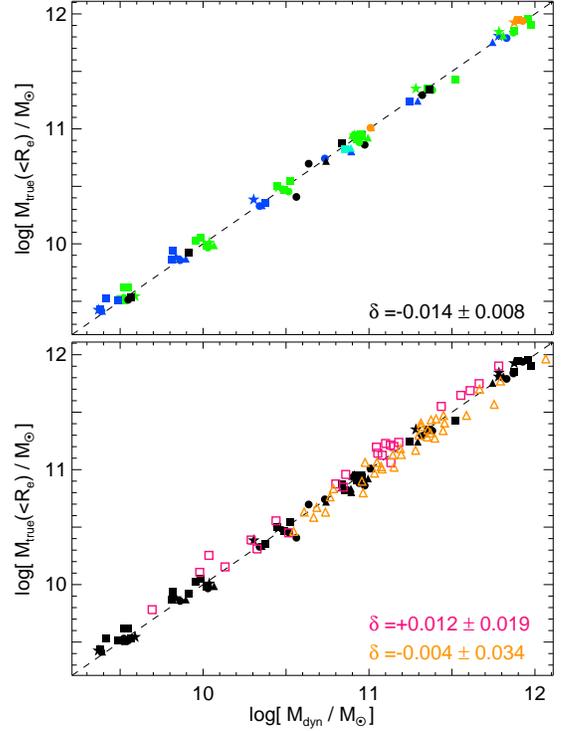}
    \caption{{\em Top:} The correlation between the dynamical mass 
    estimator $\mdyn = \mdynnorm\,\sigma^{2}\,R_{e}/G$ and 
    the true total mass $\mtrue$ (stellar plus dark matter) enclosed 
    within $R_{e}$ (as in Figure~\ref{fig:fp.pred.trueM}), 
    for the same simulations shown in our predicted 
    FP (Figure~\ref{fig:fp.pred}). Dashed line is a fit of the form 
    $\mtrue \propto \mdyn^{1+\delta}$, with the best-fit value of 
    $\delta$ shown: the best fit is indistinguishable from 
    the $\mtrue = \mdyn$ line (i.e.\ has no ``tilt''). 
    {\em Bottom:} Same, for the observed systems from Figure~\ref{fig:fp.pred.trueM}, 
    using integral modeling \citep[][magenta squares]{cappellari:fp} 
    or lensing mass maps \citep[][orange triangles]{bolton:fp,bolton:fp.update} to estimate 
    the true mass $\mtrue$. The best-fit coefficients $\mtrue\propto\mdyn^{1+\delta}$ 
    are shown for each observed sample. The simulations we use to predict the 
    FP, and the observed systems on the FP, 
    trace a nearly identical 
    $\mtrue\propto\sigma^{2}\,R_{e}$ relation consistent with the 
    traditional observational definition of the ``homology assumption.''    
    \label{fig:mdyn.mstar}}
\end{figure}

Figure~\ref{fig:mdyn.mstar} demonstrates this explicitly: we plot the
ratio of true enclosed mass to the dynamical mass estimator,
specifically for the {\em same} simulations we used to predict the FP
and its tilt.  The ratio $\mtrue/\mdyn$ is constant despite changing
dissipational fractions; fitting a power law $\mtrue \propto
\mdyn^{1+\delta}$, we obtain $\delta= -0.01 \pm 0.01$ (both
insignificant and completely negligible compared to the observed value
of the FP tilt, $\tilt\approx0.2$); i.e.\ the ``homology assumption''
is valid in the observational sense as applied to these simulations,
and there is no significant implied systematic structural or kinematic
non-homology (in the traditional sense of the term) as a function of
mass.

We compare these predictions with observed ``true'' enclosed masses
within the effective radii of observed systems, using two different
approaches for estimating the true enclosed mass (without explicitly
invoking the homology assumption). First, \citet{cappellari:fp}
estimated total masses from three-dimensional Schwarzchild modeling or
two-dimensional Jeans modeling of local ellipticals with
two-dimensional velocity field information from SAURON maps \citep[see
also][]{emsellem:sauron.rotation,
mcdermid:sauron.profiles,cappellari:anisotropy}, and used stellar
population models to estimate stellar masses (we correct these for the
choice of IMF, but otherwise do not modify them).  We also have
considered integral modeling masses from other, independent sources
\citep[e.g.][]{vandermarel:ml.models, kronawitter:ml.models,haringrix}
and obtain identical results in each case.

Figure~\ref{fig:fp.pred.trueM} plots the FP of these objects; i.e.\
the true enclosed mass determined in this manner as a function of
stellar mass. The tilt, in agreement with the simulations, is nearly
identical to that given by the dynamical mass estimator $\mdyn$.  This
is clear in Figure~\ref{fig:mdyn.mstar}, where we compare the
predicted correlation between $\mdyn$ and $\mtrue$ with these
observations (we apply the same definition of $\mdyn$ to the observed
objects as to the simulations, namely $\mdyn=
\mdynnorm\,R_{e}\,\sigma^{2}/G$).  As noted by \citet{cappellari:fp},
there is a tight correlation between $\mdyn$ and $\mtrue$ without any
significant deviation in this space from the homology assumption
($\delta=+0.012\pm0.019$).  Not only does this tight proportionality
($\delta\approx0$) agree with our simulations, but the normalization
fitted to the observed sample, i.e.\ mean ratio $\mtrue \approx
\mdynnorm\,R_{e}\,\sigma^{2}/G$, agrees with that predicted by the
simulations over the observed mass range to better than $0.03$\,dex,
further suggesting that the structural properties and profile shapes
of the simulations are in good agreement with those observed.

Second, \citet{bolton:fp,bolton:fp.update} have performed a similar exercise, but using
instead the enclosed masses determined from gravitational lensing.  We
compare their FP (i.e.\ $\mtrue(\mstar)$) in
Figure~\ref{fig:fp.pred.trueM}, and again find similar tilt to that
predicted by the simulations (using the true enclosed mass within
$R_{e}$ as $\mtrue$) and to that obtained using $\mdyn$ instead of
$\mtrue$. Figure~\ref{fig:mdyn.mstar} shows that, for these objects as
well, $\mtrue\propto\mdyn$ ($\delta=-0.004\pm0.034\approx0$), with
again a normalization (mean $\mtrue/\mdyn$) in agreement with the
simulations to within $0.02$\,dex. Note that the authors restricted
themselves to true masses and dynamical masses within $R_{e}/2$; if we
use their lens models to correct the true masses to $R_{e}$ and take
the corresponding dynamical masses at that radius, we obtain identical
results.

The simulations reproduce the observed, nearly exact proportionality
between $\mtrue$ and $\mdyn$, and therefore show the same FP behavior
as has been seen in the observations: namely that the FP tilt remains
similar regardless of whether $\mdyn$ or $\mtrue$ is considered.  This
demonstrates the point from \S~\ref{sec:diss.fx}, that non-homology
(in the general, observationally motivated sense of different
$\mtrue/\mdyn$ as a function of galaxy properties) is not a
significant contributor to the FP tilt.

\breaker
\section{Re-Mergers and FP Evolution}
\label{sec:remergers}

The predictions we seek to test, 
and the observed dependence of FP tilt on the degree
of dissipation which we have demonstrated, hold in the same manner for
both cusp and core ellipticals.  However, a number of observed
properties \citep[see e.g.][]{faber:ell.centers} suggest that core
ellipticals may be the products of subsequent ``dry'' or
spheroid-spheroid, relatively gas-free re-mergers of (generally
lower-mass) cusp ellipticals (formed directly in gas-rich mergers).
If true, do we expect the level of agreement we see?

The answer is generally yes, provided that the number of re-mergers is
not large. Numerical experiments
\citep[e.g.][]{capelato:dry.mgr.fp,dantas:dry.mgr.fp,nipoti:dry.mergers,
boylankolchin:mergers.fp,robertson:fp} indicate that remnants of
dissipationless mergers of systems which begin on the FP tend to
remain on the FP.  Even if the re-merger introduces no new tilt,
its effects would be small.  To lowest order, since the gas is mostly
exhausted, the re-merger will be dissipationless, and merging two
identical systems $M_{1}$ and $M_{2}$ on a parabolic orbit, one
expects their profiles to be roughly preserved. This leads to
the energy conservation equation
\begin{equation}
E_{f} = k\,{(M_{1}+M_{2})}\,\sigma_{f}^{2} = E_{i} = k\,M_{1}\,\sigma_{1}^{2} + k\,M_{2}\,\sigma_{2}^{2}
\end{equation}
where $\sigma_{f}$ is the velocity dispersion of the final remnant,
and $k$ is a constant that depends on the shape of the profile. From
this, one obtains the general rule that a major 1:1 merger will
approximately double $R_{e}$ while preserving $\sigma$, doubling both
$\mdyn$ and $\mstar$ \citep[e.g.][]{hausman:mergers,hernquist:phasespace}.

Relative to the FP scaling, $\mdyn\propto\mstar^{1.2}$, the remnant is
now $\sim0.06$\,dex below the FP expectation -- only
$\sim0.3-0.5\,\sigma$ ($\sigma$ being the observed FP scatter). 
So even if the system increases its mass by a
total factor of $\sim4-5$ via dry re-mergers (i.e.\ has $\sim$ two
equal-mass re-mergers or $\sim5-6$ more typical 1:3 mass ratio
re-mergers), it will move by only $\sim1\,\sigma$ with respect to the
FP.

Compare this to the impact of dissipation, which as we demonstrate in
\S~\ref{sec:obs} can change the effective radius and ratio
$\mdyn/\mstar$ (at fixed stellar mass) by nearly an order of
magnitude. We therefore expect that, unless a system has experienced
an extreme number of major dry re-mergers (expected only for the
rarest, most massive central cluster galaxies), the degree of
dissipation (i.e.\ properties we study herein) should be the dominant
factor determining the effective radii, ratio of dynamical to stellar
mass, and location with respect to the FP.

\begin{figure}
    \centering
    \plotter{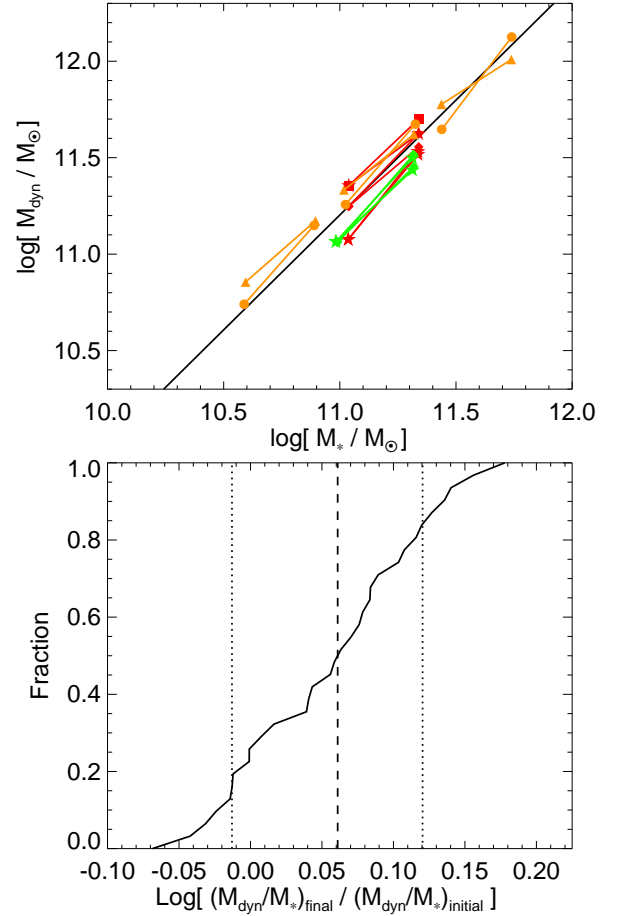}
    \caption{{\em Top:} Impact of subsequent spheroid-spheroid re-mergers 
    on the FP. We show the observed FP from Figure~\ref{fig:fp.pred} (black line), 
    with the initial and final positions of simulated systems before and after a 
    major re-merger (point style as Figure~\ref{fig:fp.pred}; lines connect each 
    re-merger to its progenitor). 
    {\em Bottom:} The cumulative distribution in relative dynamical to stellar mass 
    ratio before and after a major re-merger, in our entire ensemble of 
    re-merger simulations. Vertical dashed (dotted) line shows 
    the median ($\pm1\,\sigma$) change in $\mdyn/\mstar$. 
    Re-mergers tend to slightly increase $\mdyn/\mstar$, by almost exactly the amount 
    needed to move parallel to the FP ($0.06\,$dex), but with non-negligible 
    scatter ($\sim0.065\,$dex). 
    \label{fig:remergers}}
\end{figure}

Furthermore, it has been noted \citep{boylankolchin:mergers.fp,
boylankolchin:dry.mergers,ciotti:dry.vs.wet.mergers} that re-mergers can slightly increase
$\mdyn/\mstar$, essentially causing systems to move nearly parallel to
the FP.  We consider a sample of re-merger simulations, described in
\S~\ref{sec:sims}, in order to demonstrate this in
Figure~\ref{fig:remergers}. Specifically, these are identical (mass
ratio 1:1) re-mergers of remnants of previous gas-rich, disk-disk
mergers, merged in various orbital configurations. The details are
discussed in \S~\ref{sec:sims}, but we note that although we plot only
identical re-mergers here (for illustrative purposes), the results
scale appropriately for different mass ratios and various mixed
encounters (i.e.\ merging different spheroid remnants of similar
mass).  Plotting both the pre and post-remerger systems on the FP in
Figure~\ref{fig:remergers}, we see that they move nearly parallel to
the fitted relation ($\tilt\approx0.2$).  Quantitatively,
the median increase in dynamical to stellar mass ratio is
$\sim0.06\,$dex, precisely as needed to move the system along the
FP (to offset the $\sim0.06$\,dex offset from the FP estimated above,
if $\mdyn/\mstar$ were exactly conserved in a re-merger).

%

This effect is subtly related to e.g.\ changes in the orbital isotropy
and kinematics of the remnant in a re-merger, but primarily owes to a
real (albeit small) physical increase in the ratio of enclosed dark
matter mass within the stellar effective radius (the progenitors and
remnants obey the ``homology assumption'' as in
Figure~\ref{fig:mdyn.mstar}, and we see an almost identical effect
plotting the true enclosed mass instead of $\mdyn$).  Considering
the initial and final distributions of stellar and dark matter
particles as a function of their initial radius, it is straightforward
to understand this effect.

Despite the nearly uniform inflation of the light profile in a
re-merger, there is substantial scattering of stars (and dark matter): i.e.\ although
stars at some initial $r_{i}$ will be, at the end of the merger, at a
median $r_{f}\approx2\,r_{i}$, the distribution of final radii will be
approximately lognormal with scatter $\approx0.4\,$dex (see
\paperthree). The integrated stellar mass will be constant, but such
a scattering will tend to broaden the dark matter distribution 
in a way that slightly increases the dark matter mass {\em within} the 
stellar $R_{e}$ --
i.e.\ scatter some dark matter from near the effective radius of the
halo to both smaller and larger radii (total dark matter mass is
conserved, but the densities are slightly lower near the halo
effective radius -- which is much larger than the galaxy effective
radius which concerns us here -- and slightly higher at much smaller
and larger radii). This will slightly raise the central dark matter
density, in a way consistent with the observed (weak) change in
$\mdyn/\mstar$ in re-mergers (effectively mixing slightly more 
dark matter into the stellar distribution).  The net effect of this is that
re-mergers move systems even less with respect to the FP, and
dissipation is still the dominant factor setting $\mdyn/\mstar$.

Figure~\ref{fig:remergers} also demonstrates that although re-mergers
tend, in the mean, to increase $\mdyn/\mstar$ such that they move
parallel to the FP, there is significant scatter in the change in
$\mdyn/\mstar$ in re-mergers (in detail, the median increase is
$0.06\,$dex, but there is a $1\,\sigma$ scatter of $\sim0.065$\,dex
about this median). We therefore expect that growing a population of
ellipticals by a factor $\sim2$ in re-mergers (equivalent to a single
1:1 or three 1:3 re-merger for most of the objects in the
population) will contribute $\sim0.06$\,dex intrinsic scatter in
$\mdyn(\mstar)$. Compared to the intrinsic scatter (in terms of
sightline-averaged quantities) of $\sim0.08$\,dex, this is
significant, yielding a final intrinsic scatter of $\sim0.10$\,dex.
However, after accounting for the additional $\sim0.1$\,dex
sightline-to-sightline scatter, and observational effects, the
contributed scatter is negligible compared to the final observed
$\sim0.18\,$dex scatter in $\mdyn(\mstar)$.  That the observed scatter
is not much larger does suggest there has not been dramatic
re-merging, but the constraint is weak -- it would take $\sim4-5$
major (mass ratio 1:1) dry mergers (or $\sim10$ more likely 1:3
dry mergers) to noticeably increase the observed FP scatter at the
massive end.

\breaker
\section{Discussion}
\label{sec:discuss}

We have developed and implemented a set of observational tests of the
theoretical proposal that different degrees of dissipation are
responsible for the tilt in the fundamental plane, 
(suggested in e.g.\ 
\citet{djorgovski:fp.tilt,kormendy:dissipation,bender:ell.kinematics,
ciotti:kinematic.nonhomology} 
and developed in numerical simulations in 
\citet{bekki:fp.origin.tsf,onorbe:diss.fp.model,
robertson:fp} and \citet{dekelcox:fp}). 
With measurements
of the surface brightness profiles of ellipticals of sufficient
accuracy and high dynamic range, we demonstrated in
\paperone-\paperthree\ that it is possible to design an empirical
decomposition which reliably separates the dissipational (original
merger-driven starburst) and dissipationless (scattered stars from the
original stellar disks) components of galaxy light profiles. Applying
this to a large sample of observed ellipticals, we study here how
their FP correlations relate to the mass fractions in the
dissipational (or ``extra light''/starburst) component.

We show that systems with larger dissipational fractions have smaller
effective radii for their stellar masses and lower ratios of total
(stellar plus dark matter) mass to stellar mass within their effective
radii (equivalently, lower $\mdyn/\mstar$) -- i.e.\ that dissipation
can and does move objects onto the FP, in the manner 
predicted by \citet{robertson:fp}.  More important, we show that
{\em without the effects of a systematic dependence of the degree of
dissipation on mass, there is no significant FP ``tilt''}. In other
words, systems with the same dissipational/starburst fractions, even
over a wide range in mass ($\sim2-3\,$dex), exhibit a constant ratio
$\mtrue\propto\mstar$.  Dissipation not only {\em can} tilt systems on
the FP, it is {\em required} to explain the FP tilt.

The tilt originates because of an expected systematic dependence of
dissipational content on mass, arising because lower-mass disks (the
progenitors of lower-mass ellipticals in this scenario) tend to be
more gas rich. Therefore, lower-mass systems will, on average, undergo
more dissipational mergers. This appears to be true even out to
redshifts $z\sim2-3$, so that the exact formation times of the systems
are not especially important.  Quantitatively, if we convolve the
expected dependence of dissipational fraction on mass (from e.g.\ the
dependence of gas fraction on disk mass, or from the clear mean
systematic trend we see in the fits to the observed ellipticals) with
the dependence of $\mdyn/\mstar$ on dissipational fraction, we obtain
a prediction for the FP which is ``tilted'' in a manner similar to
that observed, $\mdyn\propto\mstar^{1.2}$.  

Of course, merger-induced starbursts may not be the only source of 
dissipation; for example, stellar mass loss may replenish the gas supply and 
lead to new dissipational bursts \citep[see e.g.][]{ciottiostriker:recycling}. 
Moreover, the merger history and series of induced dissipational events may be 
more complex than a single or couple of idealized major mergers 
\citep[see e.g.][]{kobayashi:pseudo.monolithic,naab:etg.formation}. For 
our purposes, however, all dissipational star formation will appear similar when observed 
and have the same effects -- we are essentially measuring the integrated amount of 
dissipation, and testing the idea that the presence of this dissipation can explain the 
difference between spheroid and disk scaling relations. 
In any case, the result is the same, and 
the agreement with disk gas fractions over the redshift range of interest 
suggests star-forming galaxies are viable candidates for the ultimate progenitor systems of 
ellipticals (however complex 
the details of the morphological transformation may be). 
In short, the systematic dependence
of the degree of dissipation on mass in elliptical formation is both {\em necessary}
and {\em sufficient} to explain the FP tilt.

\subsection{Possible Sources of Tilt: Why Dissipation is the Only Viable Possibility}
\label{sec:discuss:alt}

This conclusion is, to our knowledge, consistent with all existing
observational constraints on the FP.  We note that repeating our
experiments with alternative IMFs affects only the normalizations of
the quoted relations; considering alternative stellar evolution models
such as those of \citet{starburst99} or \citet{maraston:ssps},
compared to the default calibration from \citet{BC03} in
\citet{bell:mfs} makes no difference, since ellipticals are almost
all relatively old and the model differences focus on young stellar
populations.  Assuming, therefore, that there is not some dramatic
error in the stellar population models or the observed effective
radii, then there are only a limited number of possible explanations
for the FP ``tilt.'' \\

{\em Non-Homology:} Technically, this refers to systems being not
perfectly self-similar. This must be broken in any model in which
there is a dependence or ``tilt'' in the physical mass within some
radius and the stellar mass $\mstar$. In practice, the more practical 
meaning of ``homology'' is that systems have the same ratio of
dynamical mass to true enclosed total mass within their effective
radii; i.e.\ dynamical mass is a good proxy for true mass without a
systematic dependence on other galaxy properties. In other words,
$\mtrue = k\,R_{e}\,\sigma^{2}/G$ where $k\approx$\,constant ($\mtrue
\propto R_{e}\,\sigma^{2}/G$ is expected on dimensional grounds; $k$
is the integral argument which depends on e.g.\ the details of profile
shape and kinematics).  There are two general possibilities:

{\em (a) Kinematic Non-Homology:} In this case, the meaning of
$\sigma$ changes with mass, either because of different contributions
of rotationally supported components, or because of varying isotropy.
Essentially all studies have found that this effect does not
contribute to the FP tilt \citep[see e.g.][and references
therein]{gerhard:giant.ell.dynamics,riciputi:rotation.fp.tilt,cappellari:anisotropy,
nipoti:homology.from.mp};
specifically that there is no correlation between FP residuals and
orbital anisotropy or rotation (within the elliptical population). Moreover, 
\citet{ciotti:kinematic.nonhomology} demonstrated that reproducing 
the observed tilt with such trends would require models with internally  
inconsistent or unphysical orbit structure. 

{\em (b) Structural Non-Homology:} Alternatively, the shape of the
light profile could change sufficiently, such that e.g.\ even if two
systems were both spherical with isotropic velocity dispersion
tensors, the integral term relating $\mtrue$ and $\mdyn$ is
significantly different. (Note that this is not really independent of {\em (a)}; 
the dominant effect of a change in the mass profile shape is generally to 
change the central velocity dispersion, and correspondingly $\mdyn$, 
rather than dramatically changing e.g.\ the physical meaning of $R_{e}$, and 
differences in mass profile shape will generically require some kinematic 
non-homology.)
There has been debate regarding this
possibility, as some observational studies
\citep[e.g.][]{prugniel:fp.non-homology, trujillo:non-homology} have
argued that the observed dependence of galaxy Sersic index on mass is
sufficient to drive the tilt in the FP via structural non-homology.
Others have noted, however, that integral models which allow for
multiple components
or do not explicitly assume a constant $M/L$ with
radius do {\em not} find such non-homology
\citep[e.g.][]{carollo95:dm,rix97:dm,
gerhard:giant.ell.dynamics,gerhard03:mdyn,padmanabhan:mdyn.mstar.tilt,
cappellari:fp}. Moreover, invoking structural non-homology 
can generally explain only differences in {\em central} velocity dispersion (the 
predicted FP tilt from such dynamical models disappears rapidly as 
$\sigma$ is measured at somewhat larger fractions of $R_{e}$); the 
fact that the FP tilt is not dramatically reduced using velocity dispersions 
measured in larger apertures (even out to $R_{e}$) rules out most classes 
of such models \citep{bender:velocity.structure,ciotti:kinematic.nonhomology,
simien:kinematics.1,simien:kinematics.6,emsellem:sauron.rotation.data,
cappellari:fp,nipoti:homology.from.mp}.

Our analysis is able to explain {\em both} sides of this debate.
Fitting galaxy light profiles in a simplified manner, to a single
Sersic index, does {\em appear} to yield a significant dependence of
Sersic index on stellar mass, luminosity, or size. In \paperone, we
demonstrate that this is true of our simulations -- i.e.\ when
analyzed in the same manner as the observations, they appear to yield
an identical dependence of Sersic index on mass to that found in
\citet{prugniel:fp.non-homology} and \citet{trujillo:non-homology} 
(less massive systems, chosen in the same manner as those here to 
have higher typical dissipational fractions, have lower fitted Sersic 
indices over the same observational dynamic range; see also 
\paperthree) --
even though (at fixed degree of dissipation), the systems are
effectively homologous. Why, then, does this occur?

The answer owes to the fact that a single Sersic index is {\em not} a
physically robust description of the light profile of an elliptical
over its entire extent, and it also fails to include the dark matter
distribution. Such a fit mixes the outer dissipationless component of
the remnant with the inner, compact starburst remnant.  All the
observed samples which appear to find a dependence of Sersic index on
mass, when re-analyzed based on our physically motivated two-component
decomposition, in fact reveal that the outer, dissipationless
Sersic-like component is {\em self-similar} (see \papertwo\ and
\paperthree); what changes with mass or radius is the mass fraction in
the central dissipational component.  Furthermore, because of the true
multi-component nature of galaxy light profiles, fitting a single
Sersic index yields different answers as a function of the dynamic
range employed: for fixed observing conditions, this yields an
apparent correlation of Sersic index with galaxy mass or size
\citep[see
e.g.][]{padmanabhan:mdyn.mstar.tilt,boylankolchin:mergers.fp}.
Because of the role of e.g.\ the mass fraction at large 
radii and subtle differences in the second derivatives 
of profile shape near $R_{e}$ in driving a fitted Sersic index, 
the observed dependence of Sersic index on mass 
reflects a complex mix of real differences in galaxy merger history 
(the fact that low-mass systems have undergone more dissipational 
mergers, and that high-mass systems are increasingly likely to have undergone 
subsequent re-mergers, which will conserve the FP but scatter some small 
stellar mass to larger radii, leading to larger fitted Sersic indices) 
and observational differences in dynamic range, but little 
significant non-homology.

We have demonstrated here (\S~\ref{sec:diss.fx}) and in \papertwo\
that differences in the degree of dissipation (i.e.\ in the strength
of the central ``extra light'' component) -- while capable of driving
substantial differences in the {\em fitted} Sersic index when the
galaxy is fit to a single Sersic law -- do {\em not} in fact drive
traditional non-homology. They are not a large enough
fraction of the total mass to dramatically alter the structural
integrals. This should not be surprising: the dependence of
$\mtrue/\mdyn$ on Sersic index (in models where the mass distribution
follows a pure Sersic law) is primarily driven by the behavior at {\em
large} radii -- large Sersic indices asymptotically approach the
power-law behavior $I(r)\propto r^{-2}$, which implies a divergent
mass at large $r$, and an implied effective radius
$R_{e}\rightarrow\infty$ and rapidly 
increasing central velocity dispersion. This is precisely where observed profiles
are in fact self-similar (in a mean sense; there is considerable
variation object-to-object), with at most a weak dependence on
formation history (i.e.\ median difference $\Delta n_{s}\sim1$ for
those which have undergone major ``dry'' mergers; much less than is
needed to explain the FP tilt), as we demonstrate in both observations
and simulations in \papertwo\ and \paperthree.  

Further, at these radii, the mass density is increasingly dark
matter-dominated: the structure integrals only depend significantly on
the Sersic index $n_{s}$ if the dark matter profile is also described
by the same Sersic index -- i.e.\ if the {\em total} $M/L$ is a
constant function of radius, whereas essentially all kinematic data at
large radii indicates this is not the case (dark matter is required to
explain the kinematics at large radii). As we have shown, and as
demonstrated in observational kinematic modeling
\citep[e.g.][]{padmanabhan:mdyn.mstar.tilt,cappellari:fp} and lensing
studies \citep{bolton:fp,bolton:fp.update}, the dark matter distribution is relatively
insensitive to the details of the stellar mass distribution shape, and
therefore the {\em actual} non-homology driven by even large changes
in the stellar $n_{s}$ is much less than would be calculated assuming
the dark matter followed the same profile \citep[if it were not so,
the observed large dispersion in $n_{s}$ values would necessarily
yield much larger scatter about the FP than is observed; see
e.g.][]{bertin:weak.homology}.

The implication that a change in $n_{s}$ should lead to significant
structural non-homology is therefore misleading: the observed
dependence of $n_{s}$ on mass owes mainly to issues of finite dynamic
range and varying dissipational fractions, none of which give rise to
significant structural non-homology.  However, extrapolating to large
radii based on these estimated $n_{s}$ values and the {\em incorrect}
assumptions (1) that a single Sersic law is a physically meaningful
parameterization of the galaxy light profile, and (2) that dark matter
traces the stellar matter in a strict one-to-one fashion, implies
(incorrectly) some structural non-homology.

Moreover, many studies indicate that dynamical mass is proportional to
true enclosed mass (without requiring any homology assumption); i.e.\
that there cannot be a large contribution to tilt from {\em any} form
of traditional non-homology, whether kinematic or structural.  There
are now independent lines of support for this conclusion:
\citet{cappellari:fp} \citep[see also][]{vandermarel:ml.models,
kronawitter:ml.models,haringrix} estimate true enclosed masses within
$R_{e}$ based on two and three-integral modeling, from two-dimensional
velocity maps of observed ellipticals. Alternatively,
\citet{bolton:fp,bolton:fp.update} measure strong lensing gravitational masses.  In
both cases, the authors find $\mtrue\propto\mdyn$, without any
significant dependence on mass -- in other words, the FP is unchanged
regardless of whether the true total mass enclosed in $R_{e}$ is used,
or whether the dynamical mass proxy $\mdyn$ is used. It appears that at most, 
at the $\sim3\,\sigma$ limits on the observed $\mtrue-\mdyn$ relation 
(and based on the fitted $\mtrue-\mstar$ relations in Figure~\ref{fig:fp.pred.trueM}), 
these traditional forms of non-homology may contribute $\sim1/4$ the observed tilt. 
 \\

{\em Genuine Change in $M/L$:} 
The observations therefore imply that the FP must reflect a genuine physical
difference as a function of mass: namely, that low-mass ellipticals
have a higher ratio of stellar or baryonic mass to total (baryonic
plus dark matter) mass within the stellar $R_{e}$, relative to
high-mass ellipticals.  All means of achieving this end can be
classified into one of three categories:

{\em (a) Varying Global Baryon Fractions:} 
One could imagine that the stellar and halo mass distributions of
different ellipticals are separately self-similar, but that the {\em
total} ratio of stellar to halo mass changes as a function of galaxy
mass. In other words, there are no structural changes implied: one
decreases the halo mass relative to galaxy mass in lower-mass systems.

While this may seem plausible at the highest galaxy masses (it is well
established that in bright clusters, the total stellar to dark matter
mass ratio decreases with mass), this is relevant over much larger
scales than the stellar $R_{e}$ of the central galaxy.  Moreover, over
most of the mass range of interest, the known trends in global stellar
to dark matter fractions are {\em opposite} to that needed to
explain the FP tilt.

At masses below $\sim\mstar$, where the FP is observed to be
continuous (recall, the observed FP and its tilt extend to systems
with masses $< 10^{-2}\,\mstar$), all $\Lambda$CDM models
\citep[e.g.][]{conroy:monotonic.hod,
zentner:substructure.sam.hod,zheng:hod,
valeostriker:monotonic.hod,shankar06,vandenbosch:concordance.hod}
and observational constraints
\citep[][]{eke:groups,yang:obs.clf,mandelbaum:mhalo,
weinmann:obs.hod,wang:sdss.hod} {\em require} that the ratio of {\em
total} dark matter halo mass to stellar mass is {\em higher} in
lower-mass systems -- i.e.\ that star formation is less efficient in
low-mass systems.  This is contrary to the effect desired here, and
demonstrates that the FP tilt cannot owe to a simple global change in
baryon fraction.  Indirectly, however, the cosmological trend of
stellar to dark matter mass is in fact important -- this lower star
formation efficiency in low-mass systems means that they have larger
gas fractions when they undergo mergers, which we show does give rise
to the FP tilt.

If the global baryonic mass ratio is unchanging at small radii (or
changes in the opposite sense needed to tilt the FP), then the only
other possibility is that the size/shape of the halo and stellar
distributions vary relative to one another: i.e.\ one of the two
components is made more or less compact, relative to the other,
altering the ratio of stellar to total mass within the stellar
$R_{e}$.  Recall, the radii of interest are small relative to the halo
virial radii, and contain only a small fraction of the total halo dark
matter mass, so regardless of the total halo to stellar mass ratio,
changing the compactness of one component relative to the other can
make a significant difference.  There two possibilities for this
relative contraction/expansion:

{\em (b) Baryons are Fixed, Halos Contract:} In this scenario, the
galaxy stellar mass distributions are ``scale free'' (i.e.\ do not
change owing to external factors), but the dark matter halos are less
compact in low-mass ellipticals, so that within the stellar $R_{e}$
(i.e.\ the central regions of the halo) the total dark matter mass
fraction enclosed is lower. This is quickly ruled out: all
cosmological models
\citep[e.g.][]{nfw:profile,bullock:concentrations,wechsler:concentration,
dolag:concentrations,kuhlen:concentrations,neto:concentrations} and
observations from e.g.\ weak lensing and X-ray mass measurements
\citep{buote:obs.concentrations,schmidt:obs.concentrations,
comerford:obs.concentrations} find that halo concentration is a weak,
{\em decreasing} function of galaxy mass, the opposite of the effect
desired.

Furthermore, if this were the case, one would expect to see it in
progenitor disks, as well (the halos are insensitive to the morphological
transformation of their central galaxies) -- however, the baryonic
Tully-Fisher relation and constraints on the dark matter halos of
disks \citep[e.g.][and references therein]{persic90,persic96,belldejong:tf,mcgaugh:tf}
reflect the expected cosmological trends (namely, baryon fractions and
halo sizes scaling as predicted in $\Lambda$CDM, in a weaker and
opposite sense from that necessary if the FP tilt were to be explained
in this manner).  In short, effects {\em (a)} and {\em (b)} would
predict that disks should follow the same FP scalings as ellipticals
(modulo possible normalization offsets) -- while in fact, they obey
different scalings with, in many cases, an {\em opposite} qualitative
sense \citep[e.g.\ their size-mass, velocity-mass, surface
brightness-mass, and FP scalings;][]{fj76,kormendy:dissipation,shen:size.mass}.

\begin{figure}
    \centering
    \scaleup
    \plotter{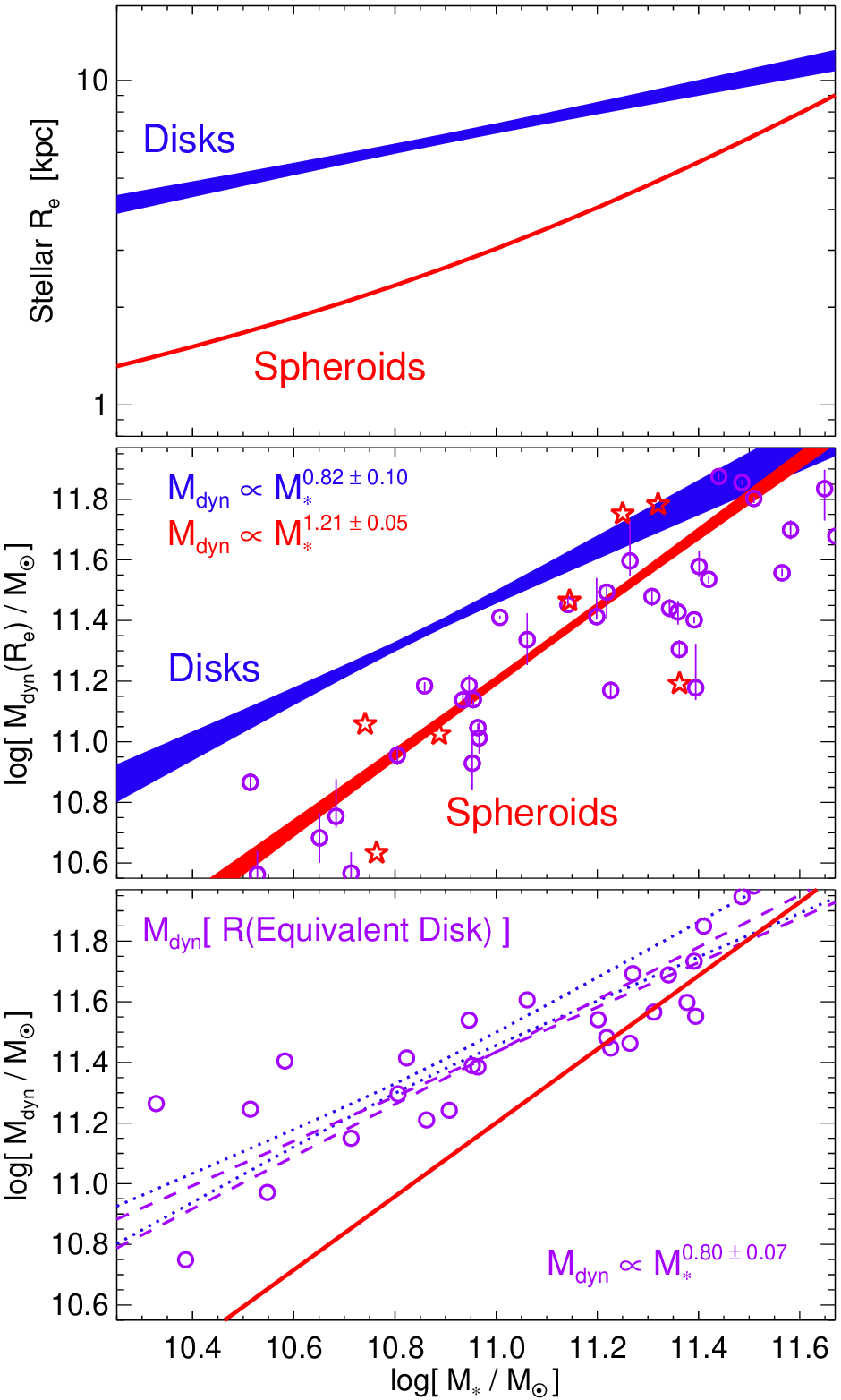}
    \caption{{\em Top:} Mean stellar size-mass relation of ellipticals, from Figure~\ref{fig:fp.projections}, 
    compared to that of disks (from \citet{shen:size.mass}). 
    {\em Middle:} FP of spheroids and disks -- i.e.\ ratio of $\mdyn$ (evaluated at $R_{e}$) 
    to $\mstar$. We show the observed ellipticals from Figure~\ref{fig:fp.pred}, 
    and the best-fit power law $\mdyn\propto\mstar^{1+\tilt}$ 
    to both relations (with corresponding uncertainties). Disk correlation is 
    from the data in \citet{belldejong:tf}. 
    At a given stellar mass, (low mass) spheroids have more compact stellar mass distributions, 
    and less total mass ($\mdyn$) enclosed within their stellar $\re$. 
    {\em Bottom:} Same, but measuring $\mdyn$ for the same observed ellipticals 
    (from {\em middle})
    at the mean expected radius of an equivalent (similar stellar mass) disk (from {\em top}). 
    The best-fit relation to these data is shown; it is indistinguishable from the relation for 
    disks. At the radii of their equivalent disks, ellipticals and disks have the {\em same} 
    enclosed total (dark matter plus stellar) masses. The distribution of dark matter 
    is {\em not} significantly different -- the smaller sizes and $\mdyn$ of ellipticals 
    must reflect a contraction of the baryonic material relative to the 
    dark matter, as predicted to occur in dissipative mergers. 
    \label{fig:equiv.disks}}
\end{figure}

Figure~\ref{fig:equiv.disks} demonstrates this explicitly. We show the
stellar size-mass relation of ellipticals (from
Figure~\ref{fig:fp.projections}), compared to that of disks (from
\citet{shen:size.mass}, but for our purposes here different sources
agree well). Modulo a small normalization offset (owing to the profile
shape), ellipticals would necessarily obey the same correlation if
they were formed in purely dissipationless mergers 
\citep[see also][]{ciotti.van.albada:msig.constraints.on.gal.form}.  Obviously,
ellipticals at low mass (where we empirically estimate and
theoretically expect dissipation to be important) are much more
compact in their {\em stellar} distributions than disks.  We also
compare the FP of both types of objects; i.e.\ dynamical mass $\mdyn$
(measured within $R_{e}$ of the stellar light) versus stellar mass. We
estimate the relation for disks based on the data and best-fit
relations in \citet{belldejong:tf} 
\citep[see also][who construct a similar correlation]{persic90,persic96}. 
Note that, for our purposes here,
it makes little difference whether we plot the baryonic or stellar
mass (there is almost no difference for ellipticals, owing to their
small gas content, and for disks, the relations fall within the quoted
uncertainties in either case).  It also makes no significant
difference whether we use the same virial constant $k$ to estimate
$\mdyn$ for both disks and ellipticals, or attempt to make some
correction for either profile shapes or the use of a velocity
dispersion as opposed to a circular velocity (the difference is small,
a factor $\lesssim2$, and we are ultimately interested in the
qualitative scalings). 

It is clear (in agreement with previous work) 
that the scaling of $\mdyn/\mstar$ in disks is opposite
that of ellipticals: there is either no tilt or {\em inverse} tilt in
the FP (i.e.\ $\mdyn/\mstar$ is the same or {\em higher} in low-mass
disks).  For ellipticals where we have kinematic data as a function of
radius, we can test whether the difference in the disk and elliptical
scalings owes to effects {\em (a)} or {\em (b)}; we do so by
evaluating their enclosed mass $\mdyn$ not at the observed stellar
effective radius of the spheroid, but at the radius of an equivalent
disk (i.e.\ at the mean $R_{e}$ of a disk of the same stellar
mass). The resulting trend of $\mdyn$ versus $\mstar$ is
indistinguishable from that of observed disks -- i.e.\ by considering
elliptical properties at their equivalent disk radii (radii they would
have in the absence of dissipation), we effectively remove the tilt of
the FP, and recover the observed correlations of disks.

At the radius of a disk of similar mass (equivalently, at the radius
the elliptical would have, if it were the product of a purely
dissipationless merger of stellar disks), ellipticals have the same
enclosed total mass as equivalent disks. In other words, at the same
(equivalent dissipationless) radius, disks and ellipticals of the same
mass have the same dark matter mass content and distribution. 
In general, it is observationally well-established that the 
FP correlations 
of ellipticals become less distinct from those of disks as 
their properties are measured at larger radii. These
correlations clearly rule out scenarios {\em (a)} and {\em (b)} above:
if case {\em (b)} were true, disks and ellipticals should obey a
similar stellar size-mass relation, and at the $R_{e}$ of an
equivalent disk, ellipticals should still have much lower $\mdyn$ than 
spirals. If case {\em (a)} were true (i.e.\ both stellar and dark matter were more
compact in ellipticals, but with lower total dark matter to stellar
mass ratios), then we would again expect much lower $\mdyn$ at the
equivalent disk radius in ellipticals.

{\em (c) Halos are Fixed, Baryons Contract:} 
The only remaining possibility is our conclusion in this paper:
namely, that the FP tilt arises because lower-mass spheroids have more
compact {\em stellar} mass distributions, relative to their halos
(equivalently, relative to what their stellar mass distribution would
be in the absence of dissipation).  We have demonstrated that this
outcome is a natural consequence of dissipation in mergers --
regardless of the initial scalings obeyed by progenitor disks, the
fact that low mass disks are more gas-rich implies that, on average,
there will be more dissipation in their mergers, yielding more compact
baryonic remnants (while having little effect on the halo), and
therefore increasing the ratio of stellar to dark matter mass inside
the stellar $R_{e}$ in lower-mass systems. We further demonstrate that
this reproduces precisely the observed tilt and scalings of elliptical
properties with mass, while being consistent with all other
observational constraints on the FP, including the ``homology
constraint,'' that $\mtrue\propto\mdyn$.

We have also shown that dissipation is {\em necessary} in achieving
this. It has been known for some time that dissipationless mergers
cannot alter the phase-space density of ellipticals relative to their
disk progenitors; consequently, ellipticals produced through
dissipationless mergers obey the same scaling relations as their
spiral progenitors (modulo normalization offsets), in stark contrast
to the observed FP scalings. We demonstrate this explicitly: systems
with the same dissipational fraction have the same ratio of total to
stellar mass within $R_{e}$; i.e.\ do not internally exhibit any FP
``tilt.''  Furthermore, we show that dissipation is the dominant factor
determining the effective radii of ellipticals at fixed mass, even
allowing for differences in formation and merger history -- therefore
if the explanation for the FP invokes any systematic change in
elliptical sizes, it {\em must} involve dissipation.  Only when the
observed dependence of dissipation on mass is included is the observed
tilt recovered.

\subsection{Additional Predictions}
\label{sec:discuss:pred}

We have extensively considered the role of dissipation in setting the
FP tilt, effectively changing the ratio of $\mdyn/\mstar$.  To the
extent that other properties also trace the degree of dissipation, we
predict that these should similarly correlate with $\mdyn/\mstar$.  In
\paperone\ and \papertwo\ we develop an extensive set of predictions
for elliptical properties that relate to the degree of dissipation in
the spheroid-forming merger, and show how these relate to, e.g., the
observed extra light (i.e.\ the tracer of the degree of dissipation in
the spheroid forming merger-induced starburst).  We refer to those
papers for details and a large number of observational proxies of the
degree of dissipation which can be used to further test the ideas
herein.

To the extent that the degree of dissipation in the original
spheroid-forming merger reflects the gas fractions of the progenitor
disks, it must also reflect the progenitor star formation history.
Broadly speaking, disks with more extended star formation histories
would be expected to have larger gas fractions at the time of their
merger, with younger stellar population ages and lower
$\alpha$-enhancement. If we ignore the effect of the merger on these
stellar populations (a reasonable assumption if the system is observed
at times significantly later than the merger, and if the mass fraction
formed in the merger-induced starburst is not large, which are true
for most of the observed systems of interest here), then these should
be reflected in the stellar populations of the elliptical remnant. 
Preliminary observational comparisons from \citet{graves:prep} 
appear to support these predictions, and can provide 
powerful independent tests and constraints 
for models of dissipation in spheroid formation: we therefore 
outline some relevant quantitative predictions.

Consider the following highly simplified toy model: identical
progenitor disks with initial gas fraction $\fgas=1$ follow an
exponential, closed-box star formation history with time scale $\tau$,
i.e.\ $\dot{M}_{\ast}\propto\exp{(-t/\tau)}$, and merge at time
$t_{m}$, when the remaining gas is rapidly consumed in a central
starburst. The gas fraction at the time of the merger (and
correspondingly, the dissipational fraction in the starburst) will be
$\fsb=\exp{(-t_{m}/\tau)}$ (giving $\tau = t_{m}/\ln{(1/\fsb)}$), and
the mass-weighted mean formation time of the stars will be $t_{\rm
form} = \tau\,[1-\exp{(-t_{m}/\tau)}] = t_{m}\,
(1-\fsb)/\ln{(1/\fsb)}$.  For systems with a similar redshift range of
their last major merger (similar $t_{m}$, expected for systems of
similar mass), their stellar formation times should therefore
correlate with the starburst fraction $\fsb$ (both depending
implicitly on the pre-merger star formation timescale $\tau$).

We have shown in \S~\ref{sec:obs} how $\mdyn/\mstar$ is predicted to
scale with the dissipational fraction $\fsb$ (this can be roughly
approximated as $\mdyn/\mstar\sim(\fsb/0.2)^{-1/2}$ over the range of
interest); combining the two, this yields an expected correlation
between formation time and $x\equiv\mdyn/\mstar$ of the form $t_{\rm
form} = 0.5\,t_{m}\, (1-0.2\,x^{-2})/\ln{(2.24\,x)}$.  Observed at
$z=0$ (i.e.\ with age $t_{\rm H}-t_{\rm form}$), ellipticals with
larger $x=\mdyn/\mstar$ at fixed mass (similar $t_{m}$) should be
older -- for a typical $t_{m}\sim10\,{\rm Gyr}$, systems with
$\mdyn/\mstar\approx 2$ are predicted to be $\sim 2\,$Gyr older than
systems with $\mdyn/\mstar\approx1$.

A shorter star formation history also implies higher
$\alpha$-enrichment in the progenitors. If we adopt the correlation
for simple star formation models in \citet{thomas05:ages},
$[\alpha/{\rm Fe}]\approx 1/5-1/6\,\log{\Delta\,t}$ (where $\Delta\,t$
is the star formation timescale for a Gaussian burst, but for our
purposes here can be replaced by $\tau$ modulo a conversion constant),
and the scalings above, then we obtain the result (independent of the
merger time $t_{m}$) that ellipticals with higher $\mdyn/\mstar$
should be more $\alpha$-enriched. Specifically, objects with
$\mdyn/\mstar\approx2$ are predicted to have $[\alpha/{\rm Fe}]$
values $\approx 0.04-0.05$ higher than systems with $\mdyn/\mstar
\approx 1$.

Predictions for the absolute metallicities are more ambiguous (but see \papertwo). 
In general, the trend of total/mean metallicity with mass will be dominated 
by the metallicity of pre-merger disks, 
which observations show (excluding the most gas-rich disks, where self-enrichment 
in the merger will dominate the final total metallicity) trace a similar  
mass-metallicity correlation to ellipticals \citep{gallazzi:ssps}. 
In the absence of outflows or recycling, the metallicity 
would be the same for any systems with 
the same total accreted gas content and stellar mass (with no 
dependence on the ``starburst'' content at fixed mass). 
However, to the extent that outflows in low-mass systems are 
responsible for the mass-metallicity relation (as is generally believed), 
then the detailed interplay between these outflows and merger-induced starbursts 
will be important. In broad terms, if 
outflow strengths and velocities are similar, then we 
expect dissipational star formation at the center of 
the galaxy to retain a higher metal content in comparison to the same star formation in a more
extended disk (since escape velocities from the galactic center and densities 
leading to radiative losses in the outflows are higher). In this case, at fixed mass, 
more dissipational (lower $\mdyn/\mstar$, higher surface brightness) systems should have 
slightly higher metallicities than their less dissipational counterparts (and, as 
demonstrated in \paperone, this should be correlated with the effects above -- 
at fixed mass, the dependence of both quantities on dissipation should give rise to 
an inverse correlation between metallicity and stellar population age or $\alpha$-enrichment). 
Experimenting with e.g.\ different degrees of dissipation, outflow strengths, and 
initial disk metalliticies in our simulations, we estimate 
this to be a relatively small ($\sim0.1$\,dex) effect -- not sufficient to dramatically effect the 
global mass-metallicity relation, but potentially visible in detailed studies. 

There is one important caveat here: systems of the same mass might
also have had more gas-rich progenitors because they formed from mergers
at very early times (i.e.\ had a different merger time $t_{m}$, in the
toy model illustrated above), making them older and more
$\alpha$-enriched. However, cosmological estimates
\citep[e.g.][]{hopkins:groups.ell} suggest this process is not
dominant at a given stellar mass -- i.e.\ systems of comparable stellar
mass (and correspondingly similar total halo masses) tend to have
similar merger histories. Specifically, the relatively large scatter
in star formation history and disk gas fractions at fixed mass and
redshift (a factor $\sim2$ in $\fgas$) is larger than the scatter
introduced by the combination of a scatter in formation times and the
systematic cosmological evolution of disk gas fractions with time.
Furthermore, systems with such early mergers will usually have
multiple subsequent mergers at later times, so they will grow
significantly in mass and have their effective radii substantially
modified by these additional processes (such that they should and will
be compared to different systems at $z=0$).

Considering higher order effects, we demonstrate in \papertwo\ that
the strength of stellar population gradients is correlated with the
degree of dissipation in the original spheroid-forming gas-rich
merger, and show in \paperthree\ that this holds even for remnants of
subsequent gas-poor ``dry'' re-mergers. The most useful gradients in
this sense are metallicity gradients -- stellar age gradients and
(especially) color gradients evolve strongly with time even in a
fixed, passively evolving elliptical (owing to the change in relative
$M/L$ for young and old stellar populations) and as such are more
ambiguous, and gradients in $\alpha$-enhancement are more sensitive to
the gradients and overall star formation histories in the pre-merger
disks. Metallicity gradients are, on the other hand, generally
dominated by the degree of dissipation and imprinted in the gas-rich
merger, and are not sensitive to the star formation history of the
pre-merger disks, making them a more robust diagnostic for our
purposes. At fixed mass, stronger gradients indicate more dissipation,
and so we predict that, at fixed mass, ellipticals with higher
$\mdyn/\mstar$ should have weaker metallicity gradients (see
\papertwo).

\subsection{Summary}
\label{sec:discuss:summary}

We have demonstrated from observations that the tilt of the FP owes to
differential degrees of dissipation as a function of mass.  Lower mass
disks are more gas-rich, so their mergers are more dissipational: a
larger fraction of the remnant mass is formed in a dissipational,
merger-induced compact central starburst in the final stages of a
major merger. This yields a remnant with a more compact stellar mass
distribution, i.e.\ smaller $R_{e}$ relative to their progenitor
disks, in lower-mass ellipticals.  The dark matter distribution is
only weakly affected -- implying that the stellar distribution in
low-mass ellipticals is more compact, relative to the dark matter,
than in equivalent disks or higher-mass ellipticals. Consequently,
relatively less dark matter mass is enclosed within the stellar
effective radius in low-mass ellipticals, so the ratio of enclosed
mass $\mdyn/\mstar$ is an increasing function of mass
($\mdyn/\mstar\propto\mstar^{\tilt}$). This is the ``tilt'' in the FP.
Given the observed, quantitative dependence of gas fractions on mass,
the tilt is predicted to be exactly that observed, $\tilt\approx0.2$.

Using a new empirical method, we robustly estimate the amount of
dissipation involved in the formation a given elliptical.
Specifically, with data of sufficient quality, we separate the
observed surface brightness profile into an outer, violently relaxed
component, which was established in a dissipationless manner, and an
inner ``starburst remnant'' or ``extra light'' component.  We
demonstrate in \paperone, \papertwo\ and \paperthree\ that
observations of both evolved ellipticals and recent merger remnants
support the proposal that this compact nuclear mass component a good
proxy for the true mass formed in a dissipational starburst. Using
this proxy, we show that the observed sizes of ellipticals, at fixed
mass, depend strongly on the degree of dissipation involved in
their formation (more so than even
e.g.\ the number of mergers in their formation history).  
Correspondingly, we show that the ratio of total to
stellar mass within the stellar effective radius, $\mdyn/\mstar$, is a
function of dissipation, both globally and at fixed mass (in the sense
that elliptical sizes and $\mdyn/\mstar$ are decreasing functions of
the amount of dissipation).  These observed dependences are highly
significant ($P_{\rm null}\lesssim10^{-7}$).  Motivated by this, we
show that by removing the mean systematic dependence of dissipation on
mass, we can empirically remove the tilt of the FP.  Considering
ellipticals with the same dissipational extra light fractions, we show
that they obey a relation $\mdyn\propto\mstar$ (i.e.\ ellipticals with
the same extra light content have the same ratio of dynamical to
stellar mass within $R_{e}$, independent of mass).

In the proposed dissipational models of the FP
\citep{bekki:fp.origin.tsf,onorbe:diss.fp.model,robertson:fp,dekelcox:fp}, the tilt of the
FP, and its projected correlations (e.g.\ the steepness of the stellar
size-mass correlation of ellipticals relative to that of disks),
arise because low-mass disks are more gas-rich, and therefore
low-mass mergers and ellipticals will have (on average) systematically
higher degrees of dissipation, and therefore smaller (relative)
$R_{e}$ and $\mdyn/\mstar$. If we consider e.g.\ simulations that obey
the observed correlation between disk gas content and mass (as opposed
to being dissipationless, or having all the same gas fractions
independent of mass -- neither of which is consistent with
observations), then the FP predicted has exactly the observed
tilt. Equivalently, the observed mean dissipational fractions of
ellipticals, as a function of mass, agree well with the observed gas
fractions of progenitor disks of the same masses, over the redshift
range $z\sim0-2$. In other words, {\em dissipation is both necessary
and sufficient to explain the FP tilt and differences between disk and
elliptical scaling relations}.

To our knowledge, this is the first {\em explicit} observational test
of these theoretical models.  We further demonstrate that other
mechanisms cannot be responsible for the majority of the FP tilt. For
example, observations have demonstrated that the ``homology
assumption,'' namely that $\mtrue\propto\sigma^{2}\,R_{e}$, is valid,
and we show that simulations predict this -- the homology breaking
introduced by dissipation is negligible.  In other words, the FP tilt
reflects the ratio of stellar to true mass enclosed within $R_{e}$:
$\mtrue/\mstar\propto\mstar^{\alpha}$ (this ratio is {\em not}
constant), rather than an ``apparent'' effect.  We also show that, if
we measure elliptical properties at the radius of an equivalent disk,
the tilt of the FP is removed: within the radii of disks of the same
mass, ellipticals have {\em the same} total and dynamical masses.
That is, the dependence of $\mdyn/\mstar$ cannot be driven primarily
by changes in the dark matter distribution at fixed baryonic
properties, nor by changes in the total dark matter to stellar mass
ratio (integrated over the entire halo).  The variation in
$\mdyn/\mstar$ within $R_{e}$ {\em must} predominantly reflect the
change in size of the baryonic component: low-mass ellipticals have
much more compact stellar distributions than similar-mass disks, and
therefore have less enclosed dark matter within that stellar $R_{e}$,
as predicted in dissipational theories.


Together, these observational tests represent an important vindication
of the ``merger hypothesis,'' that ellipticals are formed by the
gas-rich mergers of disk galaxies, and models for the origin of the 
FP in dissipational major mergers \citep{robertson:fp,dekelcox:fp}. 
We explicitly demonstrate that,
regardless of subsequent gas-poor (spheroid-spheroid or ``dry'')
re-mergers, the location of systems with respect to the FP and e.g.\
elliptical size-mass and velocity dispersion-mass relations is
primarily determined by the amount of dissipation involved in their
formation: i.e.\ the gas content involved in the original,
spheroid-forming merger. Gas rich mergers cannot be ignored in the
formation of ellipticals.  Not only have we demonstrated that the FP
is consistent with the merger hypothesis, but that (given the
systematic dependence of disk gas fractions on mass), a FP tilted in
the manner observed is a necessary prediction of the theory.

We have also shown that elliptical sizes, inferred dissipational
fractions, and the FP are completely consistent with the formation of
ellipticals in mergers of disks with similar properties (sizes, gas
fractions, dark matter halo masses and sizes) to those observed in
{\em low-redshift} ($z\sim0-1$) disks. In other words, the sizes of
ellipticals and their FP correlations {\em do not require elliptical
progenitors to be more compact than observed, low-redshift
disks}. Dissipation is sufficient to explain the differences in their
densities and sizes. The fact that, within the radius of an equivalent
disk, ellipticals obey the same correlation between total and stellar
mass actually implies that their dark matter halos (and presumably
those of their progenitors) are {\em not} significantly more compact
than those of low-redshift disks.

This is important for the viability of the merger hypothesis, given
the observations indicating that disk and halo sizes do not evolve
strongly with redshift
\citep{ravindranath:disk.size.evol,barden:disk.size.evol,
trujillo:size.evolution,zirm:drg.sizes}.  This is not to say that
elliptical sizes might not evolve with redshift \citep[which is easily
possible if e.g.\ disk gas fractions systematically evolve;
see][]{khochfar:size.evolution.model,hopkins:bhfp.theory}, nor that
ellipticals all formed at low redshift (indeed, if the disk size
evolution is weak, then ellipticals form could rapidly at relatively early times 
and still resemble the products of low-redshift
disks). However, it does imply that exotic progenitors -- progenitors
not found in the local universe -- are not required to explain the
observed correlations, surface brightness profiles (see \papertwo\ and
\paperthree), or kinematics \citep[see][]{cox:kinematics} of typical
local ellipticals.

\acknowledgments We thank John Kormendy, Tod Lauer, and Barry 
Rothberg for providing observations and suggestions for the content 
herein, and thank Brant Robertson, Marijn Franx and Sandy Faber for 
helpful discussions throughout the development of this 
paper. We also thank our referee, Luca Ciotti, for helpful advice on the 
content and discussion herein. This work
was supported in part by NSF grants ACI 96-19019, AST 00-71019, AST
02-06299, and AST 03-07690, and NASA ATP grants NAG5-12140,
NAG5-13292, and NAG5-13381. Support for 
TJC was provided by the W.~M.\ Keck 
Foundation.

\bibliography{/Users/phopkins/Documents/lars_galaxies/papers/ms}

\begin{thebibliography}{200}
\expandafter\ifx\csname natexlab\endcsname\relax\def\natexlab#1{#1}\fi

\bibitem[{{Avila-Reese} {et~al.}(2008){Avila-Reese}, {Zavala}, {Firmani}, \&
  {Hern{\'a}ndez-Toledo}}]{avilareese:baryonic.tf}
{Avila-Reese}, V., {Zavala}, J., {Firmani}, C., \& {Hern{\'a}ndez-Toledo},
  H.~M. 2008, \aj, in press, arXiv:0807.0636 [astro-ph], 807

\bibitem[{{Balcells} {et~al.}(2007){Balcells}, {Graham}, \&
  {Peletier}}]{balcells:bulge.xl}
{Balcells}, M., {Graham}, A.~W., \& {Peletier}, R.~F. 2007, \apj, 665, 1084

\bibitem[{{Barden} {et~al.}(2005)}]{barden:disk.size.evol}
{Barden}, M., {et~al.} 2005, \apj, 635, 959

\bibitem[{{Barnes}(1988)}]{barnes:disk.halo.mergers}
{Barnes}, J.~E. 1988, \apj, 331, 699

\bibitem[{{Barnes} \& {Hernquist}(1996)}]{barneshernquist96}
{Barnes}, J.~E., \& {Hernquist}, L. 1996, \apj, 471, 115

\bibitem[{{Barnes} \& {Hernquist}(1991)}]{barnes.hernquist.91}
{Barnes}, J.~E., \& {Hernquist}, L.~E. 1991, \apjl, 370, L65

\bibitem[{{Bekki}(1998)}]{bekki:fp.origin.tsf}
{Bekki}, K. 1998, \apj, 496, 713

\bibitem[{{Bell} \& {de Jong}(2000)}]{belldejong:disk.sfh}
{Bell}, E.~F., \& {de Jong}, R.~S. 2000, \mnras, 312, 497

\bibitem[{{Bell} \& {de Jong}(2001)}]{belldejong:tf}
---. 2001, \apj, 550, 212

\bibitem[{{Bell} {et~al.}(2003){Bell}, {McIntosh}, {Katz}, \&
  {Weinberg}}]{bell:mfs}
{Bell}, E.~F., {McIntosh}, D.~H., {Katz}, N., \& {Weinberg}, M.~D. 2003, \apjs,
  149, 289

\bibitem[{{Bender}(1988)}]{bender:88.shapes}
{Bender}, R. 1988, \aap, 193, L7

\bibitem[{{Bender} {et~al.}(1992){Bender}, {Burstein}, \&
  {Faber}}]{bender:ell.kinematics}
{Bender}, R., {Burstein}, D., \& {Faber}, S.~M. 1992, \apj, 399, 462

\bibitem[{{Bender} {et~al.}(1993){Bender}, {Burstein}, \&
  {Faber}}]{bender:ell.kinematics.a4}
---. 1993, \apj, 411, 153

\bibitem[{{Bender} {et~al.}(1987){Bender}, {Doebereiner}, \&
  {Moellenhoff}}]{bender:87.a4}
{Bender}, R., {Doebereiner}, S., \& {Moellenhoff}, C. 1987, \aap, 177, L53

\bibitem[{{Bender} {et~al.}(1988){Bender}, {Doebereiner}, \&
  {Moellenhoff}}]{bender:data}
---. 1988, \aaps, 74, 385

\bibitem[{{Bender} {et~al.}(1994){Bender}, {Saglia}, \&
  {Gerhard}}]{bender:velocity.structure}
{Bender}, R., {Saglia}, R.~P., \& {Gerhard}, O.~E. 1994, \mnras, 269, 785

\bibitem[{{Bender} {et~al.}(2007)}]{bender:06}
{Bender}, R., {et~al.} 2007, \apj, in preparation

\bibitem[{{Bernardi} {et~al.}(2003)}]{bernardi:correlations}
{Bernardi}, M., {et~al.} 2003, \apj, 125, 1849

\bibitem[{{Bertin} {et~al.}(2002){Bertin}, {Ciotti}, \& {Del
  Principe}}]{bertin:weak.homology}
{Bertin}, G., {Ciotti}, L., \& {Del Principe}, M. 2002, \aap, 386, 149

\bibitem[{{Binggeli} {et~al.}(1985){Binggeli}, {Sandage}, \&
  {Tammann}}]{binggeli:vcc}
{Binggeli}, B., {Sandage}, A., \& {Tammann}, G.~A. 1985, \aj, 90, 1681

\bibitem[{{Bolton} {et~al.}(2007){Bolton}, {Burles}, {Treu}, {Koopmans}, \&
  {Moustakas}}]{bolton:fp}
{Bolton}, A.~S., {Burles}, S., {Treu}, T., {Koopmans}, L.~V.~E., \&
  {Moustakas}, L.~A. 2007, \apjl, 665, L105

\bibitem[{{Bolton} {et~al.}(2008){Bolton}, {Treu}, {Koopmans}, {Gavazzi},
  {Moustakas}, {Burles}, {Schlegel}, \& {Wayth}}]{bolton:fp.update}
{Bolton}, A.~S., {Treu}, T., {Koopmans}, L.~V.~E., {Gavazzi}, R., {Moustakas},
  L.~A., {Burles}, S., {Schlegel}, D.~J., \& {Wayth}, R. 2008, \apj, in press
  arXiv:0805.1932 [astro-ph], 805

\bibitem[{{Borriello} \& {Salucci}(2001)}]{borriello01}
{Borriello}, A., \& {Salucci}, P. 2001, \mnras, 323, 285

\bibitem[{{Borriello} {et~al.}(2003){Borriello}, {Salucci}, \&
  {Danese}}]{borriello03}
{Borriello}, A., {Salucci}, P., \& {Danese}, L. 2003, \mnras, 341, 1109

\bibitem[{{Boylan-Kolchin} {et~al.}(2005){Boylan-Kolchin}, {Ma}, \&
  {Quataert}}]{boylankolchin:mergers.fp}
{Boylan-Kolchin}, M., {Ma}, C.-P., \& {Quataert}, E. 2005, \mnras, 362, 184

\bibitem[{{Boylan-Kolchin} {et~al.}(2006){Boylan-Kolchin}, {Ma}, \&
  {Quataert}}]{boylankolchin:dry.mergers}
---. 2006, \mnras, 369, 1081

\bibitem[{{Bruzual} \& {Charlot}(2003)}]{BC03}
{Bruzual}, G., \& {Charlot}, S. 2003, \mnras, 344, 1000

\bibitem[{{Bullock} {et~al.}(2001){Bullock}, {Kolatt}, {Sigad}, {Somerville},
  {Kravtsov}, {Klypin}, {Primack}, \& {Dekel}}]{bullock:concentrations}
{Bullock}, J.~S., {Kolatt}, T.~S., {Sigad}, Y., {Somerville}, R.~S.,
  {Kravtsov}, A.~V., {Klypin}, A.~A., {Primack}, J.~R., \& {Dekel}, A. 2001,
  \mnras, 321, 559

\bibitem[{{Buote} {et~al.}(2007){Buote}, {Gastaldello}, {Humphrey},
  {Zappacosta}, {Bullock}, {Brighenti}, \&
  {Mathews}}]{buote:obs.concentrations}
{Buote}, D.~A., {Gastaldello}, F., {Humphrey}, P.~J., {Zappacosta}, L.,
  {Bullock}, J.~S., {Brighenti}, F., \& {Mathews}, W.~G. 2007, \apj, 664, 123

\bibitem[{{Busha} {et~al.}(2005){Busha}, {Evrard}, {Adams}, \&
  {Wechsler}}]{busha:halomass}
{Busha}, M.~T., {Evrard}, A.~E., {Adams}, F.~C., \& {Wechsler}, R.~H. 2005,
  \mnras, 363, L11

\bibitem[{{Caon} {et~al.}(1994){Caon}, {Capaccioli}, \&
  {D'Onofrio}}]{caon:profiles}
{Caon}, N., {Capaccioli}, M., \& {D'Onofrio}, M. 1994, \aaps, 106, 199

\bibitem[{{Caon} {et~al.}(1990){Caon}, {Capaccioli}, \& {Rampazzo}}]{caon90}
{Caon}, N., {Capaccioli}, M., \& {Rampazzo}, R. 1990, \aaps, 86, 429

\bibitem[{{Capelato} {et~al.}(1995){Capelato}, {de Carvalho}, \&
  {Carlberg}}]{capelato:dry.mgr.fp}
{Capelato}, H.~V., {de Carvalho}, R.~R., \& {Carlberg}, R.~G. 1995, \apj, 451,
  525

\bibitem[{{Cappellari} {et~al.}(2006)}]{cappellari:fp}
{Cappellari}, M., {et~al.} 2006, \mnras, 366, 1126

\bibitem[{{Cappellari} {et~al.}(2007)}]{cappellari:anisotropy}
---. 2007, \mnras, 379, 418

\bibitem[{{Carlberg}(1986)}]{carlberg:phase.space}
{Carlberg}, R.~G. 1986, \apj, 310, 593

\bibitem[{{Carollo} {et~al.}(1995){Carollo}, {de Zeeuw}, {van der Marel},
  {Danziger}, \& {Qian}}]{carollo95:dm}
{Carollo}, C.~M., {de Zeeuw}, P.~T., {van der Marel}, R.~P., {Danziger}, I.~J.,
  \& {Qian}, E.~E. 1995, \apjl, 441, L25

\bibitem[{{Chabrier}(2003)}]{chabrier:imf}
{Chabrier}, G. 2003, \pasp, 115, 763

\bibitem[{{Ciotti} {et~al.}(1996){Ciotti}, {Lanzoni}, \&
  {Renzini}}]{ciotti:kinematic.nonhomology}
{Ciotti}, L., {Lanzoni}, B., \& {Renzini}, A. 1996, \mnras, 282, 1

\bibitem[{{Ciotti} {et~al.}(2007){Ciotti}, {Lanzoni}, \&
  {Volonteri}}]{ciotti:dry.vs.wet.mergers}
{Ciotti}, L., {Lanzoni}, B., \& {Volonteri}, M. 2007, \apj, 658, 65

\bibitem[{{Ciotti} \& {Ostriker}(2007)}]{ciottiostriker:recycling}
{Ciotti}, L., \& {Ostriker}, J.~P. 2007, \apj, 665, 1038

\bibitem[{{Ciotti} \& {van
  Albada}(2001)}]{ciotti.van.albada:msig.constraints.on.gal.form}
{Ciotti}, L., \& {van Albada}, T.~S. 2001, \apjl, 552, L13

\bibitem[{{Comerford} \& {Natarajan}(2007)}]{comerford:obs.concentrations}
{Comerford}, J.~M., \& {Natarajan}, P. 2007, \mnras, 379, 190

\bibitem[{{Conroy} {et~al.}(2006){Conroy}, {Wechsler}, \&
  {Kravtsov}}]{conroy:monotonic.hod}
{Conroy}, C., {Wechsler}, R.~H., \& {Kravtsov}, A.~V. 2006, \apj, 647, 201

\bibitem[{{C{\^o}t{\'e}} {et~al.}(2006)}]{cote:virgo}
{C{\^o}t{\'e}}, P., {et~al.} 2006, \apjs, 165, 57

\bibitem[{{Courteau} {et~al.}(2007){Courteau}, {Dutton}, {van den Bosch},
  {MacArthur}, {Dekel}, {McIntosh}, \& {Dale}}]{courteau:disk.scalings}
{Courteau}, S., {Dutton}, A.~A., {van den Bosch}, F.~C., {MacArthur}, L.~A.,
  {Dekel}, A., {McIntosh}, D.~H., \& {Dale}, D.~A. 2007, \apj, 671, 203

\bibitem[{{Covington} {et~al.}(2008){Covington}, {Dekel}, {Cox}, {Jonsson}, \&
  {Primack}}]{covington:diss.size.expectation}
{Covington}, M., {Dekel}, A., {Cox}, T.~J., {Jonsson}, P., \& {Primack}, J.~R.
  2008, \mnras, 384, 94

\bibitem[{{Cox} {et~al.}(2006{\natexlab{a}}){Cox}, {Di Matteo}, {Hernquist},
  {Hopkins}, {Robertson}, \& {Springel}}]{cox:xray.gas}
{Cox}, T.~J., {Di Matteo}, T., {Hernquist}, L., {Hopkins}, P.~F., {Robertson},
  B., \& {Springel}, V. 2006{\natexlab{a}}, \apj, 643, 692

\bibitem[{{Cox} {et~al.}(2006{\natexlab{b}}){Cox}, {Dutta}, {Di Matteo},
  {Hernquist}, {Hopkins}, {Robertson}, \& {Springel}}]{cox:kinematics}
{Cox}, T.~J., {Dutta}, S.~N., {Di Matteo}, T., {Hernquist}, L., {Hopkins},
  P.~F., {Robertson}, B., \& {Springel}, V. 2006{\natexlab{b}}, \apj, 650, 791

\bibitem[{{Dantas} {et~al.}(2003){Dantas}, {Capelato}, {Ribeiro}, \& {de
  Carvalho}}]{dantas:dry.mgr.fp}
{Dantas}, C.~C., {Capelato}, H.~V., {Ribeiro}, A.~L.~B., \& {de Carvalho},
  R.~R. 2003, \mnras, 340, 398

\bibitem[{{Dav{\'e}} {et~al.}(1999){Dav{\'e}}, {Hernquist}, {Katz}, \&
  {Weinberg}}]{dave:lyalpha}
{Dav{\'e}}, R., {Hernquist}, L., {Katz}, N., \& {Weinberg}, D.~H. 1999, \apj,
  511, 521

\bibitem[{{Davis} {et~al.}(1985){Davis}, {Cawson}, {Davies}, \&
  {Illingworth}}]{davis:85}
{Davis}, L.~E., {Cawson}, M., {Davies}, R.~L., \& {Illingworth}, G. 1985, \aj,
  90, 169

\bibitem[{{de Lucia} \& {Blaizot}(2007)}]{delucia:sam}
{de Lucia}, G., \& {Blaizot}, J. 2007, \mnras, 375, 2

\bibitem[{{de Vaucouleurs}(1948)}]{devaucouleurs}
{de Vaucouleurs}, G. 1948, Annales d'Astrophysique, 11, 247

\bibitem[{{Dekel} \& {Cox}(2006)}]{dekelcox:fp}
{Dekel}, A., \& {Cox}, T.~J. 2006, \mnras, 370, 1445

\bibitem[{{Desroches} {et~al.}(2007){Desroches}, {Quataert}, {Ma}, \&
  {West}}]{desroches:scaling.law.curvature}
{Desroches}, L.-B., {Quataert}, E., {Ma}, C.-P., \& {West}, A.~A. 2007, \mnras,
  377, 402

\bibitem[{{Di Matteo} {et~al.}(2005){Di Matteo}, {Springel}, \&
  {Hernquist}}]{dimatteo:msigma}
{Di Matteo}, T., {Springel}, V., \& {Hernquist}, L. 2005, \nat, 433, 604

\bibitem[{{di Serego Alighieri} {et~al.}(2005)}]{alighieri:fp.evolution}
{di Serego Alighieri}, S., {et~al.} 2005, \aap, 442, 125

\bibitem[{{Djorgovski} \& {Davis}(1987)}]{dd87:fp}
{Djorgovski}, S., \& {Davis}, M. 1987, \apj, 313, 59

\bibitem[{{Djorgovski} {et~al.}(1988){Djorgovski}, {de Carvalho}, \&
  {Han}}]{djorgovski:fp.tilt}
{Djorgovski}, S., {de Carvalho}, R., \& {Han}, M.-S. 1988, in Astronomical
  Society of the Pacific Conference Series, Vol.~4, The Extragalactic Distance
  Scale, ed. S.~{van den Bergh} \& C.~J. {Pritchet}, 329--341

\bibitem[{{Dolag} {et~al.}(2004){Dolag}, {Bartelmann}, {Perrotta},
  {Baccigalupi}, {Moscardini}, {Meneghetti}, \&
  {Tormen}}]{dolag:concentrations}
{Dolag}, K., {Bartelmann}, M., {Perrotta}, F., {Baccigalupi}, C., {Moscardini},
  L., {Meneghetti}, M., \& {Tormen}, G. 2004, \aap, 416, 853

\bibitem[{{Doyon} {et~al.}(1994){Doyon}, {Wells}, {Wright}, {Joseph}, {Nadeau},
  \& {James}}]{Doyon94}
{Doyon}, R., {Wells}, M., {Wright}, G.~S., {Joseph}, R.~D., {Nadeau}, D., \&
  {James}, P.~A. 1994, \apjl, 437, L23

\bibitem[{{Dressler} {et~al.}(1987){Dressler}, {Lynden-Bell}, {Burstein},
  {Davies}, {Faber}, {Terlevich}, \& {Wegner}}]{dressler87:fp}
{Dressler}, A., {Lynden-Bell}, D., {Burstein}, D., {Davies}, R.~L., {Faber},
  S.~M., {Terlevich}, R., \& {Wegner}, G. 1987, \apj, 313, 42

\bibitem[{{Eke} {et~al.}(2004)}]{eke:groups}
{Eke}, V.~R., {et~al.} 2004, \mnras, 355, 769

\bibitem[{{Emsellem} {et~al.}(2004)}]{emsellem:sauron.rotation.data}
{Emsellem}, E., {et~al.} 2004, \mnras, 352, 721

\bibitem[{{Emsellem} {et~al.}(2007)}]{emsellem:sauron.rotation}
---. 2007, \mnras, 379, 401

\bibitem[{{Erb} {et~al.}(2006){Erb}, {Steidel}, {Shapley}, {Pettini}, {Reddy},
  \& {Adelberger}}]{erb:lbg.gasmasses}
{Erb}, D.~K., {Steidel}, C.~C., {Shapley}, A.~E., {Pettini}, M., {Reddy},
  N.~A., \& {Adelberger}, K.~L. 2006, \apj, 646, 107

\bibitem[{{Faber} \& {Jackson}(1976)}]{fj76}
{Faber}, S.~M., \& {Jackson}, R.~E. 1976, \apj, 204, 668

\bibitem[{{Faber} {et~al.}(1997){Faber}, {Tremaine}, {Ajhar}, {Byun},
  {Dressler}, {Gebhardt}, {Grillmair}, {Kormendy}, {Lauer}, \&
  {Richstone}}]{faber:ell.centers}
{Faber}, S.~M., {Tremaine}, S., {Ajhar}, E.~A., {Byun}, Y.-I., {Dressler}, A.,
  {Gebhardt}, K., {Grillmair}, C., {Kormendy}, J., {Lauer}, T.~R., \&
  {Richstone}, D. 1997, \aj, 114, 1771

\bibitem[{{Ferrarese} {et~al.}(2006)}]{ferrarese:profiles}
{Ferrarese}, L., {et~al.} 2006, \apjs, 164, 334

\bibitem[{{Gallazzi} {et~al.}(2006){Gallazzi}, {Charlot}, {Brinchmann}, \&
  {White}}]{gallazzi06:ages}
{Gallazzi}, A., {Charlot}, S., {Brinchmann}, J., \& {White}, S.~D.~M. 2006,
  \mnras, 370, 1106

\bibitem[{{Gallazzi} {et~al.}(2005){Gallazzi}, {Charlot}, {Brinchmann},
  {White}, \& {Tremonti}}]{gallazzi:ssps}
{Gallazzi}, A., {Charlot}, S., {Brinchmann}, J., {White}, S.~D.~M., \&
  {Tremonti}, C.~A. 2005, \mnras, 362, 41

\bibitem[{{Genzel} {et~al.}(2001){Genzel}, {Tacconi}, {Rigopoulou}, {Lutz}, \&
  {Tecza}}]{Genzel01}
{Genzel}, R., {Tacconi}, L.~J., {Rigopoulou}, D., {Lutz}, D., \& {Tecza}, M.
  2001, \apj, 563, 527

\bibitem[{{Gerhard}(2003)}]{gerhard03:mdyn}
{Gerhard}, O. 2003, in The Mass of Galaxies at Low and High Redshift, ed.
  R.~{Bender} \& A.~{Renzini}, 62--+

\bibitem[{{Gerhard} {et~al.}(2001){Gerhard}, {Kronawitter}, {Saglia}, \&
  {Bender}}]{gerhard:giant.ell.dynamics}
{Gerhard}, O., {Kronawitter}, A., {Saglia}, R.~P., \& {Bender}, R. 2001, \aj,
  121, 1936

\bibitem[{{Goto}(2005)}]{goto:e+a.merger.connection}
{Goto}, T. 2005, \mnras, 357, 937

\bibitem[{{Graves} {et~al.}(2008)}]{graves:prep}
{Graves}, J., {et~al.} 2008, \apj, in preparation

\bibitem[{{Gunn}(1987)}]{gunn87}
{Gunn}, J.~E. 1987, in Nearly Normal Galaxies. From the Planck Time to the
  Present, ed. S.~M. {Faber}, 455--464

\bibitem[{{H{\"a}ring} \& {Rix}(2004)}]{haringrix}
{H{\"a}ring}, N., \& {Rix}, H.-W. 2004, \apjl, 604, L89

\bibitem[{{Hausman} \& {Ostriker}(1978)}]{hausman:mergers}
{Hausman}, M.~A., \& {Ostriker}, J.~P. 1978, \apj, 224, 320

\bibitem[{{Hernquist}(1989)}]{hernquist.89}
{Hernquist}, L. 1989, \nat, 340, 687

\bibitem[{{Hernquist}(1990)}]{hernquist:profile}
---. 1990, \apj, 356, 359

\bibitem[{{Hernquist}(1993)}]{hernquist:sph.cautions}
---. 1993, \apj, 404, 717

\bibitem[{{Hernquist} {et~al.}(1993){Hernquist}, {Spergel}, \&
  {Heyl}}]{hernquist:phasespace}
{Hernquist}, L., {Spergel}, D.~N., \& {Heyl}, J.~S. 1993, \apj, 416, 415

\bibitem[{{Heymans} {et~al.}(2006)}]{heymans:mhalo-mgal.evol}
{Heymans}, C., {et~al.} 2006, \mnras, 371, L60

\bibitem[{{Hibbard} \& {Yun}(1999)}]{hibbard.yun:excess.light}
{Hibbard}, J.~E., \& {Yun}, M.~S. 1999, \apjl, 522, L93

\bibitem[{{Hopkins} {et~al.}(2008{\natexlab{a}}){Hopkins}, {Cox}, {Dutta},
  {Hernquist}, {Kormendy}, \& {Lauer}}]{hopkins:cusps.ell}
{Hopkins}, P.~F., {Cox}, T.~J., {Dutta}, S.~N., {Hernquist}, L., {Kormendy},
  J., \& {Lauer}, T.~R. 2008{\natexlab{a}}, \apj, submitted, arXiv:0805.3533
  [astro-ph], 805

\bibitem[{{Hopkins} {et~al.}(2008{\natexlab{b}}){Hopkins}, {Cox}, {Kere{\v s}},
  \& {Hernquist}}]{hopkins:groups.ell}
{Hopkins}, P.~F., {Cox}, T.~J., {Kere{\v s}}, D., \& {Hernquist}, L.
  2008{\natexlab{b}}, \apjs, 175, 390

\bibitem[{{Hopkins} {et~al.}(2008{\natexlab{c}}){Hopkins}, {Cox}, {Younger}, \&
  {Hernquist}}]{hopkins:disk.survival}
{Hopkins}, P.~F., {Cox}, T.~J., {Younger}, J.~D., \& {Hernquist}, L.
  2008{\natexlab{c}}, \apj, submitted, arXiv:0806.1739 [astro-ph], 806

\bibitem[{{Hopkins} {et~al.}(2008{\natexlab{d}}){Hopkins}, {Hernquist}, {Cox},
  {Dutta}, \& {Rothberg}}]{hopkins:cusps.mergers}
{Hopkins}, P.~F., {Hernquist}, L., {Cox}, T.~J., {Dutta}, S.~N., \& {Rothberg},
  B. 2008{\natexlab{d}}, \apj, 679, 156

\bibitem[{{Hopkins} {et~al.}(2008{\natexlab{e}}){Hopkins}, {Hernquist}, {Cox},
  {Keres}, \& {Wuyts}}]{hopkins:cusps.evol}
{Hopkins}, P.~F., {Hernquist}, L., {Cox}, T.~J., {Keres}, D., \& {Wuyts}, S.
  2008{\natexlab{e}}, \apj, submitted

\bibitem[{{Hopkins} {et~al.}(2007{\natexlab{a}}){Hopkins}, {Hernquist}, {Cox},
  {Robertson}, \& {Krause}}]{hopkins:bhfp.theory}
{Hopkins}, P.~F., {Hernquist}, L., {Cox}, T.~J., {Robertson}, B., \& {Krause},
  E. 2007{\natexlab{a}}, \apj, 669, 45

\bibitem[{{Hopkins} {et~al.}(2008{\natexlab{f}}){Hopkins}, {Hernquist}, {Cox},
  {Younger}, \& {Besla}}]{hopkins:disk.heating}
{Hopkins}, P.~F., {Hernquist}, L., {Cox}, T.~J., {Younger}, J.~D., \& {Besla},
  G. 2008{\natexlab{f}}, \apj, submitted, arXiv:0806.2861 [astro-ph], 806

\bibitem[{{Hopkins} {et~al.}(2008{\natexlab{g}}){Hopkins}, {Lauer}, {Cox},
  {Hernquist}, \& {Kormendy}}]{hopkins:cores}
{Hopkins}, P.~F., {Lauer}, T.~R., {Cox}, T.~J., {Hernquist}, L., \& {Kormendy},
  J. 2008{\natexlab{g}}, \apj, submitted, arXiv:0806.2325 [astro-ph], 806

\bibitem[{{Hopkins} {et~al.}(2007{\natexlab{b}}){Hopkins}, {Lidz}, {Hernquist},
  {Coil}, {Myers}, {Cox}, \& {Spergel}}]{hopkins:clustering}
{Hopkins}, P.~F., {Lidz}, A., {Hernquist}, L., {Coil}, A.~L., {Myers}, A.~D.,
  {Cox}, T.~J., \& {Spergel}, D.~N. 2007{\natexlab{b}}, \apj, 662, 110

\bibitem[{{James} {et~al.}(1999){James}, {Bate}, {Wells}, {Wright}, \&
  {Doyon}}]{James99}
{James}, P., {Bate}, C., {Wells}, M., {Wright}, G., \& {Doyon}, R. 1999,
  \mnras, 309, 585

\bibitem[{{Jiang} \& {Kochanek}(2007)}]{jiang:lensing.baryon.frac}
{Jiang}, G., \& {Kochanek}, C.~S. 2007, \apj, 671, 1568

\bibitem[{{Jorgensen} {et~al.}(1993){Jorgensen}, {Franx}, \&
  {Kjaergaard}}]{jorgensen:fp.scatter}
{Jorgensen}, I., {Franx}, M., \& {Kjaergaard}, P. 1993, \apj, 411, 34

\bibitem[{{Jorgensen} {et~al.}(1996){Jorgensen}, {Franx}, \&
  {Kjaergaard}}]{jorgensen:fp}
---. 1996, \mnras, 280, 167

\bibitem[{{Joseph} \& {Wright}(1985)}]{joseph85}
{Joseph}, R.~D., \& {Wright}, G.~S. 1985, \mnras, 214, 87

\bibitem[{{Kannappan}(2004)}]{kannappan:gfs}
{Kannappan}, S.~J. 2004, \apjl, 611, L89

\bibitem[{{Katz} {et~al.}(1996){Katz}, {Weinberg}, \&
  {Hernquist}}]{katz:treesph}
{Katz}, N., {Weinberg}, D.~H., \& {Hernquist}, L. 1996, \apjs, 105, 19

\bibitem[{{Khochfar} \& {Silk}(2006)}]{khochfar:size.evolution.model}
{Khochfar}, S., \& {Silk}, J. 2006, \apjl, 648, L21

\bibitem[{{Kobayashi}(2004)}]{kobayashi:pseudo.monolithic}
{Kobayashi}, C. 2004, \mnras, 347, 740

\bibitem[{{Kormendy}(1977)}]{kormendy77:correlations}
{Kormendy}, J. 1977, \apj, 218, 333

\bibitem[{{Kormendy}(1985)}]{kormendy:spheroidal1}
---. 1985, \apj, 295, 73

\bibitem[{{Kormendy}(1987)}]{kormendy:spheroidal2}
{Kormendy}, J. 1987, in Nearly Normal Galaxies. From the Planck Time to the
  Present, ed. S.~M. {Faber}, 163--174

\bibitem[{{Kormendy}(1989)}]{kormendy:dissipation}
---. 1989, \apjl, 342, L63

\bibitem[{{Kormendy}(1999)}]{kormendy99}
{Kormendy}, J. 1999, in Astronomical Society of the Pacific Conference Series,
  Vol. 182, Galaxy Dynamics - A Rutgers Symposium, ed. D.~R. {Merritt},
  M.~{Valluri}, \& J.~A. {Sellwood}, 124--+

\bibitem[{{Kormendy} {et~al.}(2008){Kormendy}, {Fisher}, {Cornell}, \&
  {Bender}}]{jk:profiles}
{Kormendy}, J., {Fisher}, D.~B., {Cornell}, M.~E., \& {Bender}, R. 2008, \apj,
  submitted

\bibitem[{{Kormendy} \& {Freeman}(2004)}]{kormendyfreeman:scaling}
{Kormendy}, J., \& {Freeman}, K.~C. 2004, in IAU Symposium, Vol. 220, Dark
  Matter in Galaxies, ed. S.~{Ryder}, D.~{Pisano}, M.~{Walker}, \&
  K.~{Freeman}, 377--+

\bibitem[{{Kormendy} {et~al.}(2005){Kormendy}, {Gebhardt}, {Fisher}, {Drory},
  {Macchetto}, \& {Sparks}}]{kormendy:05}
{Kormendy}, J., {Gebhardt}, K., {Fisher}, D.~B., {Drory}, N., {Macchetto},
  F.~D., \& {Sparks}, W.~B. 2005, \aj, 129, 2636

\bibitem[{{Kormendy} \& {Sanders}(1992)}]{kormendysanders92}
{Kormendy}, J., \& {Sanders}, D.~B. 1992, \apjl, 390, L53

\bibitem[{{Kronawitter} {et~al.}(2000){Kronawitter}, {Saglia}, {Gerhard}, \&
  {Bender}}]{kronawitter:ml.models}
{Kronawitter}, A., {Saglia}, R.~P., {Gerhard}, O., \& {Bender}, R. 2000, \aaps,
  144, 53

\bibitem[{{Kuhlen} {et~al.}(2005){Kuhlen}, {Strigari}, {Zentner}, {Bullock}, \&
  {Primack}}]{kuhlen:concentrations}
{Kuhlen}, M., {Strigari}, L.~E., {Zentner}, A.~R., {Bullock}, J.~S., \&
  {Primack}, J.~R. 2005, \mnras, 357, 387

\bibitem[{{Laine} {et~al.}(2003){Laine}, {van der Marel}, {Lauer}, {Postman},
  {O'Dea}, \& {Owen}}]{laine:03}
{Laine}, S., {van der Marel}, R.~P., {Lauer}, T.~R., {Postman}, M., {O'Dea},
  C.~P., \& {Owen}, F.~N. 2003, \aj, 125, 478

\bibitem[{{Lake} \& {Dressler}(1986)}]{LakeDressler86}
{Lake}, G., \& {Dressler}, A. 1986, \apj, 310, 605

\bibitem[{{Lauer}(1985)}]{lauer:85}
{Lauer}, T.~R. 1985, \apjs, 57, 473

\bibitem[{{Lauer} {et~al.}(1995){Lauer}, {Ajhar}, {Byun}, {Dressler}, {Faber},
  {Grillmair}, {Kormendy}, {Richstone}, \& {Tremaine}}]{lauer:95}
{Lauer}, T.~R., {Ajhar}, E.~A., {Byun}, Y.-I., {Dressler}, A., {Faber}, S.~M.,
  {Grillmair}, C., {Kormendy}, J., {Richstone}, D., \& {Tremaine}, S. 1995,
  \aj, 110, 2622

\bibitem[{{Lauer} {et~al.}(2005)}]{lauer:centers}
{Lauer}, T.~R., {et~al.} 2005, \aj, 129, 2138

\bibitem[{{Lauer} {et~al.}(2007{\natexlab{a}})}]{lauer:bimodal.profiles}
---. 2007{\natexlab{a}}, \apj, 664, 226

\bibitem[{{Lauer} {et~al.}(2007{\natexlab{b}})}]{lauer:massive.bhs}
---. 2007{\natexlab{b}}, \apj, 662, 808

\bibitem[{{Leitherer} {et~al.}(1999)}]{starburst99}
{Leitherer}, C., {et~al.} 1999, \apjs, 123, 3

\bibitem[{{Liu} {et~al.}(2005){Liu}, {Zhou}, {Ma}, {Wu}, {Yang}, {Li}, \&
  {Chen}}]{liu:05}
{Liu}, Y., {Zhou}, X., {Ma}, J., {Wu}, H., {Yang}, Y., {Li}, J., \& {Chen}, J.
  2005, \aj, 129, 2628

\bibitem[{{Mandelbaum} {et~al.}(2006){Mandelbaum}, {Seljak}, {Kauffmann},
  {Hirata}, \& {Brinkmann}}]{mandelbaum:mhalo}
{Mandelbaum}, R., {Seljak}, U., {Kauffmann}, G., {Hirata}, C.~M., \&
  {Brinkmann}, J. 2006, \mnras, 368, 715

\bibitem[{{Maraston}(2005)}]{maraston:ssps}
{Maraston}, C. 2005, \mnras, 362, 799

\bibitem[{{McDermid} {et~al.}(2006)}]{mcdermid:sauron.profiles}
{McDermid}, R.~M., {et~al.} 2006, \mnras, 1312

\bibitem[{{McGaugh}(2005)}]{mcgaugh:tf}
{McGaugh}, S.~S. 2005, \apj, 632, 859

\bibitem[{{Mihos} \& {Hernquist}(1994{\natexlab{a}})}]{mihos:cusps}
{Mihos}, J.~C., \& {Hernquist}, L. 1994{\natexlab{a}}, \apjl, 437, L47

\bibitem[{{Mihos} \& {Hernquist}(1994{\natexlab{b}})}]{mihos:method}
---. 1994{\natexlab{b}}, \apj, 437, 611

\bibitem[{{Mihos} \& {Hernquist}(1994{\natexlab{c}})}]{mihos:starbursts.94}
---. 1994{\natexlab{c}}, \apjl, 431, L9

\bibitem[{{Mihos} \& {Hernquist}(1996)}]{mihos:starbursts.96}
---. 1996, \apj, 464, 641

\bibitem[{{Naab} {et~al.}(2007){Naab}, {Johansson}, {Ostriker}, \&
  {Efstathiou}}]{naab:etg.formation}
{Naab}, T., {Johansson}, P.~H., {Ostriker}, J.~P., \& {Efstathiou}, G. 2007,
  \apj, 658, 710

\bibitem[{{Navarro} {et~al.}(1996){Navarro}, {Frenk}, \& {White}}]{nfw:profile}
{Navarro}, J.~F., {Frenk}, C.~S., \& {White}, S.~D.~M. 1996, \apj, 462, 563

\bibitem[{{Neto} {et~al.}(2007)}]{neto:concentrations}
{Neto}, A.~F., {et~al.} 2007, \mnras, 381, 1450

\bibitem[{{Nipoti} {et~al.}(2003){Nipoti}, {Londrillo}, \&
  {Ciotti}}]{nipoti:dry.mergers}
{Nipoti}, C., {Londrillo}, P., \& {Ciotti}, L. 2003, \mnras, 342, 501

\bibitem[{{Nipoti} {et~al.}(2008){Nipoti}, {Treu}, \&
  {Bolton}}]{nipoti:homology.from.mp}
{Nipoti}, C., {Treu}, T., \& {Bolton}, A.~S. 2008, \mnras, in press
  arXiv:0806.0570 [astro-ph], 806

\bibitem[{{O{\~n}orbe} {et~al.}(2005){O{\~n}orbe}, {Dom{\'{\i}}nguez-Tenreiro},
  {S{\'a}iz}, {Serna}, \& {Artal}}]{onorbe:diss.fp.model}
{O{\~n}orbe}, J., {Dom{\'{\i}}nguez-Tenreiro}, R., {S{\'a}iz}, A., {Serna}, A.,
  \& {Artal}, H. 2005, \apjl, 632, L57

\bibitem[{{O'Shea} {et~al.}(2005){O'Shea}, {Nagamine}, {Springel}, {Hernquist},
  \& {Norman}}]{oshea:sph.tests}
{O'Shea}, B.~W., {Nagamine}, K., {Springel}, V., {Hernquist}, L., \& {Norman},
  M.~L. 2005, \apjs, 160, 1

\bibitem[{{Ostriker}(1980)}]{ostriker80}
{Ostriker}, J.~P. 1980, Comments on Astrophysics, 8, 177

\bibitem[{{Padmanabhan} {et~al.}(2004)}]{padmanabhan:mdyn.mstar.tilt}
{Padmanabhan}, N., {et~al.} 2004, New Astronomy, 9, 329

\bibitem[{{Pahre} {et~al.}(1998){Pahre}, {Djorgovski}, \& {de
  Carvalho}}]{pahre:nir.fp}
{Pahre}, M.~A., {Djorgovski}, S.~G., \& {de Carvalho}, R.~R. 1998, \aj, 116,
  1591

\bibitem[{{Peletier} {et~al.}(1990){Peletier}, {Davies}, {Illingworth},
  {Davis}, \& {Cawson}}]{peletier:profiles}
{Peletier}, R.~F., {Davies}, R.~L., {Illingworth}, G.~D., {Davis}, L.~E., \&
  {Cawson}, M. 1990, \aj, 100, 1091

\bibitem[{{Persic} \& {Salucci}(1988)}]{persic88}
{Persic}, M., \& {Salucci}, P. 1988, \mnras, 234, 131

\bibitem[{{Persic} \& {Salucci}(1990)}]{persic90}
---. 1990, \mnras, 245, 577

\bibitem[{{Persic} {et~al.}(1996{\natexlab{a}}){Persic}, {Salucci}, \&
  {Stel}}]{persic96:data}
{Persic}, M., {Salucci}, P., \& {Stel}, F. 1996{\natexlab{a}}, Astrophysical
  Letters Communications, 33, 205

\bibitem[{{Persic} {et~al.}(1996{\natexlab{b}}){Persic}, {Salucci}, \&
  {Stel}}]{persic96}
---. 1996{\natexlab{b}}, \mnras, 281, 27

\bibitem[{{Postman} \& {Lauer}(1995)}]{postmanlauer:95}
{Postman}, M., \& {Lauer}, T.~R. 1995, \apj, 440, 28

\bibitem[{{Prugniel} \& {Simien}(1997)}]{prugniel:fp.non-homology}
{Prugniel}, P., \& {Simien}, F. 1997, \aap, 321, 111

\bibitem[{{Ravindranath} {et~al.}(2004)}]{ravindranath:disk.size.evol}
{Ravindranath}, S., {et~al.} 2004, \apjl, 604, L9

\bibitem[{{Riciputi} {et~al.}(2005){Riciputi}, {Lanzoni}, {Bonoli}, \&
  {Ciotti}}]{riciputi:rotation.fp.tilt}
{Riciputi}, A., {Lanzoni}, B., {Bonoli}, S., \& {Ciotti}, L. 2005, \aap, 443,
  133

\bibitem[{{Rix} {et~al.}(1997){Rix}, {de Zeeuw}, {Cretton}, {van der Marel}, \&
  {Carollo}}]{rix97:dm}
{Rix}, H.-W., {de Zeeuw}, P.~T., {Cretton}, N., {van der Marel}, R.~P., \&
  {Carollo}, C.~M. 1997, \apj, 488, 702

\bibitem[{{Robertson} {et~al.}(2006{\natexlab{a}}){Robertson}, {Bullock},
  {Cox}, {Di Matteo}, {Hernquist}, {Springel}, \&
  {Yoshida}}]{robertson:disk.formation}
{Robertson}, B., {Bullock}, J.~S., {Cox}, T.~J., {Di Matteo}, T., {Hernquist},
  L., {Springel}, V., \& {Yoshida}, N. 2006{\natexlab{a}}, \apj, 645, 986

\bibitem[{{Robertson} {et~al.}(2006{\natexlab{b}}){Robertson}, {Cox},
  {Hernquist}, {Franx}, {Hopkins}, {Martini}, \& {Springel}}]{robertson:fp}
{Robertson}, B., {Cox}, T.~J., {Hernquist}, L., {Franx}, M., {Hopkins}, P.~F.,
  {Martini}, P., \& {Springel}, V. 2006{\natexlab{b}}, \apj, 641, 21

\bibitem[{{Robertson} {et~al.}(2006{\natexlab{c}}){Robertson}, {Hernquist},
  {Cox}, {Di Matteo}, {Hopkins}, {Martini}, \&
  {Springel}}]{robertson:msigma.evolution}
{Robertson}, B., {Hernquist}, L., {Cox}, T.~J., {Di Matteo}, T., {Hopkins},
  P.~F., {Martini}, P., \& {Springel}, V. 2006{\natexlab{c}}, \apj, 641, 90

\bibitem[{{Rothberg} \& {Joseph}(2004)}]{rj:profiles}
{Rothberg}, B., \& {Joseph}, R.~D. 2004, \aj, 128, 2098

\bibitem[{{Rothberg} \&
  {Joseph}(2006{\natexlab{a}})}]{rothberg.joseph:kinematics}
---. 2006{\natexlab{a}}, \aj, 131, 185

\bibitem[{{Rothberg} \&
  {Joseph}(2006{\natexlab{b}})}]{rothberg.joseph:rotation}
---. 2006{\natexlab{b}}, \aj, 132, 976

\bibitem[{{Rusin} \& {Kochanek}(2005)}]{rusin05:lensing.structure}
{Rusin}, D., \& {Kochanek}, C.~S. 2005, \apj, 623, 666

\bibitem[{{Rusin} {et~al.}(2003){Rusin}, {Kochanek}, \&
  {Keeton}}]{rusin03:lensing.structure}
{Rusin}, D., {Kochanek}, C.~S., \& {Keeton}, C.~R. 2003, \apj, 595, 29

\bibitem[{{Sanders} \& {Mirabel}(1996)}]{sanders96:ulirgs.mergers}
{Sanders}, D.~B., \& {Mirabel}, I.~F. 1996, \araa, 34, 749

\bibitem[{{Sargent} {et~al.}(1989){Sargent}, {Sanders}, \&
  {Phillips}}]{sargent89}
{Sargent}, A.~I., {Sanders}, D.~B., \& {Phillips}, T.~G. 1989, \apjl, 346, L9

\bibitem[{{Sargent} {et~al.}(1987){Sargent}, {Sanders}, {Scoville}, \&
  {Soifer}}]{sargent87}
{Sargent}, A.~I., {Sanders}, D.~B., {Scoville}, N.~Z., \& {Soifer}, B.~T. 1987,
  \apjl, 312, L35

\bibitem[{{Schmidt} \& {Allen}(2007)}]{schmidt:obs.concentrations}
{Schmidt}, R.~W., \& {Allen}, S.~W. 2007, \mnras, 379, 209

\bibitem[{{Scoville} {et~al.}(1986){Scoville}, {Sanders}, {Sargent}, {Soifer},
  {Scott}, \& {Lo}}]{scoville86}
{Scoville}, N.~Z., {Sanders}, D.~B., {Sargent}, A.~I., {Soifer}, B.~T.,
  {Scott}, S.~L., \& {Lo}, K.~Y. 1986, \apjl, 311, L47

\bibitem[{{Shankar} {et~al.}(2006){Shankar}, {Lapi}, {Salucci}, {De Zotti}, \&
  {Danese}}]{shankar06}
{Shankar}, F., {Lapi}, A., {Salucci}, P., {De Zotti}, G., \& {Danese}, L. 2006,
  \apj, 643, 14

\bibitem[{{Shapley} {et~al.}(2005){Shapley}, {Coil}, {Ma}, \&
  {Bundy}}]{shapley:z1.abundances}
{Shapley}, A.~E., {Coil}, A.~L., {Ma}, C.-P., \& {Bundy}, K. 2005, \apj, 635,
  1006

\bibitem[{{Shen} {et~al.}(2003){Shen}, {Mo}, {White}, {Blanton}, {Kauffmann},
  {Voges}, {Brinkmann}, \& {Csabai}}]{shen:size.mass}
{Shen}, S., {Mo}, H.~J., {White}, S.~D.~M., {Blanton}, M.~R., {Kauffmann}, G.,
  {Voges}, W., {Brinkmann}, J., \& {Csabai}, I. 2003, \mnras, 343, 978

\bibitem[{{Shier} \& {Fischer}(1998)}]{ShierFischer98}
{Shier}, L.~M., \& {Fischer}, J. 1998, \apj, 497, 163

\bibitem[{{Simien} \& {Prugniel}(1997)}]{simien:kinematics.1}
{Simien}, F., \& {Prugniel}, P. 1997, \aaps, 122, 521

\bibitem[{{Simien} \& {Prugniel}(2002)}]{simien:kinematics.6}
---. 2002, \aap, 384, 371

\bibitem[{{Soifer} {et~al.}(1984{\natexlab{a}})}]{soifer84a}
{Soifer}, B.~T., {et~al.} 1984{\natexlab{a}}, \apjl, 278, L71

\bibitem[{{Soifer} {et~al.}(1984{\natexlab{b}})}]{soifer84b}
---. 1984{\natexlab{b}}, \apjl, 283, L1

\bibitem[{{Springel}(2005)}]{springel:gadget}
{Springel}, V. 2005, \mnras, 364, 1105

\bibitem[{{Springel} {et~al.}(2005){Springel}, {Di Matteo}, \&
  {Hernquist}}]{springel:models}
{Springel}, V., {Di Matteo}, T., \& {Hernquist}, L. 2005, \mnras, 361, 776

\bibitem[{{Springel} \& {Hernquist}(2002)}]{springel:entropy}
{Springel}, V., \& {Hernquist}, L. 2002, \mnras, 333, 649

\bibitem[{{Springel} \& {Hernquist}(2003)}]{springel:multiphase}
---. 2003, \mnras, 339, 289

\bibitem[{{Springel} \& {Hernquist}(2005)}]{springel:spiral.in.merger}
---. 2005, \apjl, 622, L9

\bibitem[{{Thomas} {et~al.}(2005){Thomas}, {Maraston}, {Bender}, \& {Mendes de
  Oliveira}}]{thomas05:ages}
{Thomas}, D., {Maraston}, C., {Bender}, R., \& {Mendes de Oliveira}, C. 2005,
  \apj, 621, 673

\bibitem[{{Toomre}(1977)}]{toomre77}
{Toomre}, A. 1977, in Evolution of Galaxies and Stellar Populations, ed. B.~M.
  {Tinsley} \& R.~B. {Larson}, 401

\bibitem[{{Toomre} \& {Toomre}(1972)}]{toomre72}
{Toomre}, A., \& {Toomre}, J. 1972, \apj, 178, 623

\bibitem[{{Trager} {et~al.}(2000){Trager}, {Faber}, {Worthey}, \&
  {Gonz{\'a}lez}}]{trager:ages}
{Trager}, S.~C., {Faber}, S.~M., {Worthey}, G., \& {Gonz{\'a}lez}, J.~J. 2000,
  \aj, 119, 1645

\bibitem[{{Treu} {et~al.}(2005)}]{treu:fp.evolution}
{Treu}, T., {et~al.} 2005, \apj, 633, 174

\bibitem[{{Trujillo} {et~al.}(2004){Trujillo}, {Burkert}, \&
  {Bell}}]{trujillo:non-homology}
{Trujillo}, I., {Burkert}, A., \& {Bell}, E.~F. 2004, \apjl, 600, L39

\bibitem[{{Trujillo} {et~al.}(2006)}]{trujillo:size.evolution}
{Trujillo}, I., {et~al.} 2006, \apj, 650, 18

\bibitem[{{Vale} \& {Ostriker}(2006)}]{valeostriker:monotonic.hod}
{Vale}, A., \& {Ostriker}, J.~P. 2006, \mnras, 371, 1173

\bibitem[{{van den Bosch} {et~al.}(2007)}]{vandenbosch:concordance.hod}
{van den Bosch}, F.~C., {et~al.} 2007, \mnras, 376, 841

\bibitem[{{van der Marel}(1991)}]{vandermarel:ml.models}
{van der Marel}, R.~P. 1991, \mnras, 253, 710

\bibitem[{{van der Wel} {et~al.}(2005){van der Wel}, {Franx}, {van Dokkum},
  {Rix}, {Illingworth}, \& {Rosati}}]{vanderwel:fp.evolution}
{van der Wel}, A., {Franx}, M., {van Dokkum}, P.~G., {Rix}, H.-W.,
  {Illingworth}, G.~D., \& {Rosati}, P. 2005, \apj, 631, 145

\bibitem[{{van Dokkum} \& {van der Marel}(2007)}]{vandokkum:fp.evol}
{van Dokkum}, P.~G., \& {van der Marel}, R.~P. 2007, \apj, 655, 30

\bibitem[{{Vitvitska} {et~al.}(2002){Vitvitska}, {Klypin}, {Kravtsov},
  {Wechsler}, {Primack}, \& {Bullock}}]{vitvitska:spin}
{Vitvitska}, M., {Klypin}, A.~A., {Kravtsov}, A.~V., {Wechsler}, R.~H.,
  {Primack}, J.~R., \& {Bullock}, J.~S. 2002, \apj, 581, 799

\bibitem[{{von der Linden} {et~al.}(2007){von der Linden}, {Best}, {Kauffmann},
  \& {White}}]{vonderlinden:bcg.scaling.relations}
{von der Linden}, A., {Best}, P.~N., {Kauffmann}, G., \& {White}, S.~D.~M.
  2007, \mnras, 379, 867

\bibitem[{{Wang} {et~al.}(2006){Wang}, {Li}, {Kauffmann}, \& {de
  Lucia}}]{wang:sdss.hod}
{Wang}, L., {Li}, C., {Kauffmann}, G., \& {de Lucia}, G. 2006, \mnras, 371, 537

\bibitem[{{Wechsler} {et~al.}(2002){Wechsler}, {Bullock}, {Primack},
  {Kravtsov}, \& {Dekel}}]{wechsler:concentration}
{Wechsler}, R.~H., {Bullock}, J.~S., {Primack}, J.~R., {Kravtsov}, A.~V., \&
  {Dekel}, A. 2002, \apj, 568, 52

\bibitem[{{Weinmann} {et~al.}(2006){Weinmann}, {van den Bosch}, {Yang}, \&
  {Mo}}]{weinmann:obs.hod}
{Weinmann}, S.~M., {van den Bosch}, F.~C., {Yang}, X., \& {Mo}, H.~J. 2006,
  \mnras, 366, 2

\bibitem[{{Yang} {et~al.}(2005){Yang}, {Mo}, {Jing}, \& {van den
  Bosch}}]{yang:obs.clf}
{Yang}, X., {Mo}, H.~J., {Jing}, Y.~P., \& {van den Bosch}, F.~C. 2005, \mnras,
  358, 217

\bibitem[{{Younger} {et~al.}(2008){Younger}, {Hopkins}, {Cox}, \&
  {Hernquist}}]{younger:minor.mergers}
{Younger}, J.~D., {Hopkins}, P.~F., {Cox}, T.~J., \& {Hernquist}, L. 2008,
  \apj, submitted, arXiv:0804.2672 [astro-ph], 804

\bibitem[{{Zentner} {et~al.}(2005){Zentner}, {Berlind}, {Bullock}, {Kravtsov},
  \& {Wechsler}}]{zentner:substructure.sam.hod}
{Zentner}, A.~R., {Berlind}, A.~A., {Bullock}, J.~S., {Kravtsov}, A.~V., \&
  {Wechsler}, R.~H. 2005, \apj, 624, 505

\bibitem[{{Zheng} {et~al.}(2005)}]{zheng:hod}
{Zheng}, Z., {et~al.} 2005, \apj, 633, 791

\bibitem[{{Zirm} {et~al.}(2007)}]{zirm:drg.sizes}
{Zirm}, A.~W., {et~al.} 2007, \apj, 656, 66

\end{thebibliography}

\end{document}